\documentclass[a4paper,fleqn,usenatbib]{mnras}

\usepackage{newtxtext,newtxmath}

\usepackage{graphicx}
\usepackage{amsmath}
\usepackage{amssymb}
\usepackage[utf8]{inputenc}
\usepackage{graphics,epsfig,ulem,times}
\usepackage{pgfplots}
\usepackage[caption=false]{subfig}
\usepackage{xcolor}
\usepackage{xtab}

\usepackage{array}
\newcolumntype{C}[1]{>{\centering\arraybackslash}p{#1}}
\newcolumntype{?}{!{\vrule width 2.5\arrayrulewidth}}
\usepackage{multirow, makecell}
\usepackage[flushleft]{threeparttable}
\usepackage{caption}

\newcommand{\alphaCO}{\alpha_{\rm CO}}
\newcommand{\LCO}{L'_{\rm CO}}
\newcommand{\ICO}{I_{\rm CO}}
\newcommand{\CI}{[C\,{\sc i}] }
\newcommand{\CIfull}{[C\,{\sc i}]($^3$P$_1-^3$P$_0$)~}
\newcommand{\LCI}{L_{[\rm CI]}}
\newcommand{\tdep}{t_{\rm dep}}
\newcommand{\gdr}{\delta_{\rm gdr}}

\title[Molecular gas in dusty SFGs]{An ALMA/NOEMA survey of the molecular gas properties of high-redshift star-forming galaxies}

\author[J. Birkin et al.]{
Jack E.\ Birkin,$^{1}$\thanks{E-mail: jack.birkin@durham.ac.uk}
Axel Weiss,$^{2}$
J.\,L.\ Wardlow,$^{3}$
Ian Smail,$^{1}$
A.\,M.\ Swinbank,$^{1}$
\newauthor
U.\ Dudzevi{\v{c}}i{\={u}}t{\.{e}},$^{1}$
Fang Xia An,$^{4}$
Y.\ Ao,$^{5,6}$
S.\,C.\ Chapman,$^{7}$
Chian-Chou Chen,$^{8}$
\newauthor
E.\ da Cunha,$^{9}$
H.\ Dannerbauer,$^{10,11}$
B.\ Gullberg,$^{12}$
J.\,A.\ Hodge,$^{13}$
S.\ Ikarashi,$^{1}$
R.\,J.\ Ivison,$^{14}$
\newauthor
Y.\ Matsuda,$^{15,16}$
S.\,M.\ Stach,$^{1}$
F.\ Walter,$^{17}$
W.-H.\ Wang$^{7}$
and P.\ van der Werf$^{12}$
\\
$^{1}$Centre for Extragalactic Astronomy, Department of Physics, Durham University, South Road, Durham, DH1 3LE, UK\\
$^{2}$Max-Planck-Institut f{\"u}r Radioastronomie, Auf dem H{\"u}gel 69 D-53121 Bonn, Germany\\
$^{3}$Department of Physics, Lancaster University, Lancaster, LA1 4YB, UK\\
$^{4}$Inter-University Institute for Data Intensive Astronomy, University of the Western Cape, Robert Sobukwe Road, Bellville 7535, Cape Town, South Africa\\
$^{5}$Purple Mountain Observatory and Key Laboratory for Radio Astronomy, Chinese Academy of Sciences, Nanjing, China\\
$^{6}$School of Astronomy and Space Science, University of Science and Technology of China, Hefei, Anhui, China\\
$^{7}$Department of Physics and Atmospheric Science, Dalhousie University, Halifax, Halifax, NS B3H 3J5, Canada\\
$^{8}$Academia Sinica Institute of Astronomy and Astrophysics (ASIAA), No. 1, Section 4, Roosevelt Road, Taipei 10617, Taiwan\\
$^{9}$International Centre for Radio Astronomy Research, University of Western Australia, 35 Stirling Hwy, Crawley, WA 6009, Australia\\
$^{10}$Instituto de Astrofísica de Canarias (IAC), E-38205 La Laguna, Tenerife, Spain\\
$^{11}$Universidad de La Laguna, Dpto. Astrofísica, E-38206 La Laguna, Tenerife, Spain\\
$^{12}$Department of Space, Earth and Environment, Chalmers University of Technology, 41296 Gothenburg, Sweden\\
$^{13}$Leiden Observatory, Leiden University, P.O. box 9513, NL-2300 RA
Leiden, the Netherlands\\
$^{14}$European Southern Observatory, Karl Schwarzschild Strasse 2, D-85748, Garching, Germany\\
$^{15}$National Astronomical Observatory of Japan, 2-21-1 Osawa, Mitaka, Tokyo 181-8588, Japan\\
$^{16}$Department of Astronomy, School of Science, SOKENDAI (The Graduate University for Advanced Studies), Osawa, Mitaka, Tokyo 181-8588, Japan\\
$^{17}$Max-Planck-Institut f{\"u}r Astronomy, K{\"o}nigstuhl 17, D-69117 Heidelberg, Germany
}

\pubyear{2020}

\begin{document}

\setlength{\parskip}{0pt}
\label{firstpage}
\pagerange{\pageref{firstpage}--\pageref{lastpage}}
\maketitle

\begin{abstract}
We present a survey of the molecular gas in 61 submillimetre galaxies (SMGs) selected from 870\,$\mu$m continuum surveys of the COSMOS, UDS and ECDFS fields, using the Atacama Large Millimeter Array (ALMA) and the Northern Extended Millimeter Array (NOEMA). 46 $^{12}$CO ($J=$\,2--5) emission lines are detected in 45 of the targets at $z=$\,1.2--4.8, with redshifts indicating that those which are submillimetre bright and undetected/faint in the optical/near-infrared typically lie at higher redshifts, with a gradient of $\Delta z/\Delta S_{870}=$\,0.11\,$\pm$\,0.04\,mJy$^{-1}$.
We also supplement our data with literature sources to construct a statistical CO spectral line energy distribution and find the $^{12}$CO line luminosities in SMGs peak at $J_{\rm up}\sim$\,6, consistent with the Cosmic Eyelash, among similar studies.
Our SMGs lie mostly on or just above the main sequence, displaying a decrease in their gas depletion timescales $\tdep = M_{\rm gas}/{\rm SFR}$ with redshift in the range $z\sim$\,1--5 and a median of 200\,$\pm$\,50\,Myr at $z\sim$\,2.8. This coincides with an increase in molecular gas fraction $\mu_{\rm gas} = M_{\rm gas}/M_\ast$ across the same redshift range. 
Finally we demonstrate that the $M_{\rm baryon}$--$\sigma$ distribution of our SMGs is consistent with that followed by early-type galaxies in the Coma cluster, providing strong support to the suggestion that SMGs are progenitors of massive local spheroidal galaxies. On the basis of this we suggest that the SMG populations above and below an 870-$\mu$m flux limit of $S_{870}\sim$\,5\,mJy may correspond  to the division between slow- and fast-rotators seen in local early-type galaxies.
\end{abstract}




%
%
%
\section{Introduction}
\label{sec:intro}

It is believed that approximately half of all star formation and AGN activity that has ever occurred is obscured by dust \citep{puget96, dole06}, with this optical/UV light  absorbed and then re-emitted in the far-infrared \citep{blain02}. The most highly-obscured sources in the local Universe are Ultra-Luminous Infrared Galaxies (ULIRGs), galaxies with infrared luminosities greater than 10$^{12}L_\odot$, which were discovered by the {\it InfraRed Astronomy Satellite} \cite[{\it IRAS};][]{neugebauer84}. It was subsequently found that local ULIRGs typically have high star-formation rates (SFRs) $\gtrsim$\,50\,M$_\odot$yr$^{-1}$ that result from strong compression and cooling of the gas triggered by a major merger \citep[see][for a review]{sanders96}. In a cosmological context, while ULIRGs only contribute a small fraction of the global star-formation rate density (SFRD) at $z\sim0$, the picture is radically different at $z\gtrsim$\,1 \citep{magnelli13, dudzeviciute20}.  Understanding the processes which drive this population of dusty, strongly star-forming galaxies  at $z\gtrsim$\,1 is therefore  an important element in understanding galaxy evolution at high redshifts and high mass \citep{hodge20}. 

Among the high-redshift counterparts of ULIRGs are submillimetre galaxies \citep[SMGs;][]{smail97, hughes98} -- sources selected by their long-wavelength dust continuum emission, corresponding to flux densities of $\gtrsim$\,1\,mJy at 870\,$\mu$m, i.e.\ on the Rayleigh-Jeans tail of the dust spectral energy distribution (SED), where observations benefit from a negative $K$-correction. Surveys of SMGs are thus dust mass-limited across $z\sim$\,1--6, with a peak in space density at $z\sim$\,2--3 \citep{chapman05, pope06, wardlow11, weiss13}, i.e.\ around so-called ``Cosmic Noon'', at which time they are believed to account for a significant fraction of the global SFRD \citep{barger00, swinbank14, dudzeviciute20}. 

Representing a population that hosts some of the most actively star-forming systems that have ever existed, SMGs have provided a strong test of star formation and galaxy evolution models \citep{baugh05, bower06, dave10, mcalpine19, lagos20}. Their star-formation rates are typically estimated to be $\sim$\,100--1000\,M$_\odot$\,yr$^{-1}$ \citep{engel10, magnelli12a, swinbank14, dudzeviciute20} and their heavy dust obscuration results in the vast majority of their optical/UV light being re-emitted in the infrared, producing far-infrared luminosities of $\gtrsim$\,10$^{12}$--10$^{13}$\,L$_\odot$. Studies have shown that the star formation occurs in compact disks ($\sim$\,2--3\,kpc in diameter) of dust \citep{tacconi06, simpson15, ikarashi15, gullberg19, hodge19}, suggesting, like local ULIRGs, submillimetre galaxies may be triggered by  mergers or interactions \citep{mcalpine19}. It is also hypothesised that the SMG population are the progenitors of local spheroidal galaxies \citep[e.g.][]{sanders88, coppin08, simpson14}.

%
%
\begin{figure*}
    \centering
    \includegraphics[width=\linewidth]{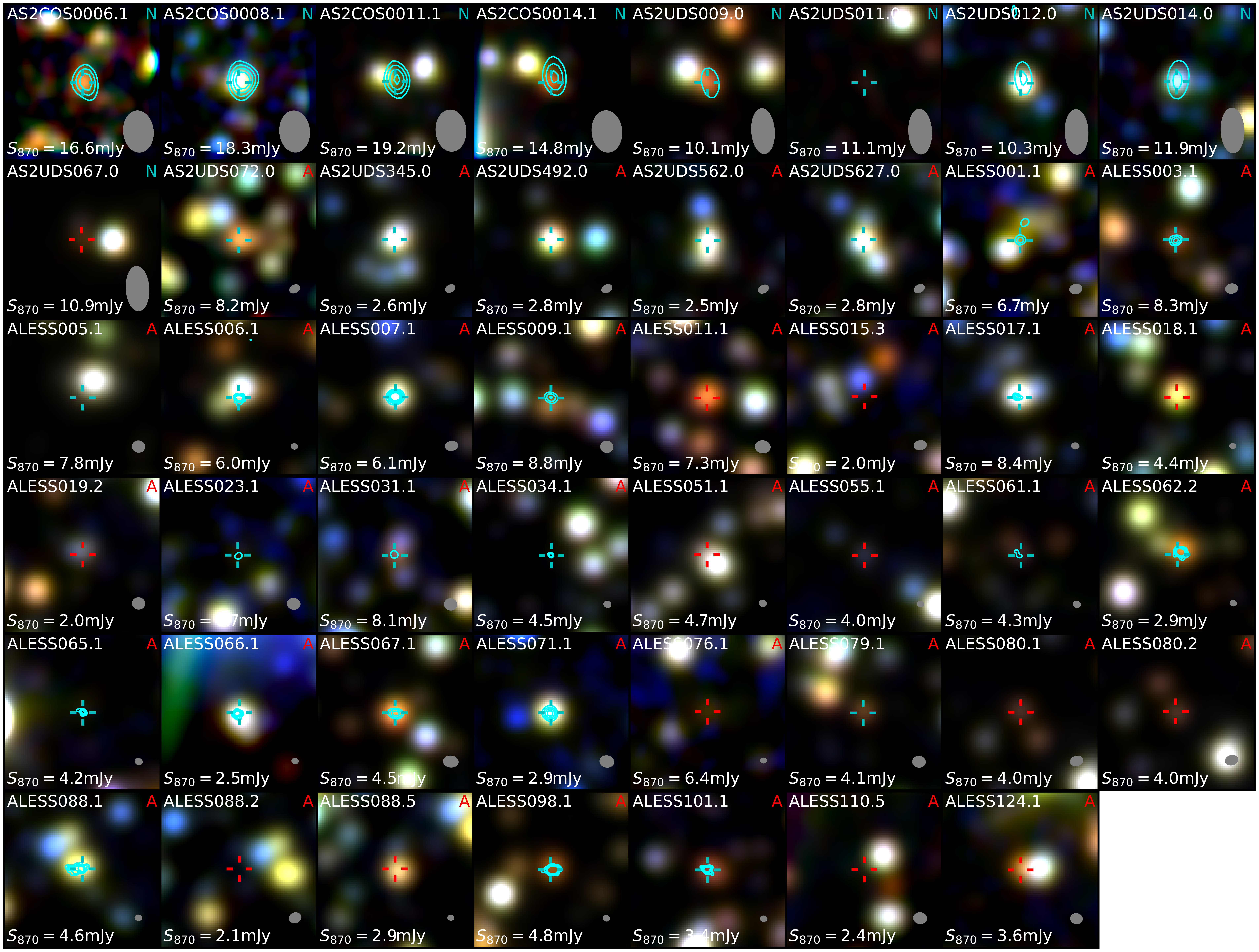}
    \caption{25$''$\,$\times$\,25$''$ ($\sim$\,200\,kpc at the median redshift of our sample) colour thumbnails composed of $K$-band, IRAC 3.6\,$\mu$m and IRAC 4.5\,$\mu$m images of the targets in our sample for which this imaging is available. We see that SMGs are in general redder than field galaxies, but this is not the case for all sources. Sources are ordered by 870\,$\mu$m flux density from top left to bottom right. The crosshair (cyan for CO-detected and red for CO non-detected) indicates the position of the 870\,$\mu$m emission detected by ALMA, with a typical beam size of $\sim$\,0.3--0.5$''$, the 870\,$\mu$m flux density of which is reported in each frame. The cyan contours represent CO emission at the 5-, 7-, 9- and 11-$\sigma$ levels. We indicate whether the target was observed with ALMA (A) or NOEMA (N) and show the synthesised beam in the top- and bottom-right corners, respectively. The ALMA/NOEMA 3\,mm beam sizes are typically 1.5$''$\,$\times$\,1$''$ and 8$''$\,$\times$\,4$''$, respectively.
}
    \label{fig:thumbnails}
\end{figure*}

Following rapid progress in the last decade, we are now in a position to study the population of SMGs statistically, with homogeneous samples of $\gtrsim$\,1000 sources having been catalogued from single-dish bolometer surveys and identified with ALMA \citep{hodge13, hatsukade16, miettinen17, cowie17, franco18, stach19}, the PdBI/NOEMA \citep{smolcic12} and SMA \citep{iono06, barger12, hill18}. Three  examples of such surveys, which are the focus of this work, are the ALMA SCUBA-2 Cosmic Evolution Survey (AS2COSMOS) \citep{simpson20}, ALMA SCUBA-2 Ultra Deep Survey (AS2UDS)  \citep{stach19} and ALMA LABOCA ECDFS Submillimetre Survey (ALESS) samples \cite{hodge13}. Analysis of the sources from such surveys has provided a wealth of information from  modelling of the multiwavelength spectral energy distributions (SEDs) of the SMGs using codes such as  {\sc magphys} \citep{dacunha15, miettinen17}, with the large sample size of AS2UDS in particular allowing us to derive robust statistical measurements of photometric redshifts, star-formation rates, infrared luminosities and many other properties \citep{dudzeviciute20}.

Two key observables needed to understand the evolution of high-redshift dust obscured galaxies are their  gas and dynamical masses: the former being the fuel for star formation, the main component of which is the molecular hydrogen H$_2$. Carbon monoxide (CO) emission is a standard tracer of H$_2$ which otherwise cannot be observed due to its lack of a permanent dipole moment, preventing any transitions from being appreciably excited in the cold interstellar medium (ISM) of SMGs \citep{solomon92, omont07, carilli13}.   Moreover, observations of   CO emission lines can provide insights into both galaxy gas masses, from the line luminosities, and also dynamical mass, from the line width -- where the CO emission has the added benefits of being relatively immune to the influences of dust obscuration and biases due to outflows or AGN activity, which plague many of the emission lines used to trace
dynamics in the restframe optical/UV \citep{swinbank06}.   

The first CO studies of SMGs were performed by \cite{frayer98, frayer99}, showing that these galaxies exhibit broad and often double-peaked CO lines, gas masses of order 10$^{10}$\,M$_\odot$, and short gas depletion timescales of $\tdep\sim$\,50\,Myr. Observations of the CO emission at high resolution showed that the SMG population displays a mix of sources with complex gas motions, indicative of  mergers, and sources with compact gas disks, which could be an indication of fuelling by steady gas accretion \citep{tacconi08, engel10, chen17}. Other early studies include \cite{greve05}, who found broad lines indicating dynamical masses of order 10$^{11}$\,M$_\odot$, \cite{daddi10}, who estimated gas fractions of $\sim$\,50--65 percent in similarly-luminous colour-selected galaxies at $z\sim$\,1.5, and \cite{ivison11}, who resolved the CO(1--0) emission from four SMGs with the Expanded Very Large Array, finding typical sizes of $\sim$\,16\,kpc.  In the first major CO survey of SMGs, \cite{bothwell13} studied the moderate-$J_{\rm up}$ CO emission in 40 SMGs with the Plateau de Bure Interferometer with 26 firm detections and six candidate detections, and used this to derive molecular gas masses, along with a median SLED for SMGs. This work provided useful constraints on the molecular emission, but the sample was limited by the reliance on sources with known spectroscopic redshifts and radio identifications, and hence was biased towards the optically-bright, lower-redshift and potentially AGN-dominated end of the population \citep{chapman05, hainline09, hainline11}.

However, one critical measurement we lack for many SMGs is precise spectroscopic redshifts \citep[although see][]{chapman05, danielson17}, an essential prerequisite for  understanding the properties of these galaxies. The current catalogue of redshifts for SMGs range from well-constrained spectroscopic redshifts for some optically-brighter sources to poorly constrained photometric redshifts for the  optically-faint/blank sources. As noted earlier,  CO emission is the most effective tracer of the gas and dynamical masses of galaxies, and the low- and mid-$J_{\rm up}$ transitions are redshifted to $\lambda\sim3$\,mm (at $z>$\,2) making them observable with (sub-)millimetre interferometers such as ALMA \citep[e.g.][]{wardlow18} and NOEMA \citep[previously the Plateau de Bure Interferometer;][]{neri03, daddi08, chapman15}, both of which have become powerful tools for  3\,mm ``blind'' scans, to determine redshifts for SMGs from their CO emission lines \citep{weiss09, swinbank10}, thanks to technological advancements allowing wider frequency coverage. For example, NOEMA has recently upgraded to a new wideband receiver and the {\it PolyFix} correlator \citep{broguiere20}, along with the addition of new antennae for greater collecting area, giving the instrument 16\,GHz of bandwidth. ALMA is the most powerful telescope of its kind, and can also achieve wider coverage with multiple frequency tunings of its 7.5\,GHz bandwidth. This means that we can search for CO emission from dust-obscured galaxies with no a priori knowledge of their redshift. As an example of the success rate of such studies, \cite{weiss13} conducted a blind 3\,mm ALMA survey of 26 strongly-lensed dusty star-forming galaxies selected at 1.4\,mm, using the South Pole Telescope (SPT), successfully detecting at least one CO, \CI or H$_2$O line in 23 of their targets \citep[also see e.g.][]{vieria13}.

With redshifts and masses for representative samples of SMGs we would be in a position  to place SMGs in the wider context of galaxy evolution.  In recent years research in this area has also begun to focus on the properties of the more ``typical'' high-redshift galaxies. These include the 
so-called ``main sequence'' population, which is defined in terms of the apparent correlation between stellar mass and star-formation rate \citep{noeske07a, whitaker12}. For submillimetre galaxies, which are usually considered to be ``starburst'' galaxies given their high infrared luminosities, it is particularly difficult to measure stellar masses  due to their heavy dust obscuration, and therefore it is not entirely clear where they lie in the SFR--$M_\ast$ plane \citep[e.g.][]{hainline11}. There is evidence, however, that some SMGs may in fact lie on the main sequence, with increasing frequency at high redshifts \citep{dacunha15, koprowski16, elbaz18, dudzeviciute20}. The implications of this for our understanding of the processes in SMGs, especially at higher redshifts, including the relative roles of triggering mechanisms in SMGs are unclear and will remain so until more sources in this regime are studied. For example, the existence of the main sequence has been interpreted to indicate that star formation in these galaxies is maintained by steady gas accretion, however more work is needed to understand this, especially as the main sequence itself is subject to selection effects \citep{hodge20}.

%
%
\begin{table}
    \centering
    \setlength\extrarowheight{1pt}
    \begin{tabular}{cccc} \hline \hline
    \noalign{\smallskip}
         \multicolumn{4}{c}{Number of targets}\\
          \hbox{~} & { \textit{{Spec-z}}} & { \textit{{Scan}}} & { Total} \\ \hline 
         AS2COSMOS & 0 & 5 & 5 \\
         AS2UDS & 4 & 13 & 17 \\
         ALESS & 26 & 13 & 39 \\ \hline
         Total sources & 30 & 31 & 61 \\ \noalign{\smallskip} \hline  \noalign{\smallskip}
         
         Median $S_{870}$ & 4.2 (2.6--6.0) & 8.8 (4.4--13.9) & 5.9 (2.8--10.5) \\
         Median $K$ & 21.2 (20.3--22.7) & 22.9 (22.1--23.7) & 22.3 (20.7--23.5) \\
         Median $V$ & 24.3 (22.9--25.4) & 26.0 (24.8--27.2) & 25.1 (23.8--26.8) \\ \noalign{\smallskip} \hline  \noalign{\smallskip}

         $N_{\rm detected, cont.}$ & 13 & 26 & 39 \\
         $N_{\rm detected, CO}$ & 19 & 26 & 45 \\ \noalign{\smallskip} \hline \hline \noalign{\smallskip}
    
    \end{tabular}
    \caption{Summary of our source selection and the 870\,$\mu$m fluxes of our subsamples. When reporting the median $S_{870}/K/V$ we also give the 16th--84th percentile ranges in brackets.}
    \label{tab:tab_select}
\end{table}

We have therefore undertaken a survey of 61 submillimetre galaxies with precise ALMA 870\,$\mu$m continuum detections in AS2COSMOS, AS2UDS and ALESS, with observations using ALMA and NOEMA in the 3\,mm band, aiming to characterise their molecular gas content and provide precise spectroscopic redshifts. To ensure our survey covers both a broad range in the population while containing statistically significant subsamples we combine two selection methods, including both a survey of typically submillimetre-bright SMGs lacking spectroscopic redshifts, which make ideal targets for blind CO scans,
and a study of generally fainter SMGs with pre-existing restframe optical/UV spectroscopic redshifts. Together these provide a sample with the wide range in 870\,$\mu$m flux ($S_{870}$) and optical/near-infrared brightness needed to study the properties of a representative and unbiased cross-section of the population. Our sample is one of the largest of its kind, and with it we take advantage of the unmatched sensitivity of ALMA/NOEMA and the wealth of multi-wavelength data available in our target fields. We will address the redshift distribution, gas excitation, dynamics and gas masses of SMGs, the evolution of their gas fractions and gas depletion timescales, along with their relation to the star-forming main sequence. As a study of similar size and intent, we will compare throughout to \cite{bothwell13}. 

The outline of this paper is as follows: in \S\ref{sec:observations} we outline the sample selection and observations carried out, along with our data reduction and analysis methods, before discussing the measurements made. In \S\ref{sec:results} we discuss the results and their implications. In \S\ref{sec:conclusions} we conclude our findings and discuss future work. Throughout this paper we use the AB magnitude system, a Chabrier IMF, and adopt a flat $\Lambda$-CDM cosmology defined by ($\Omega_\mathrm{m}$, $\Omega_\Lambda$, H$_0$) = (0.27, 0.73, 71\,km\,s$^{-1}$\,Mpc$^{-1}$).

\section{Observations and Data Analysis}
\label{sec:observations}

\subsection{Sample selection}

Our 61 targets are selected from ALMA-identified SMGs in the ALMA-SCUBA-2 Cosmic Evolution Survey \citep[AS2COSMOS;][]{simpson20}, the ALMA-SCUBA-2 Ultra Deep Survey \citep[AS2UDS;][]{stach19} and the ALMA-LABOCA ECDFS Submillimetre Survey \citep[ALESS;][]{hodge13}. These targets are divided into two subsets:
\begin{enumerate}
    \item \textbf{\textit{Scan} sample}: 30 sources were targeted with blind scans in the 3\,mm band. These sources have the brightest 870\,$\mu$m fluxes in the three SMG catalogues (see Table~\ref{tab:tab_select}), that are also expected to lie on or near the star-forming main sequence at $z\sim$\,2--5. The brightness of these sources at 870\,$\mu$m suggests that they will also be bright CO emitters, but they have poorly constrained redshifts. Therefore we have scanned the 3\,mm band to detect their CO emission using multiple tunings. Of these 30 sources, five are drawn from AS2COSMOS (with $S_{870}=$\,15--20\,mJy), 13 from AS2UDS (with $S_{870}=$\,7--13\,mJy) and 12 from ALESS (with $S_{870}=$\,2--9\,mJy). The relative brightness of the sources in part reflects the survey volume of the corresponding three fields.
    \item \textbf{\textit{Spec-z} sample}: 30 sources with restframe optical/UV spectroscopic redshifts. Four of these sources are taken from AS2UDS \citep{dudzeviciute20}, and the remaining 26 are taken from ALESS \citep{danielson17}. These sources are typically brighter in the optical and near-infrared, and fainter in the submillimetre than the {\it scan} sample (see Table~\ref{tab:tab_select}).
\end{enumerate}
A breakdown of this selection is displayed in Table~\ref{tab:tab_select}, and a list of source properties are given in Table~\ref{tab:observed}. We reiterate here that the aim of this study is to provide an analysis of the molecular gas in submillimetre galaxies, building on the work highlighted in \S\ref{sec:intro} with a large sample of high quality data. We will, for the majority of this analysis, consider the entire sample as one, noting that the wide range in photometric redshift, 870\,$\mu$m flux and $K$-band magnitude of our targets make the sample well suited for studying correlations in the properties of the population.

\subsection{Observations and data reduction}

Observations were obtained from six projects, four with ALMA and two with NOEMA/{\it PolyFix}, between 2017 and 2020. 15 targets from the {\it scan} sample, five from AS2COSMOS and ten from AS2UDS, were observed with NOEMA/{\it PolyFix} in projects S18CG and W18EL.
Targets were observed with two spectral setups of two pairs of 8\,GHz sidebands to achieve a total bandwidth of 32\,GHz covering $\sim$\,82--114\,GHz, with an integration time of 1.5\,hours per setup using the combined CD configuration which is suitable for low-resolution detection experiments. Reduction of the data was carried out using the {\sc gildas} software. The raw data were calibrated using standard pipelines, with bad visibilities flagged and removed in the process. For bandpass and flux calibration we observed J1018+055, 0906+015 and J0948+003 for AS2COSMOS sources and 0238$-$084, 0215+015 and J0217$-$083 for AS2UDS sources. Calibrated $uv$ tables were imaged using natural weighting with the {\sc mapping} routine in {\sc gildas}, and the resultant dirty cubes were outputted to {\sc fits} format for analysis with our own {\sc python} routines. 
Typical synthesised beam sizes for the NOEMA data are 8$''$\,$\times$\,4$''$ at 3\,mm, with the observations achieving a typical RMS depth of 0.7\,mJy.

The remaining 46 targets were observed with ALMA in projects 2016.1.00564.S, 2017.1.01163.S, 2017.1.01512.S and 2019.1.00337.S. 16 of the targets in the {\it scan} sample (three from AS2UDS, 13 from ALESS) were observed using five tunings to achieve 32\,GHz of bandwidth covering $\sim$\,82--114\,GHz, with integration times of $\sim$\,15 minutes per tuning. All 30 targets in the {\it spec-z} sample were observed using a single tuning centred on the frequency of the CO line expected in the 3\,mm band (ALMA band 3). Integration times ranged from $\sim$\,25--40 minutes. All of the above were carried out using the 12\,m array in compact configurations. Reduction of the data was carried out using the {\sc common astronomy software applications} \citep[{\sc casa};][]{casa} software, employing standard pipelines to produce naturally-weighted dirty cubes, which we then outputted to {\sc fits} format for analysis with our own {\sc python} routines. For bandpass and flux calibration we observed J0423$-$0120, J0238+1636, J0217$-$0820 for AS2UDS sources and J0522$-$3627, J0342$-$3007, J0317$-$2803 and J0334$-$4008 for ALESS sources. Typical synthesised beam sizes for the ALMA data are 1.5$''$\,$\times$\,1.0$''$, with the observations achieving a typical RMS depth of 0.3\,mJy.

\subsection{Line detection}
\label{sec:detections}

From our reduced datacubes we extract spectra in an aperture centred on the position of the 870\,$\mu$m emission. As our sample contains (marginally) resolved and unresolved sources we adopt two separate recipes for determining line and continuum fluxes. For sources in the {\it scan} sample, which are typically unresolved in the lower angular resolution observations, we use an aperture of diameter 1.3 times the FWHM of the synthesised beam, and multiply the flux by a factor of two, a point source correction to the flux derived from observations of the calibrators. For sources in the {\it spec-z} sample, which are higher resolution, we use an aperture of diameter three times the FWHM of the synthesised beam to ensure all the flux is captured while maintaining a high S/N. We also collapse the cubes to create a 3\,mm continuum image and check for any offset between the 870\,$\mu$m and 3\,mm continuum emission that could result in the aperture not encapsulating all of the line flux. If an offset is discovered we measure the position of the 3\,mm source and extract spectra from this position instead. This is required for six sources, with a median shift of 0.35$''$.
 
To search for CO emission from the 870\,$\mu$m-detected SMG we first estimate the noise in the cubes by extracting spectra in equivalent apertures from 100 random positions within the primary beam (masking the 3\,mm source) and calculating their RMS noise. We then generate a histogram of channel signal-to-noise ratios (S/N) in the positive and inverted spectra in order to determine a S/N cut and corresponding false positive rate. This is done using spectra that are continuum-subtracted with a running median (choosing an averaging window large enough so as not to be influenced by any line emission) and rebinned to channel widths of 300, 600 and 900\,kms$^{-1}$. We adopt S/N cuts of 4, 3.75 and 3.5 for these channel widths based on the requirement that there are no false positives in our sample. For sources in the {\it spec-z} sample we search for 3.5-$\sigma$ emission within 100\,kms$^{-1}$ of the frequency of the optical spectroscopic redshift. Following \cite{wardlow18} we also perform a blind search of the 3\,mm cubes for serendipitous CO/continuum emitters. This is done by spatially rebinning to ensure Nyquist sampling, and spectrally rebinning to channel widths of 150, 300 or 600\,kms$^{-1}$, then searching the cubes for $>$\,5-$\sigma$ channels within the primary beam.

%
%
\begin{figure*}
\begin{center}
\includegraphics[width=\textwidth]{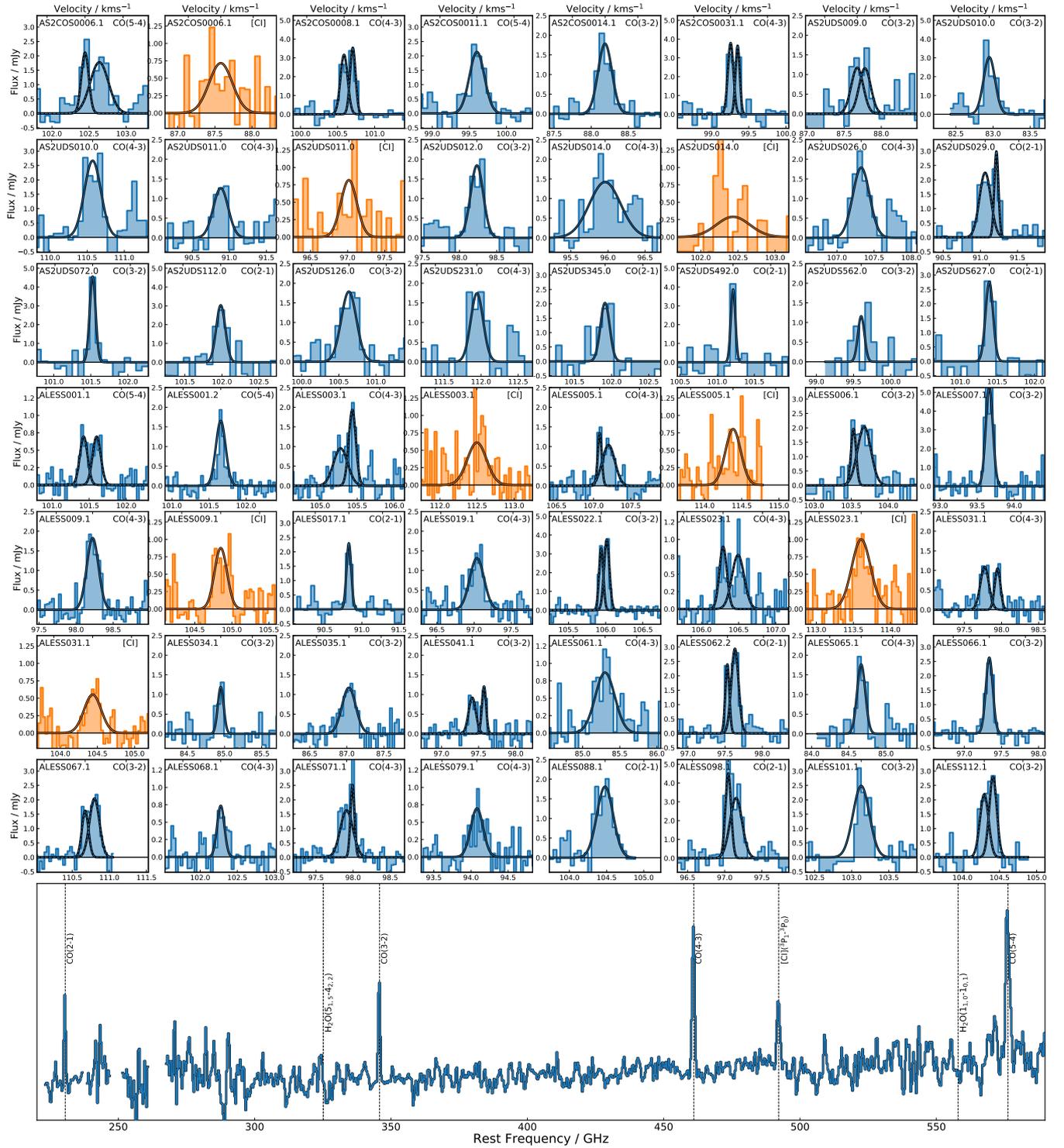}
\caption{\small{Emission line detections in the continuum-subtracted 3\,mm spectra of our sample of SMGs,  with the fit to each line overlaid. In total, we detect 56 emission lines, 46 CO lines from our 61 targets with $J_{\rm up}=$\,2--5 (blue), two CO lines in nearby ALMA-detected SMGs and eight \CIfull lines (orange).
CO is detected at high S/N (median S/N\,$\sim$\,8). We fit and plot single- and double-Gaussian profiles to each line, finding that 38\,$\pm$\,9 per cent display double-Gaussian profiles, indicative of disk dynamics or multiple components in these sources. The bottom panel shows a median composite of all CO-detected spectra in the rest frame, clearly showing the CO ladder and \CI lines. We also indicate where the rotational transitions of H$_2$O would appear, however we see no trace of these emission lines (see \S\ref{sec:properties}). All spectra are binned to a velocity resolution of $\sim$\,150\,kms$^{-1}$, and tick marks on the top axis in each panel represent $-$1500\,km\,s$^{-1}$, 0\,km\,s$^{-1}$ and 1500\,km\,s$^{-1}$ from left to right, respectively.}}
\label{fig:spectra}
\end{center}
\end{figure*}

This results in a total of 45 of our 61 targets displaying CO emission (74 per cent), 26 out of 31 (84 per cent) from the {\it scan} sample and 19 out of 30 (63 per cent) from the {\it spec-z} sample, the latter of which are typically fainter at 870\,$\mu$m (see Table~\ref{tab:tab_select}). We also detect CO in two nearby ALMA-detected SMGs not targeted in this survey, bringing the total number of CO-detected sources in our sample to 47 (and in the {\it scan} sample to 28). Three serendipitous CO emitters are uncovered. The median S/N of our CO line detections is 8.3\,$\pm$\,0.6.

\subsection{Line identification}

For the {\it scan} sample, where redshifts are not known a priori, galaxies at $z>$\,3 are expected to display either two CO lines or one CO line and the \CI($^3$P$_1-^3$P$_0$) line in our frequency coverage, in which cases identifying the detected transition is trivial. In contrast galaxies at $z\lesssim$\,3 are only expected to display one line meaning that there may be some ambiguity in identifying the transition. In the latter case we use the redshift probability distribution functions (PDFs) from SED fitting with the photometric redshift extension of the {\sc magphys} code \citep{battisti19} to determine the most likely redshift, given the observed frequency of the line. {\sc magphys} uses an energy balance technique to ensure that the UV/optical light absorbed by dust is all re-radiated in the infrared. We refer the reader to \cite{dacunha08, dacunha15} for a more comprehensive discussion of {\sc magphys} and the energy balance technique, and \cite{battisti19} for details on the photometric redshift extension of {\sc magphys}. For further details of the photometry used we refer the reader to \cite{simpson20} for sources in AS2COSMOS, \cite{dudzeviciute20} for AS2UDS and \cite{dacunha15} for ALESS.

Of the 28 {\it scan} sources in which we detect CO, one displays two CO emission lines (AS2UDS010.0) and eight display an additional \CIfull emission line, therefore nine of the 28 redshifts are unambiguous and correspond to
$J_{\rm up}=$\,4 or 5. From the 19 {\it spec-z} sources to display CO emission, 18 are detected at the expected redshift and are therefore identified unambiguously, with the remaining one source (ALESS088.1) displaying emission offset from the expected frequency by $\sim$\,3\,GHz ($\sim$\,8500\,kms$^{-1}$). Therefore a total of 27 out of 47 redshifts (57 per cent) in our sample are unambiguous.

This leaves a total of 20 sources with a single CO line. We use the   {\sc magphys} redshift PDFs to identify these 20 transitions. Firstly, we test the ability of {\sc magphys} to predict the correct line identification: we select a subsample of 16 of the 27 cases where the line transition is unambiguous and the SMGs have  $K<$\,23, where this limit is chosen to ensure this training set is matched in $K$-band brightness with the ambiguous line source sample.  We then identify the probabilities for the two most likely CO transitions based on the PDF, including a prior to weight the selection to the lower-$J_{\rm up}$ line in the event that the two lines are close in likelihood.   Based on this test we recover the correct transition for 14 out of 16 (88 per cent) of these sources.   
We then apply the same methodology to the 20 single CO line sources and estimate that these comprise: three $J_{\rm up}=$\,5, seven $J_{\rm up}=$\,4, eight $J_{\rm up}=$\,3 and two $J_{\rm up}=$\,2.  We confirm that when a higher-$J_{\rm up}$ CO line is chosen that this does not conflict with the absence of a second CO or \CI line which is predicted to be observable.
We note that the success rate from the test of PDF-based line selection would suggest that in our sample of 47 sources, with 20 ambiguous line identifications, we expect $\sim$\,2--3 redshifts to be incorrect.  We assess the impact of this on our results in the following by randomly removing 2--3 of the sources in the ambiguous sample from our analysis and we confirm that this does not change any of the claimed results outside their quoted 16--84\,th percentile confidence ranges.

\subsection{SED fitting}
\label{sec:sed}

After identifying the detected transitions we fit SEDs to our sources with the high-redshift version of {\sc magphys}, this time including our 3\,mm continuum measurement (or limit) and fixing the redshift as that measured based on our adopted CO transition, in order to derive key physical parameters of our sources. We note that this code does not take into account contributions to the continuum emission from AGN, however there is little evidence that AGN emission contaminates the optical or infrared emission of  most SMGs \citep{stach19}, particularly those with the most massive gas reservoirs, as we expect to detect here. In most cases, {\sc magphys} provides a good fit to the photometry, however we note ALESS071.1, which has an unusually high best-fit stellar mass of $M_\ast\sim$\,2$\times$\,10$^{12}$\,M$_\odot$ at the CO redshift ($z_{\rm CO}=$\,3.707, $J_{\rm up}=4$), which is secure as it agrees with the optical/UV spectroscopic redshift. The SED appears to be reasonably fit by this model, nevertheless we attempted to fit the source at redshifts corresponding to the $J_{\rm up}=$\,2, 3 and 5 transitions, but this did not improve the result. As the MIPS 24\,$\mu$m photometry does not suggest the presence of an AGN, we view it as likely that this source is lensed, or contaminated by a projected foreground source (see \ref{fig:thumbnails}). As a consequence, we flag this source in figures throughout the paper where it appears as a noticeable outlier, and confirm that it does not bias our conclusions.

We also note that for our CO sample, running {\sc magphys} with the spectroscopic redshift fixed does not result in any significant change of the parameters when compared to those found from running the photometric redshift extension of the code \citep{dacunha15, danielson17, dudzeviciute20}, although it does reduce their uncertainties.   Nevertheless, we caution that the
stellar masses derived  from the SED fitting  are likely to be subject to systematic uncertainties of a factor of $\sim$\,2--3, due to uncertainties in the constraints on the star-formation histories \citep{hainline11, dudzeviciute20}. The median properties of the whole sample found from SED fitting are $L_{\rm IR}=($5.0\,$\pm$\,1.0$)\times$\,10$^{12}$\,L$_\odot$\footnote{$L_{\rm IR}$ is measured across the range $\lambda=$\,8--1000\,$\mu$m.}, $M_\ast=(2.2\pm0.2)\times10^{11}$\,M$_\odot$, $M_{\rm dust}=($1.08\,$\pm$\,0.18$)\times$10$^{9}$\,M$_\odot$, and ${\rm SFR}=$\,400\,$\pm$\,50\,M$_\odot$\,yr$^{-1}$. The best-fit parameters for the sources are listed in Table~\ref{tab:measured}.

\subsection{Line fitting}
\label{sec:fitting}

We simultaneously fit single-/double-Gaussian profiles plus a continuum level to the lines recovered from our spectra, employing a Markov Chain Monte Carlo (MCMC) technique implemented in the \textsc{emcee} package of {\sc python} \citep{foreman-mackey13}. For sources in the {\it scan} sample, the spectral slope becomes significant over the 32\,GHz bandwidth, therefore we fit a power law continuum, rather than just a constant continuum as with the {\it spec-z} sources (which only have narrow frequency coverage). Uncertainties on the fits are calculated by fitting bootstrapped spectra and measuring the dispersion in the resultant parameter distributions. To determine whether the single- or double-Gaussian profile is the more suitable fit we compute the Akaike Information Criterion \citep[AIC;][]{akaike1974}, which penalises models for using a large number of parameters to obtain a good fit, and take the model with the lowest AIC to be the most appropriate. The continuum-subtracted spectra and line fits are shown in Fig.~\ref{fig:spectra}, and the corresponding fit parameters are tabulated in Table~\ref{tab:measured}.

We now measure the properties of our CO lines. While many of our sources are well described by Gaussian profiles, we use the intensity-weighted moments of the spectrum to obtain a profile-independent measurements \citep{bothwell13}. To ensure consistency in all measurements, we employ the same method of deriving moments  regardless of whether the line profile is deemed to be single- or double-peaked. The zeroth moment gives the intensity of the line:
\begin{equation}
M_0 = I_{\rm CO} = \int I_v{\rm d}v,
\end{equation}
where $I_v$ is the flux in a channel with velocity $v$. The first moment gives the centroid of the line:
\begin{equation}
M_1 = \Bar{v} = \dfrac{\int vI_v{\rm d}v}{\int I_v{\rm d}v}
\end{equation}
which we use to calculate the redshift. The second moment is the velocity dispersion, from which we can estimate the equivalent full-width at half-maximum (FWHM) as:
\begin{equation}
{\rm FWHM} =  2\sqrt{2\ln{2}}M_2 = 2\sqrt{2\ln{2}}~\sqrt{\dfrac{\int\left(v-\Bar{v}\right)^2I_v{\rm d}v}{\int I_v{\rm d}v}}.
\label{eq:fwhm}
\end{equation}
To calculate moments we integrate the spectra in a velocity window twice the FWHM of the Gaussian fit. We confirm this range based on simulations where we insert Gaussians with known amplitudes and linewidths at random frequencies in our spectra and attempt to recover the input value using Eq.~\ref{eq:fwhm}. Uncertainties on the second moment are estimated by resampling the spectrum with the noise spectrum, then calculating the dispersion in the recovered line widths.

We note that the CO line emission in ALESS101.1 falls onto a band gap meaning that a number of channels are missing from the line. In this case summing channels across the line results in underestimates of the linewidth and line flux, and we therefore use the properties of the Gaussian fit when deriving these quantities.

Finally, we derive the CO line luminosity of the observed transition
\begin{equation}
L'_{\rm CO,J} = 3.25 \times 10^7 I_\mathrm{CO,J} \nu_\mathrm{obs}^{-2} D_L^2 (1+z)^{-3},
\label{eq:LCO}
\end{equation}
where $L'_{\rm CO,J}$ is in units of K\,km\,s$^{-1}$ pc$^2$, $I_{\rm CO,J}$ is the velocity-integrated intensity of the line, $\nu_{\rm obs}$ is the observed frequency of the line, $D_L$ is the luminosity distance of the source in Mpc, calculated using our chosen cosmology, and $z$ is the redshift \citep{solomon05}. The \CIfull line luminosity $L'_{\rm [CI]}$ is calculated in the same way. Due to these typically being fainter lines, the frequency of the \CIfull line is fixed to be at the CO redshift when fitting Gaussians, and the \CI linewidth is fixed to be equivalent to that of the CO. We still derive the linewidth using the moments of the spectrum as with the CO (see \S\ref{sec:fitting}).  These spectra are shown in Fig.~\ref{fig:spectra}.

\subsection{Literature data}
\label{sec:literature}

In addition to the data detailed above, we also include measurements of CO and \CI luminosities and linewidths, redshifts, infrared luminosities, stellar masses and star-formation rates for unlensed SMGs from the literature. These are taken from \cite{bothwell13} and the following: \cite{alloin00}, \cite{andreani00}, \cite{aravena10}, \cite{aravena12}, \cite{calistrorivera18}, \cite{carilli10}, \cite{carilli11}, \cite{casey09}, \cite{casey11}, \cite{chapman08}, \cite{coppin12}, \cite{daddi09}, \cite{daddi10}, \cite{dannerbauer09}, \cite{genzel10}, \cite{greve03}, \cite{huynh17}, \cite{ivison11}, \cite{magdis12}, \cite{magnelli12}, \cite{riechers10}, \cite{schinnerer08}, \cite{scoville97}, \cite{stacey10}, \cite{swinbank12}, \cite{tacconi10}, \cite{walter12}, \cite{wardlow18} and \cite{yan10}.

These sources are used in Fig.~\ref{fig:SLED} and Fig.~\ref{fig:LCO}. Where appropriate we scale CO luminosities according to our adopted line ratios, and gas masses according to our adopted CO--H$_2$ conversion factor (see \S\ref{sec:gas_masses} and \S\ref{sec:evo}).

%
%
\begin{figure*}
\begin{center}
\includegraphics[width=0.95\linewidth]{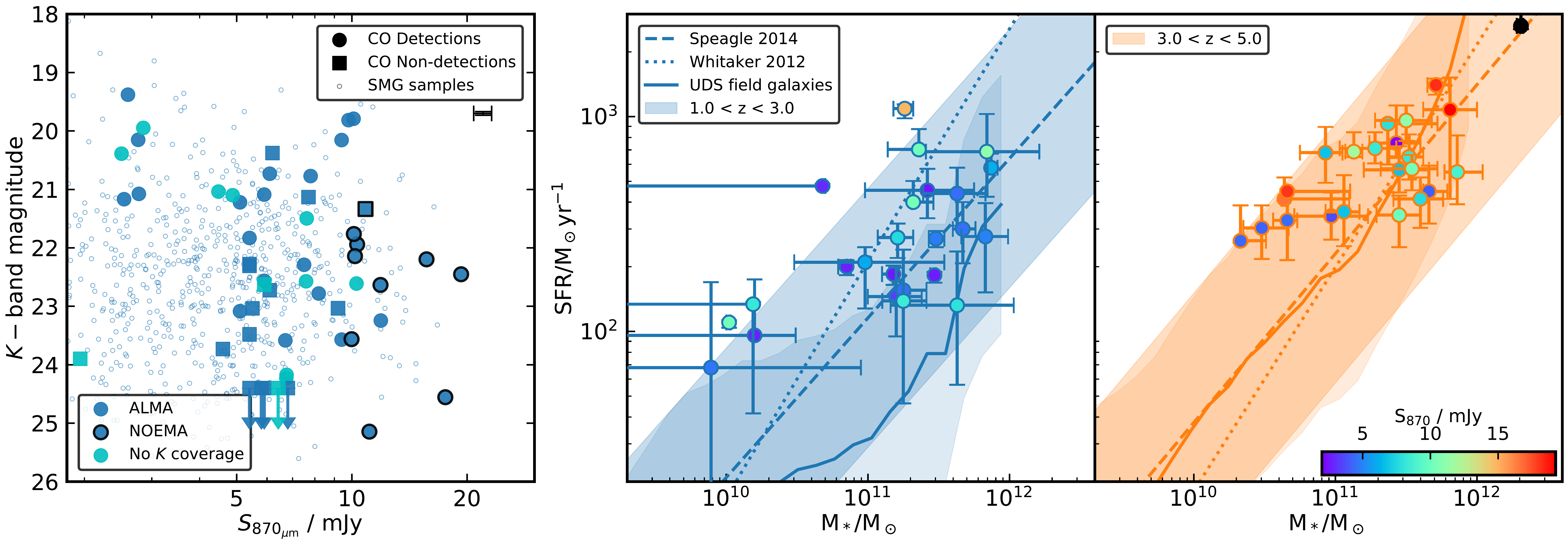}
\caption{\small{{\it Left:} $K$-band magnitude versus 870\,$\mu\mathrm{m}$ flux density for all AS2COSMOS, AS2UDS and ALESS SMGs. Filled markers represent sources targeted for this work and symbol shapes differentiate CO detections from CO non-detections. ALMA/NOEMA observations are differentiated by the symbol outline. Our sample covers the range of $K$ (median $K=$\,22.3; 16--84th percentile range 20.7--23.5) spanned by the SMG population, while we typically select sources that are bright at 870\,$\mu$m (median $S_{\rm 870}=$\,5.9\,mJy; 16--84th percentile range 2.8--10.5\,mJy). 3-$\sigma$ upper limits for non-detections are plotted, and we show a representative error bar for the whole population in the top-right corner. Four sources are undetected in the $K$-band and 14 have no $K$-band photometry. In the latter cases we estimate $K$ from the typical $K-$3.6\,$\mu$m colour at the appropriate redshift, where IRAC coverage is available (cyan points). {\it Middle and right:} The relation of our CO-detected sources to the star-forming main sequence at $z=$\,1--3 and $z=$\,3--5. We show the main sequence as predicted by three different prescriptions, and highlight a spread of a factor four in each case (0.6 dex), above which a galaxy is considered to be a starburst. 43 of our 47 CO-detected sources lie within this range. SMGs have been typically difficult to characterise with respect to the main sequence, but we show that with our precise CO redshifts we have been able to derive stellar masses and SFRs robust enough to securely place our sources on or near the main sequence, particularly at high redshift.}}
\label{fig:KS870}
\end{center}
\end{figure*}

\section{Results and Discussion}
\label{sec:results}

\subsection{CO detections}
\label{sec:properties}

We detect CO emission from a total of 47 sources: 45 of the 61 targets (74 per cent) in addition to two ALMA-identified SMGs that are close to one of the target sources. Of the 45 target SMGs to display CO emission, 26 out of 31 (84 per cent) are from the {\it scan} sample and 19 out of 30 (63 per cent) are from the {\it spec-z} sample. In total, we detect 51 CO emission lines in the range $J_{\rm up}=$\,2--5 (including three serendipitous emitters and a second CO line in AS2UDS010.0) and eight \CIfull emission lines. We overlay the CO contours of these sources onto $K$/IRAC 3.6\,$\mu$m/IRAC 4.5\,$\mu$m colour images (where imaging is available), which are displayed in Fig.\ref{fig:thumbnails}. Due to the chosen configurations of our observations we do not resolve the CO in most cases (the synthesised beam is shown in each panel), however a number of the ALESS {\it spec-z} targets were observed at higher resolution and display some structure (see e.g.\ ALESS098.1). High-resolution imaging for some of our CO sources has been presented in \cite{chen17},  \cite{calistrorivera18} and \cite{wardlow18}, with the results indicating disk dynamics and/or merger-like morphologies for the CO. The line profiles of all CO and \CI emission lines (not including serendipitous emitters), along with their single-/double-Gaussian fits, are displayed in Fig.~\ref{fig:spectra}. CO is detected with high signal-to-noise in the majority of targets, with a median S/N of 8.3\,$\pm$\,0.6, and exhibits a variety of line profiles. Our lines have a median linewidth of 540\,$\pm$\,50\,km\,s$^{-1}$, comparable with that of \cite{bothwell13}, who found a value of 510\,$\pm$\,80\,km\,s$^{-1}$. Our sources also have comparable infrared luminosities: both samples have a median $L_{\rm IR} = ($5\,$\pm$\,1$)\times$\,10$^{12}$\,L$_\odot$.

We perform a median restframe stack of all 47 spectra with CO detections to search for other weak emission lines, which is shown in Fig.~\ref{fig:spectra}\,. Other than CO emission with $J_{\rm up}=$\,2--5 and the [C{\sc i}]($^3$P$_1$-$^3$P$_0$) line, we see only weak evidence for H$_2$O(4$_{1,4}$--3$_{2,1}$) and H$_2$O(5$_{1,5}$--4$_{2,2}$) emission, for both of which we place 3-$\sigma$ limits of $L_{{\rm H}_2{\rm O}}$/$L_{\rm IR}<$\,4\,$\times$\,10$^{-3}$. \cite{jarugula19} detected these emission lines in four strongly lensed galaxies at $z\sim3$, finding $L_{{\rm H}_2{\rm O}}$/$L_{\rm IR}=$\,2.76\,$\times$\,10$^{-5}$ for their sample and literature sources, indicating that given the depth of our data we would not have detected them. These lines are also absent in the composite spectrum derived by \cite{spilker14}, although we note that the selection of these sources results in their sample having much higher infrared luminosities (median $L_{\rm IR}=$\,4.2\,$\times$\,10$^{13}$\,L$_\odot$), and correspondingly higher densities and therefore their observations would be more likely to detect the H$_2$O lines than ours.

We find that 38\,$\pm$\,9 per cent of our CO-detected sources display double-peaked line profiles according to the AIC test described in \S\ref{sec:fitting}. The median separation between peaks is 380\,$\pm$\,50\,km\,s$^{-1}$, which we interpret as evidence that these sources are typically fast rotating disks, as sources so  close in velocity would likely have already coalesced, if formed from a merger. Double-peaked emission lines could be indicative of disk dynamics, and to test this we create a simulation using a simple disk model with a rotation curve described by an arctangent model \citep{courteau97} and an exponential light profile. Assuming that our viewing angles of the sources are randomly distributed, we draw random inclination angles with a probability proportional to the sine of the angle \citep[see][]{law09}, finding the predicted fraction of AIC-classified double-peaked sources in the simulation to be consistent with that seen in our sample.  We note that some of the double-peaked sources  may be mergers instead of disks, particularly in the cases where the two peaks have very different amplitudes or line widths, although some of these sources may be disks with asymmetric light profiles. As we do not have the necessary information to distinguish between these alternatives, we consider all double-peaked sources in the same way.

We also perform a median stack of all CO-detected spectra in our SMG sample, finding no evidence for $\gtrsim$\,1000\,km\,s$^{-1}$ wings in their line profiles, indicating an absence of any significant outflows in our sample. 

%
%
\begin{figure*}
\centering
\includegraphics[width=\linewidth]{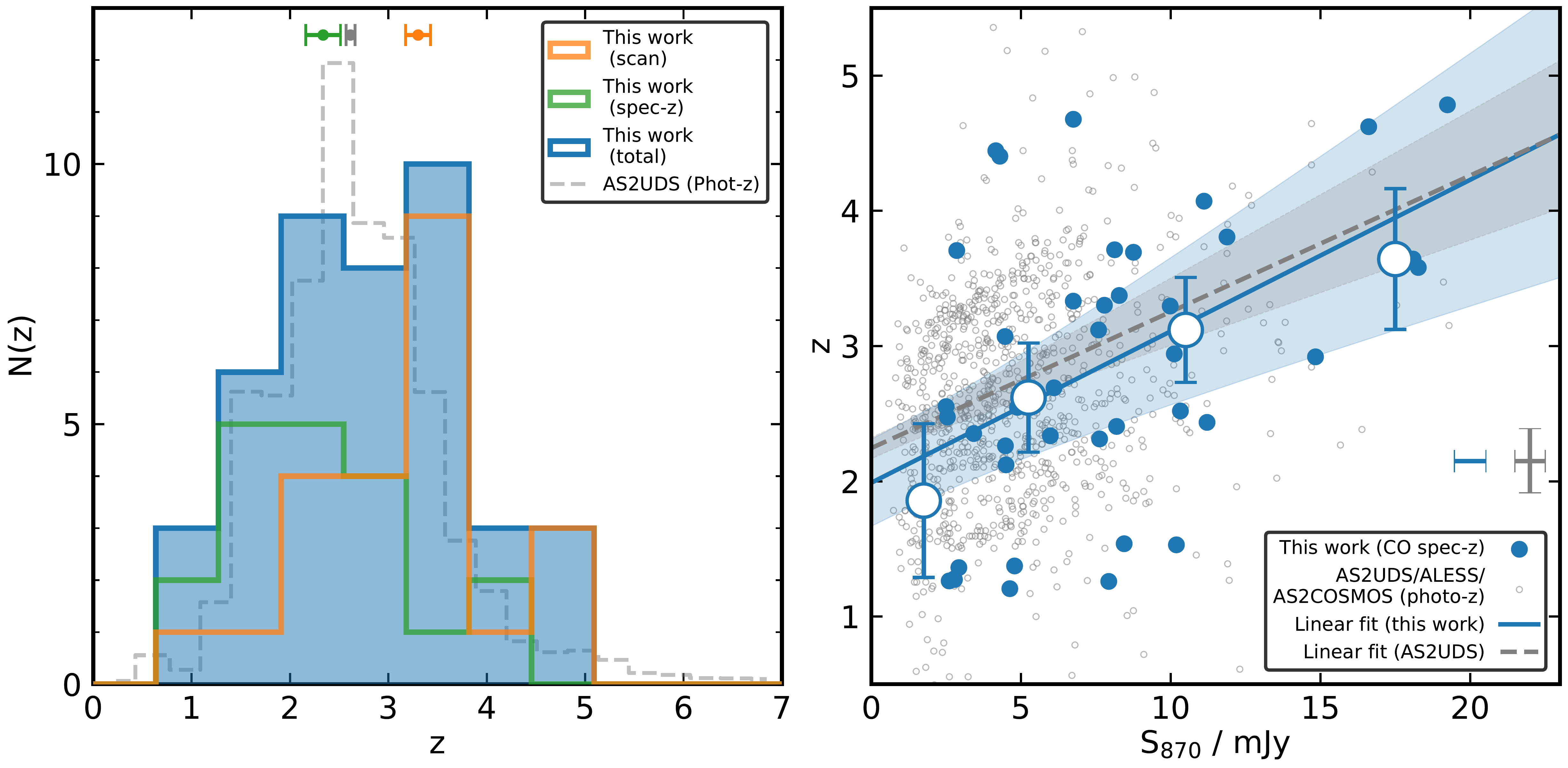}
\caption{\small{{\it Left}: The redshift distribution of our CO sample. We show both the total distribution and the distributions of the {\it scan} and {\it spec-z} subsamples, and compare with the photometric redshifts of the AS2UDS sample (scaled for clarity). The medians of each sample are shown by markers at the top of the panel. The submillimetre-bright {\it scan} sources generally lie at higher redshifts (median $z=$\,3.40\,$\pm$\,0.17) than the typically fainter {\it spec-z} sources in our sample ($z=$\,2.3\,$\pm$\,0.3), and the AS2UDS population ($z=$\,2.61\,$\pm$\,0.08). {\it Right}: redshift versus 870\,$\mu$m flux density for our CO sample and the SMGs with photometric redshifts from the AS2COSMOS, AS2UDS and ALESS surveys \protect\citep[][Ikarashi et al. 2020 in prep.]{dacunha15, dudzeviciute20}. Our CO sources, binned by $S_{870}$, are fit with a linear model of increasing redshift with $S_{870}$, giving a best fit slope of 0.11\,$\pm$\,0.04\,mJy$^{-1}$ (blue), consistent with the 0.06\,$\pm$\,0.01\,mJy$^{-1}$ measured by \protect\cite{simpson20} for AS2COSMOS and the 0.09\,$\pm$\,0.02\,mJy$^{-1}$ measured by \protect\cite{stach19} (grey dashed) for AS2UDS. The data show a modest ($\sim$\,3-$\sigma$) positive correlation, which may support a   downsizing trend (see \S\ref{sec:redshift_dist}). Representative error bars for our sample and AS2UDS are shown in the bottom right of the panel.}}
\label{fig:zdist}
\end{figure*}

\cite{wardlow18} found that 21\,$\pm$\,12 per cent of SMGs have CO-detected companion galaxies at similar velocities and within 150\,kpc in projection, suggesting gravitational interactions within these systems may act to increase the star-formation rate. We uncover three serendipitously detected line emitters using the method described in \S\ref{sec:detections}, all of which have IRAC counterparts. It is important to note that the number of such sources detected is dependent on the depth of the data, and as the bulk of our data is not as deep as that of \cite{wardlow18} we cannot compare the two results directly. However, there is no evidence that these three serendipitously detected sources are physically associated with the ALMA SMGs, as the lines display an offset of $\gg$\,1000\,km\,s$^{-1}$.  
To infer line identifications, and therefore redshifts for these three sources, we compare their IRAC colours/magnitudes with those of the AS2UDS sample and take the median redshift of the ten closest matches. The CO line properties of these sources, in addition to the aforementioned sources from \cite{wardlow18} can be found in Table~\ref{tab:serendips}. 

Fig.~\ref{fig:KS870} shows the $K$-band magnitude of our targets versus their 870\,$\mu$m flux densities.\footnote{Some sources are not in the $K$-band footprint of their respective survey field, and in these cases we estimate $K$ from the $K-$3.6\,$\mu$m colours of AS2UDS SMGs at similar redshifts, where IRAC photometry is available.} 16 of the 61 targets (26 per cent) are not detected in CO, five from the {\it scan} sample and 11 from the {\it spec-z} sample. Among the {\it scan} subsample, the detections have a median $S_{870}=$\,9.4\,$\pm$\,0.9\,mJy, whereas the non-detections have a median $S_{870}=$\,4\,$\pm$\,4\,mJy. Sources with lower 870\,$\mu$m flux densities are expected to have lower dust masses, and they are therefore also more likely to have lower gas masses, making them CO faint and so less likely to be detected. There is also a small redshift range $z\sim$\,1.8--2.0 in which no CO lines fall within the 3\,mm band, and given that $\sim$\,4 per cent of AS2UDS SMGs lie in this range, this could account for $\sim$\,1 non-detection in the {\it scan} sample. Another possibility is that one or more of these sources lies at $z>$\,5 and would therefore display $J_{\rm up}>$\,6 emission in the 3\,mm band, which may display low excitation compared to the lower-$J_{\rm up}$ transitions (we investigate the CO excitation properties in \S\ref{sec:sled}). This is unlikely to prevent us from detecting sources at $z>$\,5 however, as \cite{strandet16} found five such sources in their
ALMA 3\,mm scans. We note that in the {\it scan} sample, we detect CO in $\sim$\,92 per cent of our targets that are brighter than $S_{870}=$\,5\,mJy (22/24 detections). Therefore the non-detections in the {\it scan} sample are most likely to be SMGs at $z\sim$\,3 with faint CO lines, rather than sources that lie in the redshift gap ($z\sim$\,1.8--2.0) or beyond $z\sim$\,5. Indeed, the non-detected sources in our sample have a median photometric redshift of $z=$\,2.8$\pm$0.3.

Non-detections in the {\it spec-z} sample are likely to be due either to incorrect optical/UV spectrosopic redshifts or the faintness of the CO lines. \cite{danielson17} provide a quality factor $Q$ to describe how secure the derived redshift is, with $Q=$\,1 redshifts derived from multiple bright emission lines, $Q=$\,2 redshifts derived from one or two bright emission lines and $Q=3$ redshifts tentatively derived from one emission line and guided by the photometric redshift. Of the sources taken from this work, we detect CO in 11 of the 13 (85 per cent) sources with $Q=$\,1 redshifts, four of the nine (44 per cent) with $Q=$\,2 redshifts, and none of the five with $Q=$\,3 redshifts. Therefore we are more successful at detecting CO in the sources with secure spectroscopic redshifts, as expected. There are also two cases where sources in the {\it scan} sample have CO redshifts that rule out the spectroscopic optical/UV redshift from \cite{danielson17}, namely ALESS001.1 and ALESS003.1 which both have $Q=$\,3 redshifts. Additionally, in this subsample, as in the scans, the non-detections are marginally fainter at 870\,$\mu$m (median 4.0\,$\pm$\,0.4\,mJy) than the detections (median 4.3\,$\pm$\,0.5\,mJy), although this difference is not formally significant. Only $\sim$\,63 per cent of the {\it spec-z} sample are detected ($\sim$\,65 per cent above $S_{870}=$\,4\,mJy), mostly due to uncertain redshifts.

We also show in Fig.~\ref{fig:KS870} the position of our CO-detected SMGs in relation to the star-forming main sequence, adopting the prescription of \cite{speagle14}. We see that just four of the galaxies at $z=$\,1--3 lie over a factor of four above the main sequence (commonly used to define a starburst), and at $z=$\,3--5 all galaxies lie within a factor of four of the main sequence, owing to its proposed evolution with redshift. This plot shows that in terms of star-formation rate, our sample consists of main sequence galaxies out to $z\sim$\,4.5, albeit with high stellar masses ($\log M_\odot>11$) and high star-formation rates for the majority of the sample. While the main sequence is well studied at low redshift, our sample presents an opportunity to extend the work of lower-redshift studies such as PHIBSS \citep{tacconi18} and ASPECS \citep{walter16} to $z>$\,2 and higher gas masses. Given the uncertainties in defining the main sequence, we also show the prescription of \cite{whitaker12}, along with the observed SFR--$M_\ast$ trend in $\sim$\,300,000 UDS field galaxies \citep{dudzeviciute20}, as alternative measures. From Fig.~\ref{fig:KS870} it is clear that in the higher-redshift bin, there is marginal difference between the three main sequence tracks, while at low redshift the UDS field galaxies predict lower SFRs, which would indicate that more of our sample are starbursts than predicted by the \cite{speagle14} prescription. We note this discrepancy here, but for a clearer comparison with the literature we use the \cite{speagle14} main sequence prescription in what follows.

\subsection{Redshift Distribution}
\label{sec:redshift_dist}

Estimates of the redshift distribution of SMGs based on spectroscopic redshifts have been limited by selection biases, towards sources with
brighter optical/near-infrared counterparts and/or to those detectable counterparts in the radio or mid-infrared \citep{chapman05,danielson17}. Measurements of photometric redshifts from SED fitting to ALMA-identified
samples have been more complete \citep{dacunha15, dudzeviciute20}, but not without uncertainties, particularly in the case where sources are faint and/or the photometric coverage is poor. For example, some optically-faint sources have insufficient photometry to establish whether they are highly obscured at low redshifts or simply lie at high redshifts \citep{simpson14}. In contrast, our CO spectroscopic redshifts are precise, and our sample is not biased by
radio/MIPS identifications as well as being large enough to provide a statistically robust redshift distribution.

In Fig.~\ref{fig:zdist} we show the redshift distribution of our CO sources. The median CO redshift of our whole sample is $z=$\,3.0\,$\pm$\,0.2, and the median redshifts of the {\it scan} and {\it spec-z} samples are $z=$\,3.40\,$\pm$\,0.17 and $z=$\,2.3\,$\pm$\,0.3, respectively. Therefore the {\it spec-z} sources preferentially lie at lower redshifts, which is expected as sources typically must be brighter in the optical or near-infrared in order  to derive an restframe optical/UV spectroscopic redshift. Our results for the {\it scan} sources suggest that the optically-faint SMG population lie at higher redshifts than the median, although we find no sources in the extended tail of the photometric redshift distribution at $z>$\,5. Among the $\sim$\,1000 SMGs in AS2UDS and AS2COSMOS  only $\sim$\,1 per cent have photometric redshifts of $z>$\,5 \citep{dudzeviciute20, simpson20}, and hence this result is not surprising. This reflects the exponential decline in the number of massive gas-rich sources at high redshift, and deeper surveys may be needed to find such sources, although at least one $z>$\,5 AzTEC SMG has been detected in the COSMOS field \citep{smolcic15}.

The median redshift of the {\it scan} sample is relatively high,  approaching that reported for the 1.4\,mm-selected SPT sources observed by \cite{spilker14} ($z=$\,3.5), although this is likely a selection bias given that the {\it scan} sources were selected to be faint in the infrared or bright in the submillimetre.
Comparing with the photometric redshifts of these sources, we find a median $|z_{\rm phot}-z_{\rm CO}|/z_{\rm CO}$ of 0.17\,$\pm$\,0.05, and the median redshift of our sample as a whole is consistent with that of the AS2UDS sample, which has a median photometric redshift of
$z=$\,2.79\,$\pm$\,0.07 for a complete flux-limited sample  above $S_{870}\geq$\,3.6\,mJy \citep{dudzeviciute20}. 

In Fig.~\ref{fig:zdist} we also show the variation of redshift as a function of $S_{870}$, including our CO redshifts and photometric redshifts from AS2COSMOS/AS2UDS/ALESS as a comparison. We estimate the gradient of the trend of redshift with $S_{870}$ for the CO sample of $0.11\pm0.04$\,mJy$^{-1}$, which agrees with the estimates of $0.06\pm0.01$\,mJy$^{-1}$ and $0.09\pm0.02$\,mJy$^{-1}$ derived from trends based on photometric redshifts from the AS2COSMOS and AS2UDS samples by \cite{simpson20} and \cite{stach19}, respectively. Our gradient is not as well constrained as in the two aforementioned works due to our smaller sample size, but our spectroscopic redshifts should be more precise. These results add support for the positive correlation between $S_{870}$ and redshift that has been previously proposed in the literature \citep[e.g][]{archibald01, ivison07, smolcic12, stach19}.
This trend could be accounted for by more massive galaxies forming earlier, so-called ``downsizing'' \citep{cowie96}. Due to our selection criteria, our sample contains the galaxies with the highest dust mass (and by implication gas masses) at $z\sim$\,1--5, which also includes many of the largest
galaxies in terms of stellar mass \citep{dudzeviciute20}.  The trend we see therefore reflects an increasing upper bound on the gas and dust mass in
the most massive star-forming galaxies out to $z\sim$\,5, as we show later this is likely driven by an increasing gas fraction with redshift (see \S\ref{sec:fgas}).

\subsection{Gas Excitation}
\label{sec:sled}

The detection of CO line emission in our 3\,mm observations allows us to probe the properties of the star-forming gas in submillimetre galaxies, which, given their high dust masses and star-formation rates should be dense and highly excited. CO traces molecular clouds, with its rotational transitions being excited by collisions with H$_2$ \citep{solomon05}. An understanding of the CO excitation in SMGs is important as it provides a measure of ISM properties such as temperature and density, but it is also vital in deriving gas masses, as it is frequently necessary to estimate the CO(1--0) luminosity by extrapolating from the mid- to high-$J_{\rm up}$ transitions based on such CO spectral line energy distributions (SLED).

%
%
\begin{figure*}
\includegraphics[width=\linewidth]{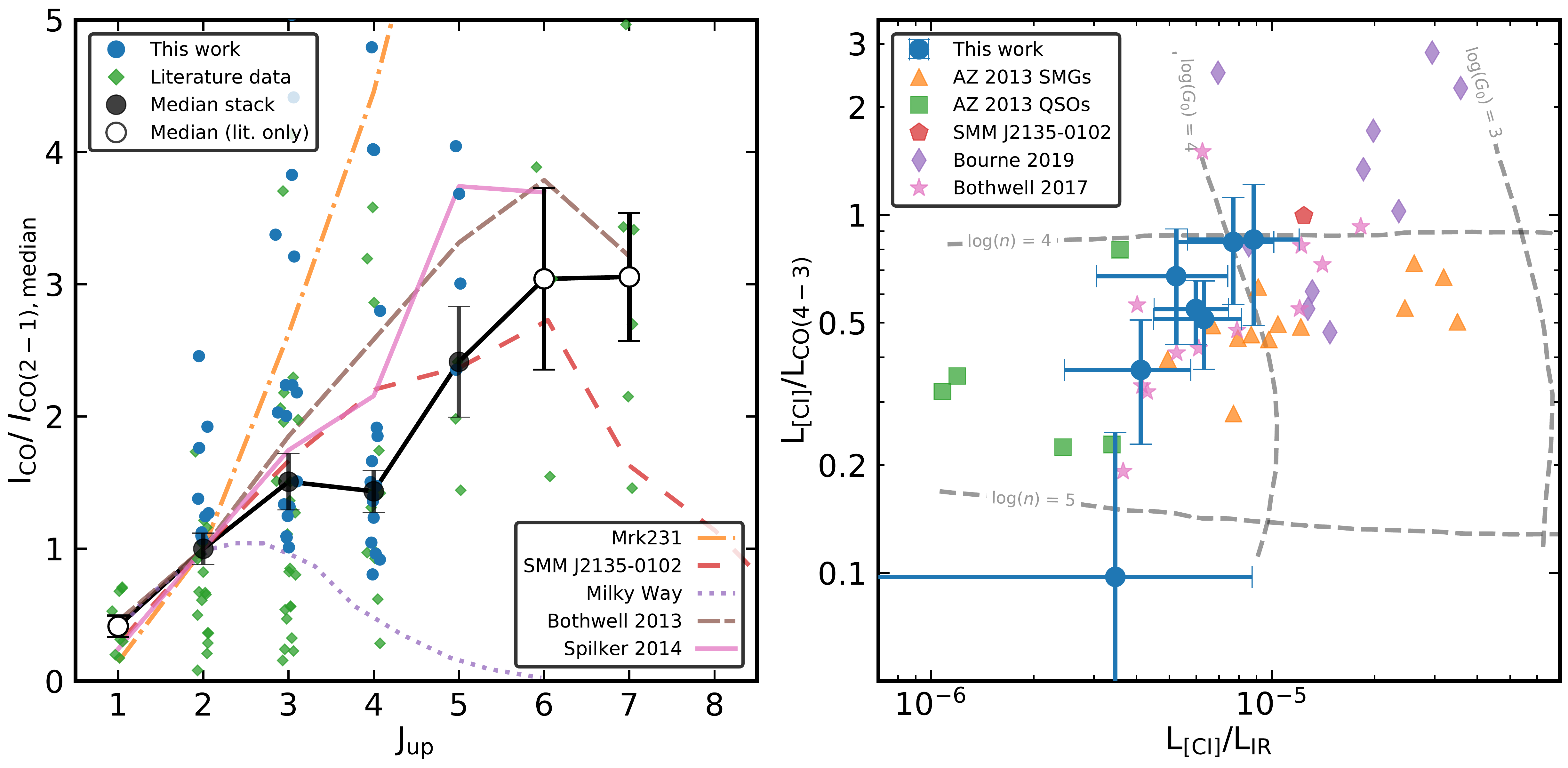}
\caption{\small{{\it Left:} A statistically derived CO SLED constructed from our CO observations of AS2COSMOS, AS2UDS and ALESS SMGs, along with a compilation of literature observations of SMGs. Our composite SLED displays an increase in excitation up to $J_{\rm up}\simeq$\,6, beyond which coverage is limited.  All $\ICO$ are normalised to the median $I_{\rm CO(2-1)}$ of their respective SLED.  We also overlay the SLEDs of the lensed SMG SMM\,J2135$-$0102 \protect\citep[the ``Cosmic Eyelash'';][]{danielson11}, the local ULIRG Markarian 231 \protect\citep[which hosts a Seyfert 1 AGN;][]{vanderwerf10} and the Milky Way \protect\citep{fixsen99}. Our SMG SLED is consistent with that of SMM\,J2135$-$0102, albeit with slightly lower excitation at $J_{\rm up}=$\,4, but appears to be somewhat cooler than the statistical SLEDs of \protect\cite{bothwell13} and \protect\cite{spilker14}, which we attribute to our sources having lower infrared luminosities than in \protect\cite{spilker14} and being at higher redshift than in \protect\cite{bothwell13}. Markarian 231 displays stronger high-$J_{\rm up}$ emission, the absence of which in the SMGs suggests that they are  typically not dominated by an AGN component. The Milky Way SLED peaks at $J_{\rm up}\sim$\,2--3 and declines rapidly beyond, indicating a much cooler and less excited ISM than in the  SMGs. Open symbols represent transitions which rely solely on the literature samples, and errors on the median are estimated from bootstrap resampling. We note that by considering only our SMG sample we derive a SLED that is consistent with the median shown in the plot for $J_{\rm up}=$\,2--5. {\it Right:} $\LCI/L_{\rm CO(4-3)}$ versus $\LCI/L_{\rm IR}$. This plot is an indicator of both gas density ($n$) and radiation field ($G_0$), and we indicate lines of constant $n$ and $G_0$ estimated from the photon dissociation region models of \protect\cite{kaufman99}. We also include SMGs and QSOs from \protect\cite{alaghband-zadeh13}, $z=$\,1 star-forming galaxies from \protect\cite{bourne19}, and SMM\,J2135$-$0102. Our seven sources are broadly consistent with having a single $n$ and $G_0$, although these sources all lie at $z\gtrsim$\,3, and we note that considering a wider variety of sources including $z\sim$\,1 star-forming galaxies and QSOs reveals a mild positive correlation between $n$ and $G_0$ suggesting a link between ISM density and activity.}}
\label{fig:SLED}
\end{figure*}

%
%
\begin{table*}
    \centering
    \begin{tabular}{C{2cm}C{3cm}C{3cm}C{3cm}}\hline\hline
\noalign{\smallskip}
     Line ratio & This work & SMM\,J2135$-$0102 & Bothwell et al.\ (2013) \\
\noalign{\smallskip} \hline \noalign{\smallskip}
     $r_{21}$ & $0.9$ (fixed) & - & $0.84\pm0.13$ \\
     $r_{31}$ & $0.60\pm0.11$ & $0.68\pm0.03$ & $0.52\pm0.09$ \\
     $r_{41}$ & $0.32\pm0.05$ & $0.50\pm0.04$ & $0.41\pm0.07$ \\
     $r_{51}$ & $0.35\pm0.08$ & $0.35\pm0.02$ & $0.32\pm0.05$ \\
     $r_{61}$ & $0.3\pm0.09$ & $0.28\pm0.02$ & $0.21\pm0.04$ \\
     $r_{71}$ & $0.22\pm0.04$ & $0.119\pm0.008$ & $0.18\pm0.04$ \\
\noalign{\smallskip}
\hline\hline
\end{tabular}
    \caption{Median CO line/brightness temperature ratios for the emission lines of 123 SMGs, comprising 47 lines from this study and a further 76 lines in similarly-selected sources from the literature (see \S\ref{sec:literature}), where $r_{J1}=L'_{\rm CO(J-J-1)}/L'_{\rm CO(1-0)}$. As CO(1--0) data is sparse for these populations, we normalise to the CO(2--1) transition and assume $r_{21}=$\,0.9. Errors are estimated from bootstrap resampling.}
    \label{tab:sled}
\end{table*}

The simplest approach to constructing a CO SLED is to observe a single source at a wide range of frequencies to detect multiple CO transitions. For example \cite{danielson11, danielson13} observed the lensed SMG SMM\,J2135$-$0102 (the ``Cosmic Eyelash''), detecting 11 separate transitions including $^{12}$CO from $J_{\rm up}=$\,1 to $J_{\rm up}=$\,9 from which they constructed a CO SLED. SMM\,J2135$-$0102 displays increasing CO line flux up to $J_{\rm up}=$\,6, beyond which it declines. \cite{papadopoulos14} carried out a similar study, observing the merger/starburst systems NGC\,6240 and Arp\,193 with {\it Herschel}/SPIRE to construct $J_{\rm up}=$\,4--13 CO SLEDs, finding Arp\,193 and NGC\,6240 to contain respectively small and large reservoirs of dense ($n\geq$\,10$^4$\,cm$^{-3}$) gas.   CO SLEDs have also been modelled numerically. \cite{lagos12} modelled the CO emission from SMGs by coupling semi-analytic models of galaxy formation with a photon-dominated region code, finding the SLED to peak at $J_{\rm up}=$\,5, although the presence of an AGN was shown to enhance the excitation beyond $J_{\rm up}=$\,6.

Where there are observations of a large sample of sources in only a few, or even just one, CO line, it is possible to build a statistical SLED \citep[][Boogaard et al.\ 2020 in prep.]{bothwell13, spilker14}. This method is subject to more uncertainties and biases, particularly in how to normalise the sources used, variations within the population and the fact that sources at different redshifts contribute to the different $J_{\rm up}$. This is therefore not a preferred method of constructing a SLED, but can still provide useful information nonetheless. \cite{bothwell13} built such a SLED from their survey of 40 SMGs, supplemented by sources from the literature, and \cite{spilker14} similarly used their 1.4\,mm-selected lensed dusty star-forming galaxies to construct a composite SLED. 

We construct our own statistical SLED for SMGs, using our 47 CO-detected sources in addition to a further 76 lines in similar sources from the literature to create a superset of 123 CO lines (see \S\ref{sec:literature} for a list of included studies). We follow a similar prescription to that used in \cite{bothwell13}, exploiting the fact that $\LCO \propto L_{\rm IR}^a$ and using these trends to normalise all $\LCO$ to the same $L_{\rm IR}$:
\begin{equation}
    \label{eq:deltaLCO}
    L'_{\rm CO,corr} = \LCO \times \left(\dfrac{\left\langle L_{\rm IR} \right\rangle}{L_{\rm IR}}\right)^a,
\end{equation}
where $L'_{\rm CO,corr}$ is the CO line luminosity a source would have at $L_{\rm IR} = \left\langle L_{\rm IR} \right\rangle$, and in this case we choose $\left\langle L_{\rm IR} \right\rangle$ to be the sample median. $a$ is the slope of the $L_{\rm CO,J}$--$L_{\rm IR}$ relation. We then convert $\LCO$ to $\ICO$ using Eq.~\ref{eq:LCO}, adopting the median redshift of the superset. \cite{bothwell13} adopt $a=$\,1 for all $J_{\rm up}$, when in reality $a$ may vary with $J_{\rm up}$ as higher-$J_{\rm up}$ transitions more closely trace  the warm star-forming gas, while low-$J_{\rm up}$ transitions trace  cooler gas \citep{greve14, rosenberg15}. We follow \cite{bothwell13} in adopting $a=$\,1, however we find that adopting $a=$\,0.6--0.8 changes the results only within the 1-$\sigma$ error bars.

We estimate $L'_{\rm CO,corr}$ using $\left\langle L_{\rm IR} \right\rangle = $\,6.1$\times$10$^{12}$\,L$_\odot$ (the median $L_{\rm IR}$ of the superset) and use Eq.~\ref{eq:LCO}, adopting $z=$\,2.51 (the median redshift of the superset), to convert them to $\ICO$, which we plot in Fig.~\ref{fig:SLED}. The median SLED is calculated from the median intensity at each $J_{\rm up}$, with bootstrapped uncertainties. We also normalise all measurements to the median $I_{\rm CO(2-1)}$ of our sample to allow a clearer comparison with other SLEDs, and the CO(2--1) transition is chosen as we have better coverage in our sample
than for the CO(1--0) transition. The SLED shows an increase in excitation up to $J_{\rm up}=$\,6, however we note that few transitions with $J_{\rm up}>$\,5 are included here, and therefore the uncertainties are much greater in this regime.  We also see a considerable scatter in the scaled line luminosities of the SMGs at each transition, suggesting a large variation in either excitation, optical depth or gas depletion timescale.  We suggest it is likely the latter factor, gas depletion, which is expected to vary rapidly in a strongly star-forming population such as SMGs, which should therefore show a wide range in CO line luminosity at a fixed far-infrared luminosity.

In Fig.~\ref{fig:SLED} we also show the median SLEDs derived by \cite{bothwell13} for 42 luminous SMGs, and by \cite{spilker14} for 1.4\,mm-selected dusty star-forming galaxies. For comparison with other single-source SLEDs, we show the SLEDs of the Milky Way \citep{fixsen99}, the aforementioned SMM\,J2135$-$0102 \citep{danielson11}, and the local ULIRG Markarian 231 \citep{vanderwerf10}. We see lower line fluxes at $J_{\rm up}=$\,3 and 4 when compared to the median SLEDs of \cite{bothwell13} and \cite{spilker14}, and our SLED is in fact closest to that of SMM\,J2135$-$0102, agreeing well in the range $J_{\rm up}=$\,2--5. Our sources do appear to display less excitation (but only at $\sim$\,2-$\sigma$ significance) at $J_{\rm up}=4$ than we would expect given the relatively constant increase in excitation seen in the other SLEDs. We note here that changing the $\LCO$--$L_{\rm IR}$ scaling from $a=$\,1 to $a=$\,0.8 results in better agreement between the two statistical SLEDs, however we use the $a=$\,1 result here, as found for local ULIRGs for $J_{\rm up}=$\,2--5 \citep[see][]{greve14}, and to remain consistent with \cite{bothwell13}. Differences between our SLED and that of \cite{spilker14} are expected as their sources are much brighter in the infrared (median 4.2$\times$10$^{13}$\,L$_\odot$, compared with our 6.1$\times$10$^{12}$\,L$_\odot$) as a result of their wide, shallow survey.

%
%
\begin{figure*}
\begin{center}
\subfloat{\includegraphics[width=\linewidth]{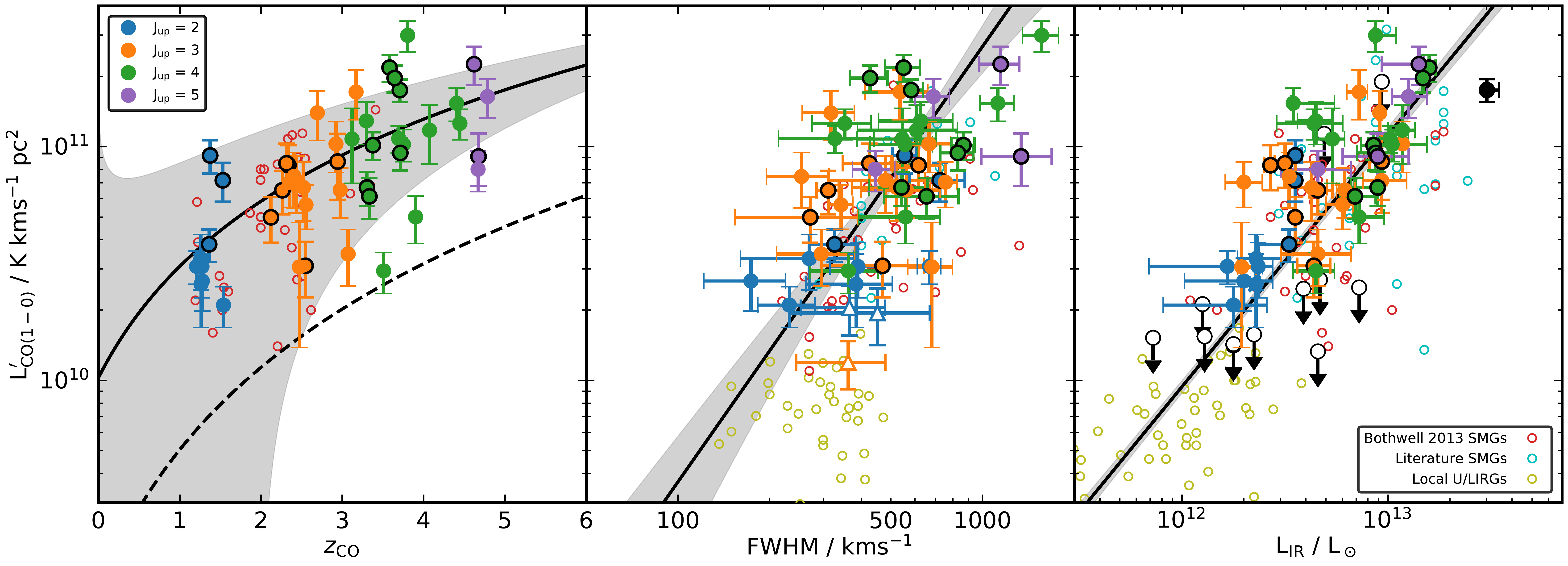}}
\caption{\small{{\it Left:} $\LCO$ versus CO redshift for our SMG sample, showing a clear trend of increasing CO luminosity with redshift which we fit with the model $\LCO\propto (1+z)^b$, finding $b=$\,1.6\,$\pm$\,0.3. This plot indicates that the gas mass in our dust-mass-selected SMGs  exhibits a steady rise with redshift, although this is partly driven by the increase in the detection limit at higher redshifts (roughly indicated by the dashed line as our data have a range of sensitivities). {\it Middle:} $\LCO$ versus FWHM for our sample along with SMGs from the literature and local ULIRGS from \protect\cite{downes98}. For our sources we indicate the transition in which the source was detected, however all sources have been corrected to $L'_{\rm CO(1-0)}$ as described in \S\ref{sec:sled}. Most  SMGs lie at the high-luminosity end of this trend, with the brightest and broadest lines indicating that they are the most massive galaxies in terms of both gas content and dynamical mass. We generally find that higher-$J_{\rm up}$ sources have larger linewidths which might suggest that the higher-redshift sources are more massive. Also included are the three serendipitous sources described in \S\ref{sec:properties}, which lie at the lower end of the trend, indicating that they may be scaled down versions of SMGs. Our data are fit with the model $\log_{10}\LCO=a\log_{10}\left({\rm FWHM/FWHM}_{\rm med}\right)+b$, with $a=$\,2.1\,$\pm$\,0.4, consistent with a rotating disk model. We also find that the median linewidth of the double-peaked sources is consistent with that of the single-peaked sources, within 1-$\sigma$ uncertainties. ULIRGs display lower line luminosities for a given linewidth, likely because their dynamical masses have an increasing contribution from their stellar component, rather than being dominated by the gas. {\it Right:} $\LCO$ versus $L_{\rm IR}$ for the same sample as in the middle panel in addition to local LIRGs. Again the SMGs lie at the extreme end of the trend, indicating large gas reservoirs and high star-formation rates. We fit all the data with the model $\log_{10}\LCO=a\log_{10}\left({L_{\rm IR}/L_{\rm IR,med}}\right)+b$, finding $a=$\,0.99\,$\pm$\,0.03. There is very little scatter in the data for a wide variety of populations, an indication that star-formation efficiency remains relatively constant for dust continuum-selected samples at $z\sim$\,1--4. Black points indicate 3-$\sigma$ upper limits on our CO non-detections.}}
\label{fig:LCO}
\end{center}
\end{figure*}

Our SMG SLED agrees with Markarian 231 at $J_{\rm up}=$\,2--3,
but at higher-$J_{\rm up}$ the latter displays much more highly-excited gas. \cite{vanderwerf10} showed that in the $J_{\rm up}\leq$\,8 regime this can be explained by heating from star formation, however above $J_{\rm up}=$\,8 the observed line ratios require X-ray heating from the galaxy's supermassive black hole. It is therefore unlikely that the moderate-$J_{\rm up}$ CO emission from most SMGs is dominated by an AGN component. By contrast, the Milky Way SLED peaks at $J_{\rm up}=$\,2--3, displaying only weak emission beyond $J_{\rm up}=$\,6. Given the close agreement we see to the  SLED of SMM\,J2135$-$0102, as measured by \cite{danielson11}, we adopt this when deriving $L_{\rm CO(1-0)}$ for our sources.

\subsubsection{[C{\sc i}]}
\label{sec:CI}

As an alternate probe of the ISM we present the \CI properties of our sample in Table~\ref{tab:CI}. Fig.~\ref{fig:SLED} shows the ratio between the \CIfull and CO(4--3) luminosity as a function of the ratio between the \CIfull and infrared luminosity. To interpret the distribution we overlay contours of gas density ($n$) and radiation field ($G_0$) predicted by the photon dissociation region (PDR) model of \cite{kaufman99}. As our \CI sample is small however, we limit ourselves to a qualitative discussion only. Our seven sources show very similar line ratios, and may even be consistent with a single value of $L_{\rm [CI]}/L_{\rm IR}$ and $L_{\rm [CI]}/L_{\rm CO(4-3)}$. The uncertainties are large however, and we are limited in that our sample contains only sources at $z\gtrsim$\,3.2, at which \CIfull can be detected in the 3\,mm band. The wider sample, including sources from \cite{bourne19} and other SMGs from \cite{alaghband-zadeh13}, displays a positive correlation. As the PDR model predicts $L_{\rm [CI]}/L_{\rm IR}$ to be anti-correlated with radiation field strength, and $L_{\rm [CI]}/L_{\rm CO(4-3)}$ to be anti-correlated with density, this suggests that sources that are denser have stronger radiation fields.

We also note that high S/N [C\,{\sc ii}] emission has been detected in a handful of AS2COSMOS, AS2UDS and ALESS sources. From this sample, these are AS2COS0006.1 (Matsuhashi et al.\ 2020 subm.), ALESS061.1 and ALESS065.1 \citep{swinbank12,gullberg18}. Further constraints on the gas density and radiation field strength can be placed by combining [C\,{\sc ii}] line fluxes with \CI line fluxes and far-infrared luminosities -- with brighter [C\,{\sc ii}] emission compared to the \CI an indicator of stronger radiation fields and lower gas densities \citep[see e.g.][]{gerin00}.

\subsection{CO(1--0) luminosities}
\label{sec:LCO}

Having established the excitation properties of the SMGs in our sample we can estimate their CO(1--0) luminosities. This will allow us to investigate how our sources fit within the $\LCO$--FWHM and $\LCO$--$L_{\rm IR}$ relations. In what follows we use $L_{\rm CO(1-0)} = L_{\rm CO,J}/r_{j1}$, adopting the $r_{j1}$ measured by \cite{danielson11} for SMM\,J2135$-$0102, to derive CO(1--0) luminosities.

\subsubsection{$\LCO$--FWHM relation}

The $\LCO$--FWHM relation is useful as it provides a measure of the correlation between the gas mass and the galaxy dynamics \citep{harris12}. Our sample has a median $L'_{\rm CO(1-0)}$ of (6.7\,$\pm$\,0.5)\,$\times10^{10}$\,K\,km\,s$^{-1}$\,pc$^{2}$ and a median FWHM of 540\,$\pm$\,40\,kms$^{-1}$, indicating more gas-rich sources than the (4.5\,$\pm$\,1.0)\,$\times10^{10}$\,K\,km\,s$^{-1}$\,pc$^{2}$ and 550\,$\pm$\,50\,km\,s$^{-1}$ found from the \cite{bothwell13} sample. Fig.~\ref{fig:LCO} shows the derived CO line luminosity as a function of line FWHM, where all line luminosities are converted to CO(1--0). For comparison we include SMGs from literature sources (see \S\ref{sec:literature}) and local ULIRGs from \cite{downes98}.  We see a steady rise in $L'_{\rm CO(1-0)}$ with redshift, suggesting a similar rising gas mass, although this is influenced by the effect of incompleteness for the less luminous sources at the highest redshifts.

The variation of $L_{\rm CO(1-0)}$ with FWHM of the CO lines in Fig.~\ref{fig:LCO} shows a 3-$\sigma$ positive correlation, indicative of increasing gas mass with dynamical mass. To interpret this we fit our data with a model of the form $\log_{10}\LCO=a\log_{10}\left({\rm FWHM/FWHM}_{\rm med}\right)+b$, using orthogonal distance regression. From this we find $a=$\,2.1\,$\pm$\,0.4 and $b=$\,10.78\,$\pm$\,0.06, with a scatter of 0.25\,dex. If the gas kinematics in our population reflects disk dynamics (see \S\ref{sec:properties}), we would expect the galaxy mass (and therefore the CO line luminosity) to increase with the square of the rotational velocity (and therefore the CO linewidth). Therefore the dynamics of the CO in our sample are consistent with rotating disks. A model of this kind was also shown to be a good fit to the sample of \cite{bothwell13}, who suggested that this implies a constant ratio between the gas and stellar dynamical contributions in CO regions. \footnote{Unlike the SMGs, the local ULIRGs in this plot show no correlation between FWHM and $\LCO$. \cite{bothwell13} suggested that this is a combination of a wide range in gas fractions, a greater contribution to the dynamics from the stellar component, or thin nuclear gas discs/rings meaning that inclination differences cause significant scatter.}

We also indicate on this plot the sources with double-peaked CO line profiles (as described in \S\ref{sec:properties}), finding these to have a median FWHM of 550\,$\pm$\,60\,km\,s$^{-1}$ compared to 520\,$\pm$\,60\,kms$^{-1}$ for the single-peaked sources, and a median $L'_{\rm CO(1-0)}$ of (7.4\,$\pm$\,0.8)\,$\times10^{10}$\,K\,km\,s$^{-1}$\,pc$^{2}$ compared to (6.3\,$\pm$\,0.8)\,$\times10^{10}$\,K\,km\,s$^{-1}$\,pc$^{2}$ for the single-peaked sources.

\subsubsection{$\LCO$--$L_{\rm IR}$ relation}

The CO emission acts as a tracer of the reservoir of gas available in SMGs to form stars, and the infrared luminosity traces the star formation currently occurring. Therefore the $\LCO$--$L_{\rm IR}$ relation indicates what fraction of the total molecular gas is being converted to new stars: the star-formation efficiency. This is analogous to the Kennicutt-Schmidt relation \citep{kennicutt98} for galaxy-integrated properties. Fig.~\ref{fig:LCO} shows the relationship between $L'_{\rm CO(1-0)}$ and $L_{\rm IR}$ for our SMG sample along with SMGs from the literature (see \S\ref{sec:literature}) and local (U)LIRGs from \cite{sanders91} and \cite{solomon97}. We fit the model $\log_{10}\LCO=a\log_{10}\left(L_{\rm IR}/L_{\rm IR,med}\right)+b$ to all data points using orthogonal distance regression, finding $a=$\,0.99\,$\pm$\,0.03 and $b=$\,10.21\,$\pm$\,0.02. The positive correlation between $L'_{\rm CO(1-0)}$ and $L_{\rm IR}$ is tight, with 0.2\,dex of scatter, and most SMGs lie at the upper end of this trend indicating massive gas reservoirs and high star-formation rates.

Whereas we find a linear slope, some authors have found the $L_{\rm CO(1-0)}$--$L_{\rm IR}$ relation to exhibit sub- or super-linear slopes. For example, \cite{greve05} found a slope of $0.62\pm0.08$ by fitting local (U)LIRGs and SMGs, although they assumed thermalised emission to convert
their moderate-$J_{\rm up}$ CO line luminosities, which would bias the result low, while \cite{genzel10} found 0.87\,$\pm$\,0.09 (note that they fitted the inverse relation, and we have converted the slope for easier comparison with ours). On the other hand, \cite{ivison11} found a super-linear slope, $a=$\,1.5\,$\pm$\,0.3 for SMGs with reliable CO(1--0) or CO(2--1) measurements, potentially indicating an additional reservoir of cool gas in some systems. We note that our conclusions are unchanged if we adopt line ratios from our own statistical SLED instead of that of SMM\,J1235$-$0102.

In theory, the slope of the $\LCO$--$L_{\rm IR}$ relation should vary with $J_{\rm up}$ as the low-$J_{\rm up}$ transitions trace the cooler gas, whereas the mid- to high-$J_{\rm up}$ transitions trace the warmer gas which is more closely linked to the star-forming regioms. For the $J_{\rm up}=$\,2--5 transitions, we find slopes of 2.7\,$\pm$\,0.4, 0.8\,$\pm$\,0.3, 1.0\,$\pm$\,0.3 and 1.1\,$\pm$\,0.4, respectively. For the $J_{\rm up}=$\,3--5 transitions, this is consistent with \cite{greve14} who performed a similar analysis on local ULIRGs. The abnormal gradient of the $J_{\rm up}=$\,2 relation may be a result of our small sample, which comprises just nine $J_{\rm up}=$\,2 detections.

It has also been suggested that the CO(5--4) emission could be a good tracer of the star-forming gas, in which case it should correlate linearly with the infrared luminosity, with \cite{daddi15} finding a slope of 0.96\,$\pm$\,0.04 for the $L_{\rm CO(5-4)}$--$L_{\rm IR}$ relation \citep[see also e.g.][]{cassata20, valentino20}. As stated in the previous paragraph we find that the four sources detected in CO(5--4) display a gradient in their trend with $L_{\rm IR}$ of $\Delta\LCO/\Delta L_{\rm IR}=$\,1.1\,$\pm$\,0.4, consistent with a linear relation between $\LCO$ and $L_{\rm IR}$, in support of the \cite{daddi15} result. To confirm that this result is not affected by our small sample size, we convert all sources detected in CO(4--3) to CO(5--4), ensuring that the correction factor is smaller and less uncertain. In this case we find a gradient of $\Delta\LCO/\Delta L_{\rm IR}=$\,1.2\,$\pm$\,0.3, also consistent with linearity.

\subsection{Gas Mass Tracers}
\label{sec:gas_masses}

As a measure of the amount of fuel available for star formation, an accurate and precise knowledge of the molecular gas content is crucial in studying any type of galaxy. From our observations we are able to compare three different tracers of the gas mass: the inferred CO(1--0) luminosity, the \CIfull luminosity and the dust mass. Furthermore we can also test three different methods of estimating dust masses: from the rest-frame 870\,$\mu$m emission, the extrapolated observed-frame 870\,$\mu$m emission and from SED modelling, all of which are similar but may have subtle differences. When estimating gas masses from these tracers all three methods are subject to a calibration factor with considerable uncertainty, therefore we focus only on the observed quantities and how well they correlate when providing a comparison. However, we will briefly discuss predicted values for gas masses using standard conversion factors.

\subsubsection{CO-H$_2$ conversion}
\label{sec:alphaCO}

Having established the excitation properties of our sample in \S\ref{sec:sled}, and therefore the CO line ratios $r_{\rm j1}$, we can calculate gas masses from the CO luminosity as
\begin{equation}
\label{eq:mgas}
M_{\rm gas} = 1.36\,\alphaCO\,r_{\rm J1}\,L'_{\rm CO,J},
\end{equation}
where $r_{\rm J1}$ represents the line ratio of the $J_{\rm up}$ transition to the CO(1--0) transition (which we adopt from SMM\,J2135$-$0102, noting that this is consistent with our statistical SLED derived in \S\ref{sec:sled}), $\alphaCO$ is the so-called CO--H$_2$ conversion factor given in units of $M_\odot$ (K\,km\,s$^{-1}$\,pc$^{2}$)$^{-1}$, $L'_{\rm CO,J}$ is the CO line luminosity of the $J_{\rm up}$ transition in units of K\,km\,s$^{-1}$\,pc$^{2}$, and the factor of 1.36 accounts for the abundance of Helium.

This method is widely employed for estimating gas masses \citep{solomon97, bolatto13} thanks to the relative ease of observing CO emission with e.g.\ ALMA, NOEMA or JVLA, although it is subject to uncertainties in correcting from the mid- and high-$J_{\rm up}$ CO transitions to the CO(1--0) luminosity, as well as in the value of $\alphaCO$, which is poorly constrained for most types of galaxies \citep[see][for a review]{carilli13}. For Milky Way and Local Group molecular clouds, multiple measurements have been made with results  in the range $\alphaCO\sim$\,1--9\,$M_\odot$ (K\,km\,s$^{-1}$\,pc$^{2}$)$^{-1}$ \citep{solomon87, leroy11, casey14}. There is good evidence of variation in $\alphaCO$ between different galaxy types and redshifts and it has been suggested that there exists a dichotomy between ``normal'' (main-sequence) star-forming galaxies and ``starburst'' galaxies. In this picture, the former behave more like disk galaxies and have a CO-H$_2$ conversion factor close to that of the Milky Way, with \cite{daddi10} most notably estimating $\alphaCO\sim$\,3.6\,M$_\odot$ (K\,km\,s$^{-1}$\,pc$^{2}$)$^{-1}$ for a handful of galaxies. For more actively star-forming galaxies, which are expected to have more turbulent interstellar media, $\alphaCO\sim$\,0.8--1\,$M_\odot$ (K\,km\,s$^{-1}$\,pc$^{2}$)$^{-1}$ is the commonly adopted value, as estimated by \cite{downes98} for local ULIRGs. In addition, $\alphaCO$ is expected to vary with metallicity \citep{bolatto13}, although some authors have found only a weak dependence \citep{sandstrom13}.

At high redshift, however,  it is very difficult to measure $\alphaCO$ and verify the appropriate value to adopt. It is not well understood how $\alphaCO$ relates to the complex physical processes that are ongoing in galaxies, and therefore attempts to constrain $\alphaCO$ are mostly empirical, often involving estimating dynamical masses and combining these with stellar masses and an assumed dark matter fraction \citep{downes98, daddi10, bothwell13, calistrorivera18}. $\alphaCO$ is then determined based on the amount of gas which accounts for the remaining mass. An additional complication is that most CO observations used for these studies trace the mid-$J_{\rm up}$ transitions in SMGs, and it is important to consider that both the CO(1--0) and H$_2$ could have a different spatial extent compared to the higher-$J_{\rm up}$ CO emission which must be taken into account when adopting sizes in the mass calculations \citep{ivison11}, as should uncertainties in the line ratios. In most cases -- in the absence of 
high-spatial-resolution intensity and kinematic maps of CO(1--0) --
the uncertainties that result from these assumptions prevent any meaningful discussion of which of the starburst or ``normal'' star-forming values of $\alphaCO$ are applicable or indeed whether these two populations have demonstrably different values of $\alphaCO$ at $z\gg$\,0.

We thus investigate whether it is feasible to roughly constrain $\alphaCO$ using our statistical sample of sources. The dynamical mass of a galaxy can be estimated both as the sum of the gas, stellar and dark matter components, and from the CO linewidth. We therefore have
\begin{equation}
M_{\rm dyn}{\rm sin}^2(i) = \dfrac{M_\ast + M_{\rm gas}}{1 - f_{\rm DM}} = C\dfrac{\sigma^2R}{G},
\label{eq:mdyn}
\end{equation}
where $M_{\rm dyn}$ is the dynamical mass of the galaxy, $i$ is the inclination angle at which it is observed, $M_\ast$ is the stellar mass, $M_{\rm gas}$ is the gas mass as calculated in Eq.~\ref{eq:mgas}, $f_{\rm DM}$ is the dark matter fraction, $\sigma$ is the circular velocity traced by the linewidth and $G$ is the gravitational constant. $C$ is dependent on the 
density profile and velocity anistropy  of the relevant galaxy population \citep{erb06, galacticdynamics, kohandel19}. All quantities correspond to their values within a radius $R$ of the galaxy centre.

We adopt
$f_{\rm DM}\sim$\,0.35 following \cite{smith19}, as found for $z\sim$\,0.12 ellipticals (which are likely to be SMG descendants, see \S\ref{sec:evo}) and
$C=$\,2.25 following \cite{galacticdynamics} which is found to be consistent with our simulated rotation curves from \S\ref{sec:properties}. By varying the parameters in our simulations, we find this to be close to a lower limit, and the value of $C=$\,3.4 adopted by \cite{erb06} is also consistent with our findings, as is the value of $C=$\,1.78 found in simulations for spiral disc galaxies by \cite{kohandel19}. While high-resolution imaging is not available for most of our sources, and therefore we have very little information on their inclinations, as a test hypothesis we make the assumption that our sample is comprised of randomly oriented disks, which we have shown to be consistent with our distribution of line profiles in \S\ref{sec:properties}. In this case we would expect a median $i=$\,57$^\circ$ and a median ${\rm sin}(i)=$\,0.79 \citep{law09}, which we adopt for all 47 sources. We choose to use an aperture $R=$\,14\,kpc to ensure we encapsulate the full extent of the CO(1--0) emission (and hence cold gas reservoirs) \citep{ivison11}.

Adopting these parameters yields a median value of $\alphaCO=$\,1.0\,$\pm$\,0.7, closer to the typical ULIRG value of $\sim$\,1 \citep{solomon97} than the  main-sequence value \citep{daddi15}, and  consistent with \cite{danielson11, danielson13} who estimated $\alphaCO\sim$\,2 for the lensed SMG SMM\,J2135$-$0102 by applying an LVG model to the CO SLED.  The value quoted above could be viewed as a lower limit, given that $C$ may take a larger value than 2.25, and $f_{\rm DM}$ may be lower than 0.35. For example, if we were to adopt $C=$\,3.4 \citep{erb06} and $f_{\rm DM}=$\,0.25 we would conclude that $\alphaCO\sim$\,4\,$\pm$\,2, consistent with both SMM\,J2135$-$0102 and the Milky Way value, as well as \cite{daddi10} who found $\alphaCO\sim$\,3.6 \citep[also shown to be consistent with main sequence galaxies by][using a similar technique to this work]{cassata20}. This underlines the uncertainties in estimating $\alphaCO$ and gas masses in high-redshift galaxies. In order to constrain $\alphaCO$ further we require resolved imaging and velocity fields of many sources, which will allow us to acquire a more complete understanding of the spatial extent of the emission, the inclination and the dark matter fraction (from rotation curves). We note that the recent spatially resolved study of \cite{calistrorivera18} obtained $\alphaCO=$\,1.1$_{-0.7}^{+0.8}$ using high-resolution imaging of the CO(3--2) emission in four SMGs at $z\sim$\,2--3, which is consistent with our estimate and suggests a low value of $\alphaCO$ may apply to SMGs. In what follows we use $\alphaCO=$\,1, making our results easier to rescale for readers, resulting in a median gas mass of $M_{\rm gas,med}=($6.7\,$\pm$\,0.5$)\times$\,10$^{10}$\,M$_{\odot}$.

\subsubsection{Gas-to-dust conversion}

The gas mass can also be estimated from the dust mass using:
\begin{equation}
M_{\rm gas} = \gdr\,M_{\rm dust},
\end{equation}
where the gas-to-dust ratio $\gdr$ is simply the ratio of gas mass to dust mass \citep{leroy11, magdis12}. The gas-to-dust ratio may vary with metallicity \citep{santini14} and redshift \citep{saintonge13} for metal-rich sources, although it is often assumed to be a constant $\gdr\sim100$ \citep{swinbank14, scoville16}. Some authors have invoked scaling relations in order to estimate the gas-phase oxygen abundance, and subsequently attempted to infer the gas-to-dust ratio, from the estimated stellar mass \citep{genzel15, tacconi18}. This method is, of course, not without very considerable uncertainties, both systematic and random, but as it relies on different assumptions to that of the CO-to-H$_2$ method it represents an independent estimate. \cite{leroy11} developed a technique to estimate $\alphaCO$ for resolved sources, assuming the gas-to-dust ratio holds constant over regions where molecular and ionised Hydrogen are in equal abundance, however for high-redshift galaxies we are mostly concerned with galaxy-integrated properties, and this approach is not feasible.

%
%
\begin{figure*}
\begin{center}
\includegraphics[width=\linewidth]{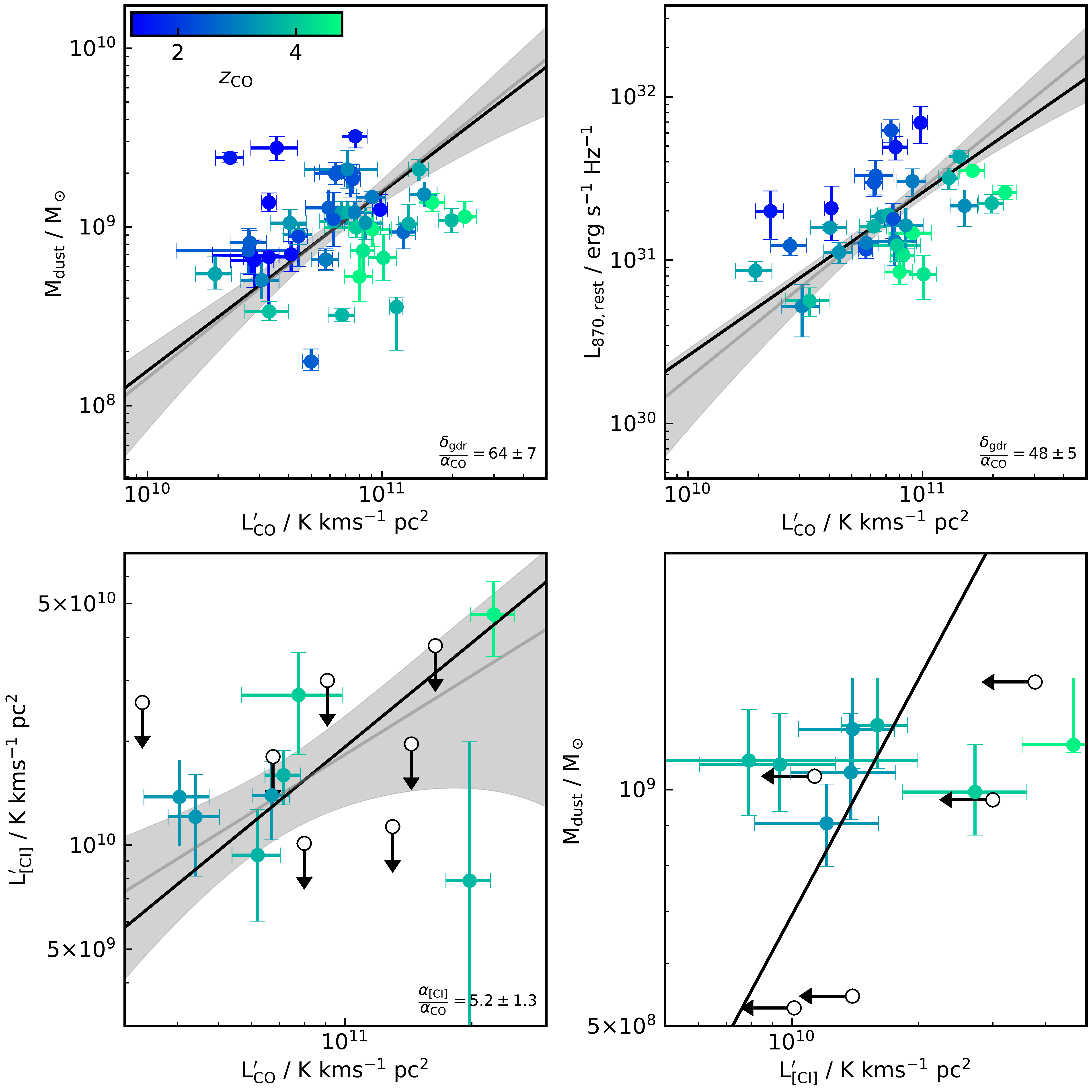}
\caption{\small{A comparison of different tracers of the gas mass in SMGs, where in all cases we perform a free fit (grey with shaded error) and a linear fit (black) in log space. {\it Top-left and right:} Dust mass from {\sc magphys} or the rest-frame 870\,$\mu$m luminosity versus CO(1--0) line luminosity. The CO(1--0) and these two measures of the dust emission appear to correlate well, with a linear model consistent with the data in both cases. Intriguingly, the rest-frame 870\,$\mu$m luminosity shows less scatter than the {\sc magphys} estimate. {\it Bottom-left:} \CIfull line luminosity versus CO line luminosity. The CO(1--0) and \CI show a weak correlation, and are consistent with a linear fit, although we are limited by our modest \CI sample size. {\it Bottom-Right:} Dust mass from {\sc magphys} versus \CIfull line luminosity. There is no correlation between the two, owing to the scatter in dust masses and the small number of \CI-detected sources, which are only detected with modest S/N. We therefore omit the free fit in this case.}}
\label{fig:gas_masses}
\end{center}
\end{figure*}

The dust mass itself can also be estimated in several ways. Firstly, dust masses are available from the {\sc magphys} SED fitting  to our sample (see \S\ref{sec:sed}), which utilises multi-band photometry from e.g.\ ALMA and {\it Herschel} \citep[][Ikarashi et al.\ 2020 in prep.]{dacunha15, dudzeviciute20}. The inclusion of shorter-wavelength photometry in this  method may result in  a bias towards warm dust, which would yield  underestimates of the dust mass \citep{scoville16}. Our dust masses from {\sc magphys} are presented in Table~\ref{tab:measured}. Secondly, the dust mass can be traced by the rest-frame 870\,$\mu$m emission \citep{dunne00}. Given that the median redshift of our sample is $z\sim$\,3, the 3\,mm continuum photometry from our ALMA/NOEMA observations probes  rest-frame 750\,$\mu$m, close to 870\,$\mu$m when compared to the rest-frame $\sim$\,220\,$\mu$m traced by the observed 870\,$\mu$m observations. This means that estimates of the rest-frame 870\,$\mu$m luminosity from the observed 3\,mm photometry are effectively independent of the spectral slope $\beta$:
\begin{equation}
M_{\rm gas} = M_{\rm dust} = \gdr \times \dfrac{L_{\rm 870,rest}}{\kappa_d(\nu)B(\nu,T_d)}
\end{equation}
where $\kappa_d$ here is the dust mass opacity coefficient (taken to be 0.077\,m$^2$kg$^{-1}$) and $B$ is the Planck function, where we adopt $T_d=$\,25\,K \citep{dunne00, scoville16}.

Finally, we can use the fact that the dust on the Rayleigh-Jeans tail is optically thin to estimate the gas mass, with the calibration proposed in \cite{scoville13}:
\begin{equation}
M_{\rm gas} = \alpha_{870}L_{870\,\mu \rm{m}} = \alpha_{870}\times1.19\times10^{27} S_{870} D_L^2,
\end{equation}
where the prefactor of 1.19$\times$10$^{27}$ is derived from measurements of low-redshift spiral and starburst galaxies (and so suffers many of the uncertainties discussed above regarding $\alphaCO$), including an extrapolation based on an assumed dust continuum slope, and the gas-to-dust ratio is built into $\alpha_{870}$. This value can be calculated for sources with observed 870\,$\mu$m measurements, i.e.\ all sources in our sample.

We compare the three methods, finding all three to correlate reasonably well, with the rest-frame 870\,$\mu$m luminosity showing the least scatter compared to the CO(1--0) luminosity. Given that we have {\sc magphys} dust masses for all CO-detected sources, whereas we only have continuum detections for around 80 per cent, and for the {\it spec-z} sample these values are less robust due to the smaller bandwidth of the observations, we use both methods when comparing tracers in the \S\ref{sec:comparison} (see Fig.~\ref{fig:gas_masses}). We do note however, that the median ratio of the $L_{\rm 870,rest}$-based mass dust estimate to the {\sc magphys} dust mass is 1.25\,$\pm$\,0.05 for the adopted dust mass opacity coefficient and dust temperature.

\subsubsection{\CI--H$_2$ conversion}

Our third and final tracer of the gas mass comes from the fine structure line of atomic carbon \citep{weiss03, papadopoulos04}:
\begin{equation}
M_{\rm gas} = 1.36\alpha_{\rm [CI]}\,L'_{\rm [CI]},
\end{equation}
where $\alpha_{\rm [CI]}$ is the [C{\sc i}]--H$_2$ conversion factor in units of M$_\odot$ (K\,km\,s$^{-1}$\,pc$^{2}$)$^{-1}$ $L'_{\rm [CI]}$ is the [C{\sc i}] line luminosity in units of K\,km\,s$^{-1}$\,pc$^{2}$, and again we include a factor of 1.36 to account for the abundance of Helium.

The \CI method benefits from the lines being optically thin which removes some of the transition ratio uncertainties that apply to estimates based on CO, and it is also expected to show smaller abundance variations as it is thought to be affected less by cosmic ray destruction \citep{papadopoulos18}. The \CI is also much easier to observe at high redshift than the low-$J_{\rm up}$ CO transitions. It has been shown that the \CI is distributed throughout molecular clouds, rather than only near their outer edges, and correlates well with the $^{13}$CO \citep{keene85}. As with the CO--H$_2$ conversion however, the \CI--H$_2$ conversion is not well understood at a theoretical level \citep{gaches19}.

\subsubsection{Comparison of tracers}
\label{sec:comparison}

We compare the gas mass estimates from the above methods in Fig.~\ref{fig:gas_masses}. The factors $\alphaCO$, $\gdr$ and $\alpha_{\rm [CI]}$ require a self-consistent calibration, and we can use our large sample of CO (and a smaller number of [C{\sc i}]) luminosities and linewidths, and dust masses to estimate the ratios of these quantities here. In Fig.~\ref{fig:gas_masses} we plot the observed quantities $L'_{\rm CO(1-0)}$, $L'_{\rm [CI]}$, $M_{\rm dust}$ and $L_{\rm 870,rest}$ against one another. In theory, if the three methods of deriving gas masses are consistent then the data should be well-described by the linear model $y=ax$ with $a$ the ratio of the two corresponding conversion factors. For example a plot of $M_{\rm dust}$ versus $L'_{\rm CO(1-0)}$ yields the ratio $\gdr/\alphaCO$. To test this we fit the model $\log_{10}(y)=a\log_{10}(x)+b$, both allowing $a$ to vary freely and fixing $a=$\,1 (meaning the two gas mass tracers scale linearly). 

In Fig.~\ref{fig:gas_masses} we see that the CO luminosity and the {\sc magphys} dust mass correlate reasonably well with one another, with the free fit having a gradient 1.0\,$\pm$\,0.3, consistent with a linear relationship. From the fixed linear fit we derive an average ratio of $\gdr/\alphaCO=$\,64\,$\pm$\,9. The data display a significant scatter, 0.42 dex, which is likely to be driven by uncertainties in the CO SLED, but may also indicate variations in $\alphaCO$ and the gas-to-dust ratio. Alternatively we use the rest-frame 870\,$\mu$m luminosity as a dust mass tracer, finding a gradient of 1.3\,$\pm$\,0.3, i.e.\ consistent with a linear trend, and with a lower scatter of 0.27 dex and a median $\gdr/\alphaCO=$\,47\,$\pm$\,5.

The \CI and CO luminosity in Fig.~\ref{fig:gas_masses} also correlate reasonably well, although we are limited by both the small number of \CI detections in our sample and their low S/N. The free fit has a gradient of 0.8\,$\pm$\,0.4 and is therefore consistent with linear scaling between the \CI and CO luminosities. The scatter is 0.25 dex, and the linear fit implies that $\alpha_{\rm [CI]}/\alphaCO=$\,5.2\,$\pm$\,1.1. 

In contrast, the \CI luminosity and {\sc magphys} dust masses in our sample do not correlate, and the data are in fact consistent with fixed dust mass with varying \CI luminosity, although again we only have a small number of \CI detections. Similarly when using the rest-frame 870\,$\mu$m luminosity as a dust mass tracer we see no correlation. Therefore we do not discuss any limit on the ratio between $\gdr$ and $\LCI$ here.

On the whole, Fig.~\ref{fig:gas_masses} shows that the CO luminosity and dust mass, whether determined by SED fitting or the rest-frame 870\,$\mu$m luminosity, are complimentary tracers of the gas mass. Before we can derive truly reliable gas mass estimates from these tracers, we must calibrate one or both methods using high-resolution imaging. We can however, compare literature measurements of $\alphaCO$ and $\gdr$ with our estimated ratio. For example, adopted the commonly-used value of $\gdr=$\,100 would be consistent with $\alphaCO\sim$\,1.6 in SMGs, roughly consistent with our value derived independently using dynamical arguments (see \S\ref{sec:alphaCO}). In contrast, if our sample is consistent with the classical $\alphaCO\sim$\,3.6 derived by \cite{daddi15} for ``normal'' star-forming galaxies, the corresponding gas-to-dust ratio would be $\gdr\sim$\,230, much higher than commonly adopted values.

\subsection{The star-forming main sequence}
\label{sec:ms}

As previously highlighted, due to their luminosities and hence relative ease of detection, SMGs are a useful laboratory for investigating the formation and evolution of massive galaxies. However, in order to fully understand the evolution of galaxies we must also target sources that are representative of the bulk population of less active galaxies across a wide redshift range -- ``normal'' or so-called ``main-sequence'' galaxies. One approach to categorising galaxies is to classify them according to the difference between their specific star-formation rate ${\rm sSFR}={\rm SFR}/M_\ast$ and the specific star-formation rate expected for a galaxy on the ``main-sequence'' ${\rm sSFR_{MS}}$ at the same stellar mass and redshift, according to some prescription. Specifically, this quantity is defined as $\Delta{\rm sSFR} = {\rm sSFR}/{\rm sSFR_{MS}}$, with $\Delta{\rm sSFR}>4$ being the arbitrary definition of a starburst galaxy. We caution however, that recent work \citep[e.g.][]{puglisi19} suggests that there is considerable variation in the properties of galaxies {\it within} the ``main-sequence'' and hence the concept of $\Delta{\rm sSFR}$, and the main-sequence more generally, may be of limited value.

%
%
\begin{figure*}
\centering
\includegraphics[width=\linewidth]{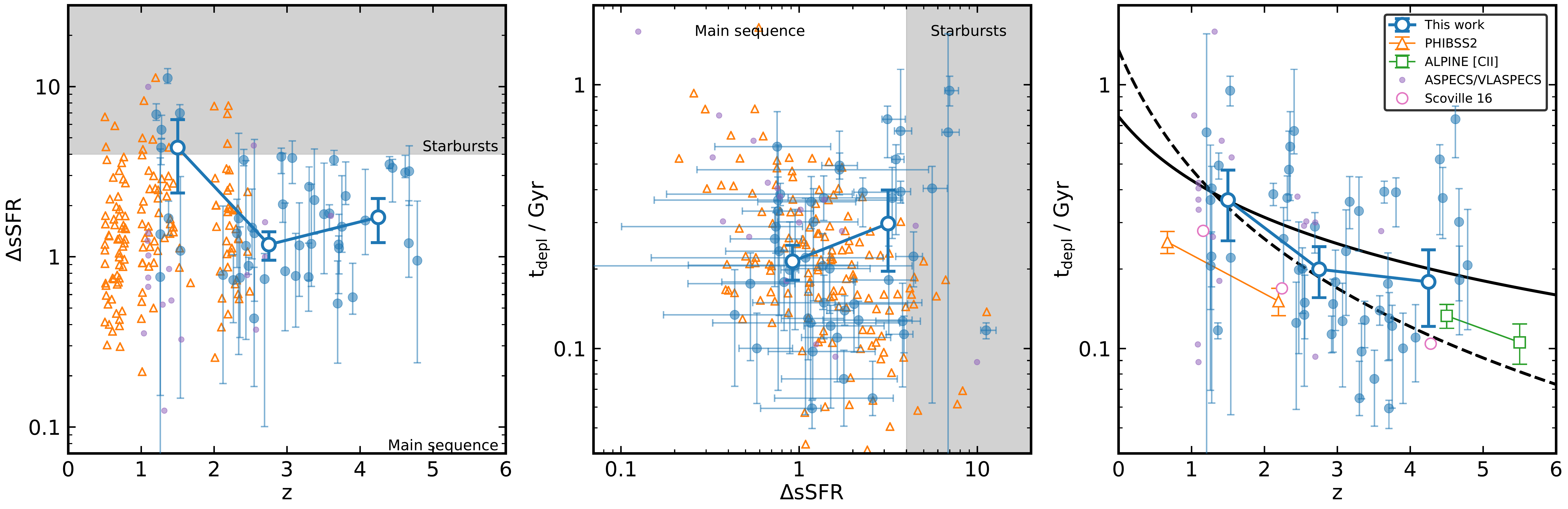}
\caption{\small{{\it Left:} Offset from the ``main-sequence'' $\Delta{\rm sSFR}={\rm sSFR}/{\rm sSFR_{MS}}$ versus redshift for our sample, using the \protect\cite{speagle14} main sequence prescription for sSFR. We indicate the region where $\Delta{\rm sSFR}>$\,4, i.e.\ the loose definition of a starburst galaxy. The majority of the SMGs lie below this region, and we also bin the data to show that in the range $z=$\,2--5 the majority of our sample is composed of apparently main sequence systems. {\it Middle:} Gas depletion timescale versus $\Delta{\rm sSFR}$. We see no significant correlation between the two properties. {\it Right:} Gas depletion timescale $\tdep=M_{\rm gas}/{\rm SFR}$ versus redshift for our SMG sample, the PHIBSS CO-detected galaxies, a compilation of $z\sim$\,0.5--3 star-forming galaxies from \protect\cite{dessauges15}, and [C{\sc ii}]-detected galaxies from the ALPINE survey \protect\citep{dessauges20}. Our SMGs are consistent with no variation or a weak decline in the range $z\sim$\,1--5, with a median of $\tdep=$\,200\,$\pm$\,50\,Myr at $z\sim$\,2.8. The dashed line shows the prediction of \protect\cite{dave12} -- $\tdep\propto(1+z)^{-1.5}$, and the solid line shows our own fit of the form $\tdep\propto(1+z)^{a}$, from which we estimate $a=-$0.53\,$\pm$\,0.14.}}
\label{fig:ASPECS}
\end{figure*}

As early CO surveys were not limited by sensitivity, the detected sources were typically extreme and therefore more often starbursts, but in recent years there has been an increased effort to target more ``normal'' galaxies, and hence to systematically study the evolution of characteristic properties such as the gas depletion timescale and gas fraction \citep{genzel15, walter16, tacconi18}, largely thanks to the improving sensitivity of ALMA and NOEMA.

With the AS2UDS sample \citep{stach19, dudzeviciute20} we have for the first time been able to establish the stellar content and star-formation rates of large and unbiased samples of reliably-identified  SMGs, with results implying the SMGs at $z=$\,1.8--3.4 typically have higher specific star-formation rates than ``normal'' galaxies, whereas at higher-redshifts SMGs may have specific
star-formation rates more similar to the (increasingly more active) bulk population of galaxies. As we have seen in Fig.~\ref{fig:KS870}, our sample contains almost exclusively SMGs within the scatter of the sequence sources at $z\sim$\,3--5, with which we can compare to galaxies from other studies. We include data from the PHIBSS1 and PHIBSS2 surveys, including 148 CO-detected main-sequence star-forming galaxies observed with PdBI/NOEMA in two samples at $z=$\,0.5--2.5 \citep{genzel15, tacconi18}, and the ALMA Spectroscopic Survey in the HUDF (ASPECS), a CO blind scan from which 22 galaxies are CO- or \CI-detected at $z=$\,0.5--3.6 \cite{walter16}. Where available we also include {[C{\sc ii}]}-detected galaxies from the ALPINE survey \citep{lefevre19}. All gas masses are scaled using our chosen SLED and an $\alphaCO=$\,1.

Fig.~\ref{fig:ASPECS} shows the evolution of $\Delta{\rm sSFR}$ with redshift for our sources \citep[using the prescription of][see Fig.~\ref{fig:KS870}]{speagle14}. We indicate the arbitrary threshold for starburst galaxies, and see that only four of our 47 (9 per cent) CO-detected sources lie in this regime, with all four in the range $z\sim$\,1--1.5.
Binning the data by redshift, we see that our sample consists of galaxies within the broad scatter of the main sequence in the range $z\sim$\,2--4, with a handful of galaxies at $z>$\,4 on the boundary between main sequence and starbursts. The PHIBSS samples \citep{tacconi18} are complementary to our own in that they are comprised of sources with similar $\Delta{\rm sSFR}$ at typically lower redshifts than we probe.

\subsubsection{Gas depletion timescale}
\label{sec:tdep}

The gas depletion timescale is given by
\begin{equation}
\tdep = \dfrac{M_{\rm gas}}{\rm SFR},
\label{eq:tdep}
\end{equation}
i.e.\ the inverse of the star-formation efficiency, assuming no replenishment of the gas in the system, and no outflows. It has been suggested that this property is mainly dependent on redshift and offset from the main sequence \citep{genzel15, tacconi18}, with the redshift dependence controlling the evolution of the main sequence itself and the $\Delta{\rm sSFR}$ dependence implying that galaxies in a starburst phase consume their gas more quickly \citep{hodge20}. In the main-sequence paradigm, determining how this property evolves leads to a better understanding of how the molecular gas fractions evolve, and is therefore the starting point for deriving scaling relations. As noted earlier, we caution that there has been shown to be considerable variation in galaxy physical properties on and off the main sequence, which calls into question the usefulness of this paradigm.

%
%
\begin{figure*}
\centering
\includegraphics[width=\linewidth]{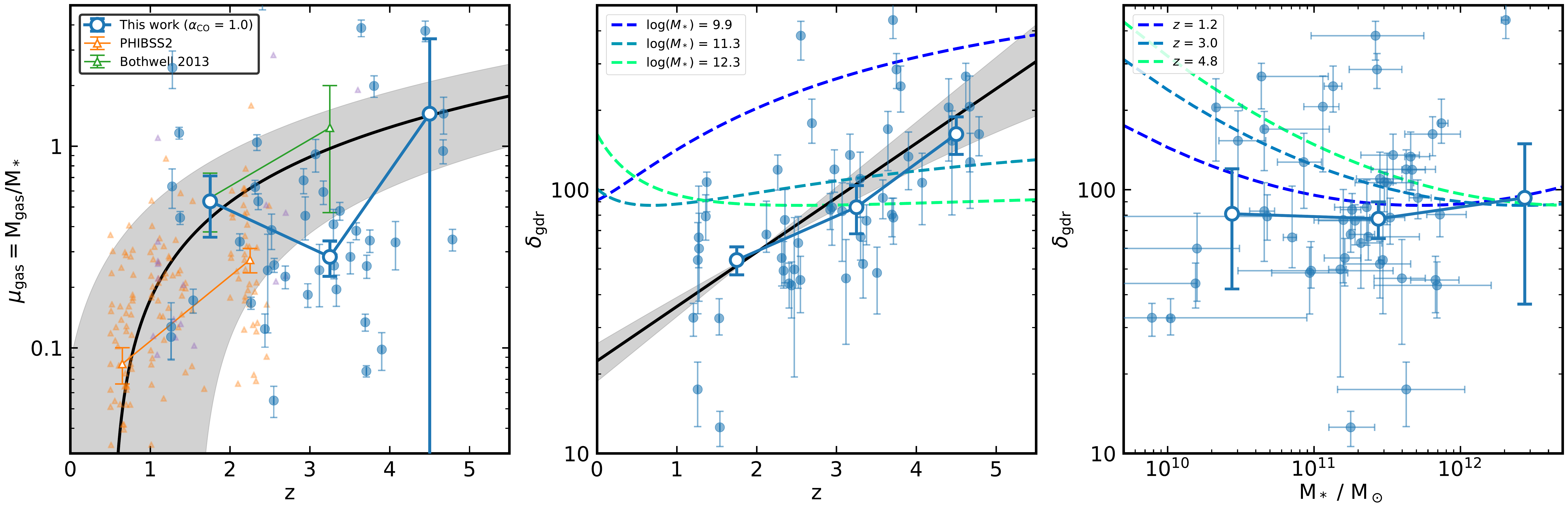}
\caption{\small{{\it Left:} Gas fraction $\mu_{\rm gas} = M_{\rm gas}/M_\ast$ versus redshift for the SMGs in our sample compared to the SMG survey of \protect\cite{bothwell13} and the typically lower-redshift PHIBSS surveys \protect\citep{tacconi18}. We show medians of our sample binned by redshift (large points).
The data show an increase in gas fraction with redshift that is rapid at low redshifts and beginning to plateau at $z\gtrsim3$.
{\it Middle:} Evolution of the gas-to-dust ratio $\gdr$ with redshift for our CO-detected sample. We fit a power law model which is consistent with a factor of $\sim$\,2 increase in $\gdr$ between $z=$\,2--5. We overlay tracks of the \protect\cite{tacconi18} prediction for the evolution of the gas-to-dust ratio with redshift at the minimum, median and maximum stellar masses of our sample. {\it Right:} Gas-to-dust ratio versus stellar mass. As with the middle panel we indicate the \protect\cite{tacconi18} prediction for the evolution of the gas-to-dust ratio, at the minimum, median and maximum redshifts of our sample.}}
\label{fig:fgas_z}
\end{figure*}

In Fig.~\ref{fig:ASPECS} we show the dependence of $\tdep$ on both $\Delta{\rm sSFR}$ and redshift separately. We find no discernible evolution of the depletion timescale with $\Delta{\rm sSFR}$ in the sample as a whole, and in fact the high-sSFR ``starbursts'' in our sample have relatively long timescales. As they are lower-redshift sources this likely reflects the evolution of the gas depletion timescale with redshift, which we also investigate in Fig.~\ref{fig:ASPECS}. The depletion timescale decreases with redshift in the range $z\sim$\,1--5, however our data are consistent with no evolution in the range $z=$\,2--5. We fit the form $\tdep\propto(1+z)^{a}$ to our data alone, finding $a=-$0.56\,$\pm$\,0.14, a much shallower dependence than the $a=-$1.5 proposed by \citep{dave12}. Additionally, when compared to the PHIBSS \citep{tacconi18} and ASPECS \citep{walter16} surveys we see that our sources exhibit longer depletion timescales, however we appear to be consistent with the [C{\sc ii}]-detected ALPINE galaxies \citep{lefevre19}. The scatter in our data is likely to be driven in part by variations in the SLEDs of individual sources (see Fig.~\ref{fig:SLED}), by the broad range in $\Delta{\rm sSFR}$ spanned by our sample and by the fact that rapidly evolving systems may naturally exhibit a wider range in $\tdep$.

\cite{tacconi18} suggest that the variation of the depletion time can be separated into the product of redshift, stellar mass and specific star-formation rate, providing an {\it Ansatz} of the form:
\begin{equation}
\log(\tdep) = A + B\log(1+z) + C\log(\Delta{\rm sSFR}) + D\log(\Delta M_\ast),
\label{eq:tacconi_tdep}
\end{equation}
where $\Delta M_\ast$ is defined as $M_\ast/$\,5\,$\times$\,10$^{10}$\,M$_\odot$ (5\,$\times$\,10$^{10}$\,M$_\odot$ is chosen as a fiducial stellar mass), and the coefficients A, B, C and D are to be determined. \cite{tacconi18} also include an optical half-light radius term, but given that they find this term to be of negligible important, and optical sizes are not useful measures of the sizes of high-redshift dust-obscured galaxies, we choose to ignore this.

From fitting this model to our data points, we find values for the coefficients of $A=-$0.29\,$\pm$\,0.7, $B=$\,0.05\,$\pm$\,0.11, $C=-$0.79\,$\pm$\,0.04 and $D=-$0.56\,$\pm$\,0.02. Hence in our sample, the gas depletion timescale is effectively independent of redshift, but strongly dependent on the offset from the ``main sequence'' and stellar mass. In contrast, \cite{tacconi18} find values of $A=$\,0.06\,$\pm$\,0.03, $B=-$0.44\,$\pm$\,0.13, $C=-$0.43\,$\pm$\,0.03 and $D=$\,0.17\,$\pm$\,0.04 when only considering CO-detected sources. Thus they find a decrease in $\tdep$ with redshift, and increase with stellar mass, which are not supported by our sample, however they also see a decrease with offset from the main sequence as we do (see Fig.~\ref{fig:ASPECS}).

Our sample displays a median $\tdep=$\,200\,$\pm$\,40\,Myr. \cite{dudzeviciute20} use a 50 per cent efficiency conversion factor when estimating the depletion timescale, but correcting for this factor their median is $\tdep=$\,292\,$\pm$\,10\,Myr for the 707 AS2UDS SMGs. Given that our sample has a median redshift of $z=$\,3.0\,$\pm$\,0.2, slightly higher than the $z=$\,2.61\,$\pm$\,0.08 for the AS2UDS sample from \cite{dudzeviciute20}, over which we see very little evolution in the depletion timescale, the difference is likely resultant from our selection of a subset of the brightest SMGs, which are therefore more active and have lower depletion timescales.

\subsubsection{Gas fraction}
\label{sec:fgas}

In addition to the gas depletion timescale, we can derive the  gas fraction
\begin{equation}
\mu_{\rm gas} = \dfrac{M_{\rm gas}}{M_\ast}
\label{eq:gas_frac}
\end{equation}
for our sources using our CO-based mass estimates and stellar masses from {\sc magphys}. The gas fraction is also expected to be a key property in galaxy evolution, following from the gas depletion timescale, describing the amount of fuel available for star formation \citep{tacconi18}. Fig.~\ref{fig:fgas_z} shows the evolution of $\mu_{\rm gas}$ with redshift, where we have included all of our CO-detected sources in addition to PHIBSS sources. The gas fraction increases with redshift, displaying a strong evolution at low redshift and a more gradual increase at high redshift \citep{geach11, tacconi18, liu19b}. As with Fig.~\ref{fig:ASPECS} we see a large amount of scatter, and by separately examining the gas and stellar masses in our sample we conclude that the trend we see in $\mu_{\rm gas}$ with redshift is driven mainly by sources at higher redshift having larger gas reservoirs. An additional explanation for the scatter could then be that these galaxies are consuming gas on very short timescales leading to wider variations in the gas fraction within the observed population.

As with the previous section, we compare our data with CO-detected main sequence galaxies from \cite{tacconi18}, who use Eq.~\ref{eq:tacconi_tdep} to derive the following model for $\mu_{\rm gas}$:
\begin{equation}
\log(\mu_{\rm gas}) = A + B(\log(1+z)-F)^\beta + C\log(\Delta{\rm sSFR}) + D\log(\Delta M_\ast),
\label{eq:tacconi_fgas}
\end{equation}
where the new parameters $F$ and $\beta$ are introduced to capture variations in main-sequence with redshift. As before we fit the model to our data, fixing $\beta=2$ \cite{tacconi18} and finding $A=-$\,0.03\,$\pm$\,0.09, $B=$\,0.4\,$\pm$\,1.6, $C=-$0.84\,$\pm$\,0.08, $D=-$0.69\,$\pm$\,0.04 and $F=$\,0.6\,$\pm$\,0.4, compared with the \cite{tacconi18} result of $A=$\,0.2\,$\pm$\,0.2, $B=-$3.4\,$\pm$\,0.8, $C=$\,0.56\,$\pm$\,0.03, $D=-$0.30\,$\pm$\,0.04 and $F=$\,0.7\,$\pm$\,0.2.   
The main differences between the two fits is that now we find that the SMGs exhibit stronger (positive) evolution in their gas fractions with redshift than the ``main sequence'' population, with comparable dependence on
offset from the ``main sequence'' and stellar mass. 

%
%
\begin{figure*}
\begin{center}
\subfloat{\includegraphics[width=\linewidth]{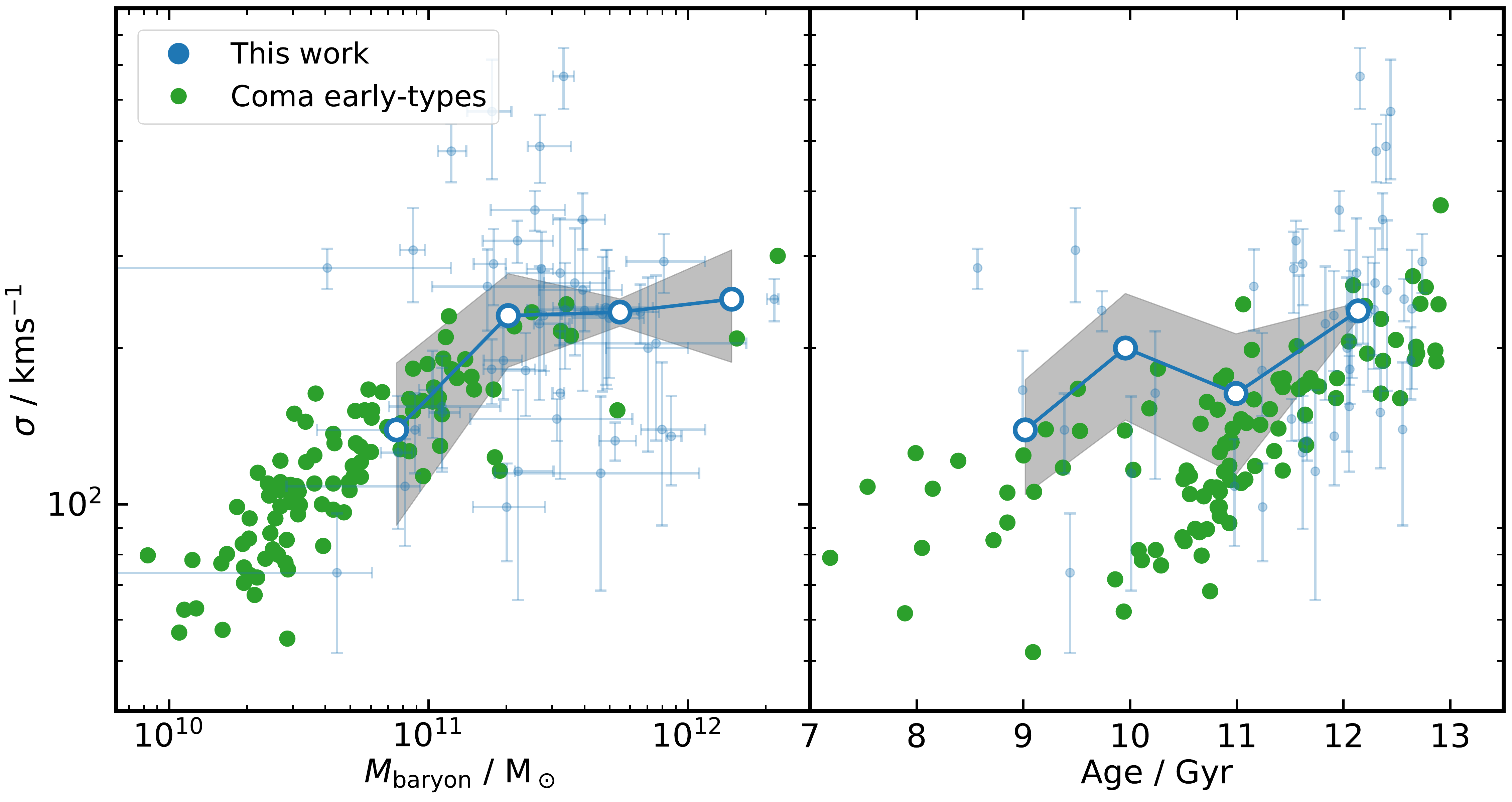}}
\caption{\small{{\it Left:} The mass--$\sigma$ relation for CO-detected SMGs in our sample and early-type galaxies in the Coma cluster from \citet{shetty20}. The open points and grey region represent the median of our sample in $M_\ast$ bins and its bootstrapped median, respectively. The SMGs are consistent with the  trend shown by that of the early-types, providing further circumstantial evidence that they could represent progenitors of such systems. We interpret the scatter in our data as an inclination angle effect, which we are unable to correct for on a case-by-case basis. 
{\it Right:} The variation in estimated stellar age as a function of velocity dispersion of the early-type galaxies in Coma from \citet{shetty20}, compared to the estimated formation ages of the SMGs in our sample, derived from the
lookback time to their observed redshift and the estimate of their expected age from the {\sc magphys} SED fitting.  We see that the trend in the SMGs roughly delineates the high-mass boundary as a function of age estimated for the early-type galaxies.
}}
\label{fig:mstarsigma}
\end{center}
\end{figure*}

\subsubsection{Gas-to-dust ratio}
\label{sec:gdr}

The relationship between the molecular gas and the dust content is encoded in the gas-to-dust ratio $\gdr$. As discussed in \S\ref{sec:gas_masses}, it is expected that the gas-to-dust ratio depends on metallicity, with more massive (and therefore probably more metal-rich) galaxies containing more dust and therefore having a lower $\gdr$ \citep{li19}. Here we wish to investigate any potential variation of $\gdr$ with redshift.
In Fig.\ref{fig:fgas_z} we show the gas-to-dust ratio of our sources adopting $\alphaCO=$\,1\,M$_\odot$ (K\,km\,s$^{-1}$\,pc$^{2}$)$^{-1}$ and using dust masses estimated from {\sc magphys} SED fitting. The SMGs display an increase in the gas-to-dust ratio by a factor $\sim$\,3 across the redshift range $z\sim$\,1--5. It also appears that galaxies with higher stellar masses have larger gas-to-dust ratios, contrary to what is expected. We note however that if we were to adopt a CO--H$_2$ conversion factor that decreases with stellar mass (and therefore metallicity), we would then find a more constant $\gdr$. We also overlay the model of \cite{tacconi18} for the evolution of the gas-to-dust ratio with redshift, adopting the minimum, median and maximum stellar mass of our sample. These models are predicted to display little-to-no evolution in the range of redshifts spanned by our sources, particularly for more massive sources, and are broadly consistent with the range in $\gdr$ we see in our sample as a function of stellar mass.

\subsection{Implications for galaxy evolution}
\label{sec:evo}

It has been proposed that submillimetre galaxies are the progenitors of massive and compact early-type galaxies in the local Universe \citep{lilly99,  simpson14, toft14}, undergoing an evolution which may proceed through a Quasi-Stellar Object (QSO) phase \citep{sanders88,swinbank06, hopkins08}. \citet{simpson14} showed that SMGs would evolve to have $z\sim$\,0 stellar masses to massive early-types (see also \citealt{dudzeviciute20}), while \cite{hodge16} demonstrated that the implied effective radii and gas surface densities of SMGs are consistent with that of the most compact massive early-types. 

We can now apply an additional test of this hypothesis using the kinematic
information from our CO survey in the context of the $M_{\rm baryon}$--$\sigma$ and $\sigma$--Age relations. We plot these in Fig.~\ref{fig:mstarsigma},  where we compare the properties of the SMGs to early-type galaxies in the Coma cluster from \citet{shetty20}.   For this comparison the baryonic masses of the SMGs comprises the sum of their stellar and gas masses, while for the (generally gas-poor) local early-types we use just their stellar masses.  For the ages of the SMGs we convert their redshifts into a lookback time and add to this the estimated ages of the systems from the {\sc magphys} SED fits to determine a crude ``formation'' age, to compare to the locally derived stellar ages from \citet{shetty20}.
We adopt $\sigma$ derived from the CO line width as our measure of the expected velocity dispersion of the descendent galaxies and because we lack individual inclination estimates for the galaxies we have to average over the population to remove the sensitivity to inclination. We therefore plot the binned median velocity dispersions as a function of stellar mass (or age) for the SMGs and a bootstrap estimate of the uncertainty in this on  Fig.~\ref{fig:mstarsigma}.  

We see that the trend in $M_{\rm baryon}$--$\sigma$ for the SMGs is a  good match for the distribution of the most massive early-type galaxies from \citet{shetty20}, not only in normalisation but also showing hints  of flattening in $\sigma$ seen at the highest masses which arises from the so-called Zone of Exclusion \citep{bender92,shetty20}. As \cite{shetty20} highlight the inflection point at masses of $\sim$\,2\,$\times$\,10$^{11}$\,M$_\odot$ corresponds to the point separating low-mass fast-rotator early-type galaxies with disks, from the more massive, round slow-rotators.  Our SMG population at $z\sim$\,3 straddle this transition, with an 870-$\mu$m flux of $S_{870}\sim $\,5\,mJy roughly corresponding to the boundary.  This flux also roughly marks the break in the SMG counts \citep{stach18} suggesting that the physical origin of this difference in the properties of early-type galaxies in the local Universe may be reflected in the properties and evolution of SMGs above and below this flux. $S_{870}\gtrsim $\,5\,mJy typically lie at higher redshifts (Fig.~\ref{fig:zdist}), having higher gas masses and gas fractions, and shorter depletion timescales (Fig.~\ref{fig:gas_masses} and \ref{fig:fgas_z}).

In terms of the $\sigma$--Age plot in Fig.~\ref{fig:mstarsigma}, we see that our derived formation ages for the SMGs tend to follow the boundary of the distribution derived for the Coma sample.  This is a result of our sample being dominated by the most massive systems as a result of our selection on dust mass. We therefore pick out the most massive galaxies formed at any epoch and so our median line tracks the upper boundary of the Coma population in this plot. 

We also note that as our CO survey is predominantly limited to the most massive gas-rich galaxies, a more sensitive survey for less luminous SMGs would likely extend to lower baryonic masses and so lower velocity dispersions and expand the overlap with the parameter
space populated by the Coma early-type galaxies in both mass and age. Nevertheless, we conclude that our CO observations indicate that the most massive SMGs are dynamically consistent with them being the progenitors of most massive compact early-type galaxies in the local Universe.  We stress that this does not preclude further, dry, merging of these systems as that is expected to predominantly influence the sizes of the galaxies, while leaving $\sigma$ relatively uneffected.

\section{Conclusions}
\label{sec:conclusions}

We have undertaken a spectroscopic survey of 61 ALMA-located SMGs in the 3\,mm band ($\nu\sim$\,82--114\,GHz) using ALMA and NOEMA to search for emission lines from the rotational transitions of CO gas. Our sample roughly divides in half: with 31  submillimetre bright, but typically optically faint/$K$-faint, SMGs lacking existing spectroscopic redshifts, and a complementary sample of 30 typically submillimetre-fainter SMGs that have optical/UV spectroscopic redshifts. For our survey we obtained complete spectral scans of the 3-mm window for the former sample, but more targetted spectral coverage of the latter. This combined sample allows us to probe a wide parameter space in the SMG population. Our main findings are as follows:
\begin{itemize}

\item CO line emission is detected in 45 of the targets, 27 of which come from blind spectral scans and 19 of which come from the targetted observations with sources with optical/UV spectroscopic redshifts, with a further two CO detections in nearby ALMA-detected SMGs. This provides a large sample of high-S/N CO detections in high-redshift galaxies.  Non-detections in the SMGs with existing
spectroscopic redshifts are likely due to inaccuracies in those redshifts, whereas we suggest non-detections in the blind spectral scan are due to
these being predominantly CO faint due to their low gas masses.
 We also uncover three serendipitous detections of CO in our datacubes at redshift $z\sim$\,1.4--2.2, although these sources are unrelated to their nearby SMG.

\item 38\,$\pm$\,9 per cent of our CO lines are best fit by double-Gaussian profiles. By simulating spatially-unresolved observations of rotation curves we show this to be consistent with a population of randomly-oriented rotating disks.  This is strong circumstantial evidence that the CO kinematics of most SMGs are dominated by the motion of gas in a rotating disk, although we stress that such disks can rapidly form during dynamical interactions and mergers.

\item The median redshift of our sample is $z=$\,3.0\,$\pm$\,0.2. We compare the redshift distribution of the optical/$K$-faint SMGs with a
flux-limited SMG sample from AS2UDS to show that these sources are found at higher redshifts than fainter SMGs, confirming previous claims of   correlation between 870\,$\mu$m flux density and redshift in this population. We measure a gradient in this correlation of 0.11\,$\pm$\,0.04\,mJy$^{-1}$, in agreement with \cite{stach19} and \cite{simpson20}. This represents potential evidence for galaxy downsizing, the phenomenon where more massive galaxies, with higher gas fractions, form earlier.

\item We study the average ISM excitation properties of SMGs by constructing a composite CO spectral line energy distribution from our own data and archival observations, finding that excitation increases with $J_{\rm up}$ up to $J_{\rm up}=$\,6. We derive line ratios that are consistent with that of SMM\,J2135$-$0102 (the ``Cosmic Eyelash'') measured in \cite{danielson11}. Using these line ratios to convert from $L'_{\rm CO,J}$ to $L'_{\rm CO(1-0)}$ we find that, as expected, our sources lie at the bright end of the $\LCO-L_{\rm IR}$ relation, with median $L'_{\rm CO(1-0)}=($6.7\,$\pm$\,0.5$)\times$\,10$^{10}$\,K\,km\,s$^{-1}$\,pc$^{2}$ and $L_{\rm IR}=($5\,$\pm$\,1$)\times$\,10$^{12}$\,L$_\odot$.

\item By combining CO line luminosities, dynamical masses estimated from CO line widths and stellar masses derived from SED fitting with {\sc magphys}, we find our sample to be consistent with a median CO--H$_2$ conversion factor of $\alphaCO=$\,1.0\,$\pm$\,0.7, assuming our sources are randomly oriented disks, but with large systematic uncertainties due to assumptions about dark matter fraction and the radial extent of the CO emission. Adopting $\alphaCO=$\,1 results in a median gas mass of $M_{\rm gas}=\,($5.3\,$\pm$\,0.7$)\times$\,10$^{10}$\,M$_\odot$. We also find a correlation between the CO line luminosity and the line width  with a power law index of 2, consistent with that expected for disk-dominated kinematics.

\item We compare the CO luminosity with two independent tracers of the molecular gas mass: the dust mass and the \CI luminosity, finding all three to correlate well where the samples are  reliable. Given the difficulty in detecting \CI emission with high significance, we suggest that the CO luminosity and dust mass, as estimated from the {\it restframe} 870-$\mu$m continuum measured in the 3-mm band, are the best correlated measures and hence the preferable choices for estimating H$_2$ masses. We use our data to estimate the average ratio between the gas-to-dust ratio and CO--H$_2$ conversion factor finding $\gdr/\alphaCO=$\,45\,$\pm$\,7 where the rest-frame 870\,$\mu$m luminosities are used to estimate the dust mass and $\gdr/\alphaCO=$\,64\,$\pm$\,9 where the dust masses from {\sc magphys} are used. However, to make reliable use of the dust mass method to estimate the gas mass also requires tighter constraints on $\alphaCO$.

\item We find that our sample is mostly comprised of galaxies whose estimated specific star-formation rates place them within the scatter of the main sequence at their respective redshifts, with the exception of a few lower-redshift starbursts. We study the properties of our sources in the context of scaling relations of the gas depletion timescale and gas fraction. At a median redshift of $z=$\,2.8, we find a median gas depletion time scale of 200\,$\pm$\,50\,Myr.

\item We use the CO line kinematics along with the estimated stellar and gas masses for our sample to demonstrate that the distribution of SMGs in the 
$M_{\rm baryon}$--$\sigma$ plane is  very similar to that of the most massive early-type galaxies in the local Universe, both in normalisation and shape.  While our selection of the highest dust mass galaxies, means that the expected age distribution of their descendants at $z\sim$\,0 matches the high-mass boundary of the distribution of Coma galaxies on the $\sigma$--Age plane. These two results provide further circumstantial evidence of a link between SMGs and the progenitors of massive early-type galaxies.  Moreover, the median trend in the SMGs spans the characteristic mass where the properties of local early-type galaxies transition from fast-rotating at lower masses to slow-rotating at higher masses.  In the SMG population this mass corresponds roughly to an 870-$\mu$m flux of $\sim$\,5\,mJy suggesting that the origin of this difference in the properties of ellipticals may be reflected in the properties and evolution of SMGs above and below this flux (which also marks a break in the SMG counts, \citealt{stach18}).

\end{itemize}

\begin{table*}
	\centering
\caption{Target details. Photometric redshifts, along with dust and stellar masses, are derived from SED fitting with {\sc magphys}. Spectroscopic redshifts are taken from \protect\cite{danielson17} (ALESS) and \protect\cite{dudzeviciute20} (AS2UDS).}
\setlength\extrarowheight{1pt}
\begin{tabular}{C{2.0cm}C{1.3cm}C{1.7cm}C{1.7cm}C{1.5cm}C{0.8cm}C{0.8cm}C{1.3cm}C{1.3cm}C{1.3cm}} \hline \hline
\thead{Source ID} & \thead{Subsample} & \thead{R.A.\ \\ (J2000)} & \thead{Dec.\ \\ (J2000)} & \thead{$S_{870}$ \\ (mJy)} & \thead{$z_\mathrm{phot}$} & \thead{$z_\mathrm{spec}$} &\thead{$M_\mathrm{dust}$ \\ (10$^{8}$\,M$_\odot$)} & \thead{$M_*$ \\ (10$^{10}$\,M$_\odot$)} & \thead{SFR\\ (M$_\odot {\rm yr}^{-1}$)} \\ \hline 
\noalign{\smallskip}
ALESS001.1 & {\it Scan} & 03:33:14.46 & $-$27:56:14.52 & 6.7 $\pm$ 0.5 & 4.78$_{-1.81}^{+2.65}$ & 4.954 & 10$_{-2}^{+3}$ & 9$_{-3}^{+3}$ & 680$_{-190}^{+200}$ \\ 
ALESS003.1 & {\it Scan} & 03:33:21.50 & $-$27:55:20.29 & 8.3 $\pm$ 0.4 & 3.88$_{-0.76}^{+0.90}$ & 4.237 & 12$_{-1}^{+2}$ & 19$_{-8}^{+8}$ & 710$_{-110}^{+130}$ \\ 
ALESS005.1 & {\it Scan} & 03:31:28.91 & $-$27:59:09.02 & 7.8 $\pm$ 0.7 & 3.67$_{-0.20}^{+0.05}$ & ... & 9$_{-1}^{+1}$ & 23.4$_{-0.5}^{+30}$ & 900$_{-300}^{+20}$ \\ 
\multicolumn{10}{c}{.....} \\
\noalign{\smallskip}
\hline \hline 
\end{tabular}
\label{tab:observed}
\end{table*}

\begin{table*}
\begin{threeparttable}
	\centering
	\caption{Line properties for sources with CO line detections. Sources in italics have single line detections where the identification of the transition relies on the photometric redshift PDF. The frequency given here is the central frequency measured from a Gaussian fit to each source. The quoted values of the velocity-integrated line intensity $I_{\rm CO}$, the redshift $z_{\rm CO}$ and the linewidth FWHM$_{\rm CO}$ are calculated from the zeroth, first and second moments of the line as described in \S\ref{sec:fitting}. 3\,mm continuum flux densities $S_{\rm 3\,mm}$ are measured from the line and continuum fit. We also quote the CO line luminosity $L'_{{\rm CO},J_{\rm up}}$ measured for the detected transition, which is indicated by the $J_{\rm up}$ column. The median fractional error on $z_{\rm CO}$ is $(2.1\pm0.3)\times10^{-4}$. S/N$_{\rm CO}$ is the single-to-noise ratio of the CO emission line, and $Q$ is the optical/UV spectroscopic redshift quality factor from \protect\cite{danielson17}.
}
    \setlength\extrarowheight{1pt}
\begin{tabular}{C{2.3cm}C{0.5cm}C{1.4cm}C{1.4cm}C{1.4cm}C{1.6cm}C{1.5cm}C{2.5cm}C{0.9cm}C{0.9cm}} \hline \hline
\thead{Source ID} & \thead{$J_{\rm up}$} & \thead{Frequency \\ (GHz)} & \thead{$I_{\rm CO}$ \\ (Jy\,km\,s$^{-1}$)} & \thead{$z_{\rm CO}$} & \thead{FWHM$_{\rm CO}$ \\ (km\,s$^{-1}$)} &\thead{$S_{\rm 3\,mm}$ \\ ($\mu$Jy)} & \thead{$L'_{{\rm CO},J_{\rm up}}$ \\ (10$^{10}$ K\,km\,s$^{-1}$\,pc$^{-2}$)} & \thead{S/N$_{\rm CO}$} & \thead{$Q$} \\ \hline 
\noalign{\smallskip}
{\it ALESS001.1}$\tnote{$\dagger$}$ & $5$ & $101.55$ & $1.0\pm0.2$ & $4.674$ & $1300 \pm 300$ & $170 \pm 5$ & $3.2 \pm 0.7$ & 5.1 & 3\\ 
{\it ALESS001.2} & $5$ & $101.66$ & $0.84\pm0.11$ & $4.669$ & $440 \pm 70$ & $98 \pm 16$ & $2.8 \pm 0.4$ & 8.0 & 3\\ 
ALESS003.1$\tnote{$\dagger$}$ & $4$ & $105.39$ & $1.08\pm0.11$ & $3.375$ & $870 \pm 80$ & $134 \pm 4$ & $3.3 \pm 0.3$ & 10.1 & 3\\ 
\multicolumn{10}{c}{.....} \\
\noalign{\smallskip}
\hline \hline 
\end{tabular}\begin{tablenotes}\footnotesize
\item[$\dagger$] Source displays a double-peaked line profile.
\end{tablenotes}
	\label{tab:measured}
\end{threeparttable}
\end{table*}

\begin{table*}
\begin{threeparttable}
	\centering
	\caption{Basic properties of the serendipitously detected CO line emitters, including sources documented by \protect\cite{wardlow18}.
}
    \setlength\extrarowheight{0pt}
\begin{tabular}{C{2.5cm}C{1.5cm}C{1.5cm}C{0.7cm}C{1.4cm}C{2cm}C{1.7cm}C{2.5cm}} \hline \hline
\thead{Source ID} & \thead{R.A.\ \\ (J2000)} & \thead{Dec.\ \\ (J2000)} & \thead{$J_{\rm up}$} & \thead{Freq.\ \\ (GHz)} & \thead{$z_{\rm CO}$} & \thead{FWHM$_{\rm CO}$ \\ (km\,s$^{-1}$)} & \thead{$L'_{{\rm CO},J_{\rm up}}$ \\ (10$^{10}$ K\,km\,s$^{-1}$\,pc$^{-2}$)} \\ \hline 
\noalign{\smallskip}
ALESS019.CO & 03:32:08.37 & $-$27:58:14.50 & 4 & 97.1 & $3.7507\pm0.0004$ & $230\pm70$ & $1.1\pm0.3$ \\
ALESS023.CO & 03:32:12.28 & $-$28:05:17.30 & 2 & 88.8 & $1.5943\pm0.0014$ & $500\pm200$ & $1.6\pm0.4$ \\
ALESS049.CO\tnote{$\dagger$} & 03:31:24.72 & $-$27:50:43.70 & 3 & 88.0 & $2.9300 \pm 0.0003$ & $550 \pm 60$ & $0.68 \pm 0.07$ \\
\multicolumn{8}{c}{.....} \\
\noalign{\smallskip}
\hline \hline
\end{tabular}
\begin{tablenotes}\footnotesize
\item[$\dagger$] Published in \cite{wardlow18}.
\end{tablenotes}

	\label{tab:serendips}
\end{threeparttable}
\end{table*}

\begin{table*}
\begin{threeparttable}
	\centering
	\caption{Properties of the \CIfull detections in our sample. When fitting the \CI emission lines we fix the redshift to be equivalent to the CO redshift, i.e.\ $z_{\rm CO} = z_{\rm [CI]}$.
    }
    \setlength\extrarowheight{0pt}
\begin{tabular}{C{2.5cm}C{0.7cm}C{1.4cm}C{1.4cm}C{2.5cm}} \hline \hline
\thead{Source ID} & \thead{$J_{\rm up}$} & \thead{$z_{\rm CO}$} & \thead{$I_{\rm [CI]}$ \\ (Jy\,km\,s$^{-1}$)} & \thead{$L'_{{\rm [CI]}}$ \\ (10$^{10}$ K kms$^{-1}$ pc$^{-2}$)} \\ \hline 
\noalign{\smallskip}
ALESS003.1 & $4$ & $3.375$ & $0.52\pm0.13$ & $1.4\pm0.4$ \\ 
ALESS005.1 & $4$ & $3.303$ & $0.47\pm0.15$ & $1.2\pm0.4$ \\ 
ALESS009.1 & $4$ & $3.694$ & $0.51\pm0.09$ & $1.6\pm0.3$ \\ 
\multicolumn{5}{c}{.....} \\
\noalign{\smallskip}
\hline \hline 
\end{tabular}
	\label{tab:CI}
\end{threeparttable}
\end{table*}

\section*{Acknowledgements}

JEB acknowledges the support of STFC studentship (ST/S50536/1). JLW acknowledges support from an STFC Ernest Rutherford Fellowship (ST/P004784/1 and ST/P004784/2). The Durham co-authors acknowledge support from STFC (ST/P000541/1) and (ST/T000244/1). Y.A. acknowledges support by NSFC grant 11933011. C.C.C. acknowledges support from the Ministry of Science and Technology of Taiwan (MOST 109-2112-M-001-016-MY3). H.D. acknowledges financial support from the Spanish Ministry of Science, Innovation and Universities (MICIU) under the 2014 Ramón y Cajal program RYC-2014-15686 and AYA2017-84061-P, the later one co-financed by FEDER (European Regional Development Funds). BG thanks the Hasselblad Foundation. JH acknowledges support of the VIDI research programme with project number 639.042.611, which is (partly) financed by the Netherlands Organisation for Scientific Research (NWO). YM acknowledges support from JSPS KAKENHI Grant (17H04831 and 17KK0098). This work is based on observations carried out under project numbers S18CG and W18EL with the IRAM NOEMA Interferometer. IRAM is supported by INSU/CNRS (France), MPG (Germany) and IGN (Spain). This paper makes use of the following ALMA data: ADS/JAO.ALMA\#2016.1.00564.S, \#2017.1.01163.S, \#2017.1.01512.S\, and \#2019.1.00337.S. ALMA is a partnership of ESO (representing its member states), NSF (USA) and NINS (Japan), together with NRC (Canada), MOST and ASIAA (Taiwan), and KASI (Republic of Korea), in cooperation with the Republic of Chile. The Joint ALMA Observatory is operated by ESO, AUI/NRAO and NAOJ.

\section*{Data availability}

The data used in this paper are available through the ALMA and IRAM/NOEMA data archives, although some are currently still subject to a proprietary period (One year for ALMA data and three years for NOEMA data). Reduced data products can be shared upon publication by request to the author.



\bibliographystyle{mnras}
\bibliography{bibliography}

\begin{thebibliography}{}
\makeatletter
\relax
\def\mn@urlcharsother{\let\do\@makeother \do\$\do\&\do\#\do\^\do\_\do\%\do\~}
\def\mn@doi{\begingroup\mn@urlcharsother \@ifnextchar [ {\mn@doi@}
  {\mn@doi@[]}}
\def\mn@doi@[#1]#2{\def\@tempa{#1}\ifx\@tempa\@empty \href
  {http://dx.doi.org/#2} {doi:#2}\else \href {http://dx.doi.org/#2} {#1}\fi
  \endgroup}
\def\mn@eprint#1#2{\mn@eprint@#1:#2::\@nil}
\def\mn@eprint@arXiv#1{\href {http://arxiv.org/abs/#1} {{\tt arXiv:#1}}}
\def\mn@eprint@dblp#1{\href {http://dblp.uni-trier.de/rec/bibtex/#1.xml}
  {dblp:#1}}
\def\mn@eprint@#1:#2:#3:#4\@nil{\def\@tempa {#1}\def\@tempb {#2}\def\@tempc
  {#3}\ifx \@tempc \@empty \let \@tempc \@tempb \let \@tempb \@tempa \fi \ifx
  \@tempb \@empty \def\@tempb {arXiv}\fi \@ifundefined
  {mn@eprint@\@tempb}{\@tempb:\@tempc}{\expandafter \expandafter \csname
  mn@eprint@\@tempb\endcsname \expandafter{\@tempc}}}

\bibitem[\protect\citeauthoryear{{Akaike}}{{Akaike}}{1974}]{akaike1974}
{Akaike} H.,  1974, IEEE Transactions on Automatic Control, \href
  {https://ui.adsabs.harvard.edu/abs/1974ITAC...19..716A} {19, 716}

\bibitem[\protect\citeauthoryear{{Alaghband-Zadeh} et~al.,}{{Alaghband-Zadeh}
  et~al.}{2013}]{alaghband-zadeh13}
{Alaghband-Zadeh} S.,  et~al., 2013, \mn@doi [\mnras] {10.1093/mnras/stt1390},
  \href {https://ui.adsabs.harvard.edu/abs/2013MNRAS.435.1493A} {435, 1493}

\bibitem[\protect\citeauthoryear{{Alloin}, {Barvainis}  \&
  {Guilloteau}}{{Alloin} et~al.}{2000}]{alloin00}
{Alloin} D.,  {Barvainis} R.,   {Guilloteau} S.,  2000, \mn@doi [\apjl]
  {10.1086/312429}, \href
  {https://ui.adsabs.harvard.edu/abs/2000ApJ...528L..81A} {528, L81}

\bibitem[\protect\citeauthoryear{{Andreani}, {Cimatti}, {Loinard}  \&
  {R{\"o}ttgering}}{{Andreani} et~al.}{2000}]{andreani00}
{Andreani} P.,  {Cimatti} A.,  {Loinard} L.,   {R{\"o}ttgering} H.,  2000,
  \aap, \href {https://ui.adsabs.harvard.edu/abs/2000A&A...354L...1A} {354, L1}

\bibitem[\protect\citeauthoryear{{Aravena} et~al.,}{{Aravena}
  et~al.}{2010}]{aravena10}
{Aravena} M.,  et~al., 2010, \mn@doi [\apj] {10.1088/0004-637X/718/1/177},
  \href {https://ui.adsabs.harvard.edu/abs/2010ApJ...718..177A} {718, 177}

\bibitem[\protect\citeauthoryear{{Aravena} et~al.,}{{Aravena}
  et~al.}{2012}]{aravena12}
{Aravena} M.,  et~al., 2012, \mn@doi [\mnras]
  {10.1111/j.1365-2966.2012.21697.x}, \href
  {https://ui.adsabs.harvard.edu/abs/2012MNRAS.426..258A} {426, 258}

\bibitem[\protect\citeauthoryear{{Archibald}, {Dunlop}, {Hughes}, {Rawlings},
  {Eales}  \& {Ivison}}{{Archibald} et~al.}{2001}]{archibald01}
{Archibald} E.~N.,  {Dunlop} J.~S.,  {Hughes} D.~H.,  {Rawlings} S.,  {Eales}
  S.~A.,   {Ivison} R.~J.,  2001, \mn@doi [\mnras]
  {10.1046/j.1365-8711.2001.04188.x}, \href
  {https://ui.adsabs.harvard.edu/abs/2001MNRAS.323..417A} {323, 417}

\bibitem[\protect\citeauthoryear{{Barger}, {Cowie}  \& {Richards}}{{Barger}
  et~al.}{2000}]{barger00}
{Barger} A.~J.,  {Cowie} L.~L.,   {Richards} E.~A.,  2000, \mn@doi [\aj]
  {10.1086/301341}, \href
  {https://ui.adsabs.harvard.edu/abs/2000AJ....119.2092B} {119, 2092}

\bibitem[\protect\citeauthoryear{{Barger}, {Wang}, {Cowie}, {Owen}, {Chen}  \&
  {Williams}}{{Barger} et~al.}{2012}]{barger12}
{Barger} A.~J.,  {Wang} W.~H.,  {Cowie} L.~L.,  {Owen} F.~N.,  {Chen} C.~C.,
  {Williams} J.~P.,  2012, \mn@doi [\apj] {10.1088/0004-637X/761/2/89}, \href
  {https://ui.adsabs.harvard.edu/abs/2012ApJ...761...89B} {761, 89}

\bibitem[\protect\citeauthoryear{{Battisti} et~al.,}{{Battisti}
  et~al.}{2019}]{battisti19}
{Battisti} A.~J.,  et~al., 2019, \mn@doi [\apj] {10.3847/1538-4357/ab345d},
  \href {https://ui.adsabs.harvard.edu/abs/2019ApJ...882...61B} {882, 61}

\bibitem[\protect\citeauthoryear{{Baugh}, {Lacey}, {Frenk}, {Granato}, {Silva},
  {Bressan}, {Benson}  \& {Cole}}{{Baugh} et~al.}{2005}]{baugh05}
{Baugh} C.~M.,  {Lacey} C.~G.,  {Frenk} C.~S.,  {Granato} G.~L.,  {Silva} L.,
  {Bressan} A.,  {Benson} A.~J.,   {Cole} S.,  2005, \mn@doi [\mnras]
  {10.1111/j.1365-2966.2004.08553.x}, \href
  {https://ui.adsabs.harvard.edu/abs/2005MNRAS.356.1191B} {356, 1191}

\bibitem[\protect\citeauthoryear{{Bender}, {Burstein}  \& {Faber}}{{Bender}
  et~al.}{1992}]{bender92}
{Bender} R.,  {Burstein} D.,   {Faber} S.~M.,  1992, \mn@doi [\apj]
  {10.1086/171940}, \href
  {https://ui.adsabs.harvard.edu/abs/1992ApJ...399..462B} {399, 462}

\bibitem[\protect\citeauthoryear{{Binney} \& {Tremaine}}{{Binney} \&
  {Tremaine}}{2008}]{galacticdynamics}
{Binney} J.,  {Tremaine} S.,  2008, {Galactic Dynamics: Second Edition}

\bibitem[\protect\citeauthoryear{{Blain}, {Smail}, {Ivison}, {Kneib}  \&
  {Frayer}}{{Blain} et~al.}{2002}]{blain02}
{Blain} A.~W.,  {Smail} I.,  {Ivison} R.~J.,  {Kneib} J.~P.,   {Frayer} D.~T.,
  2002, \mn@doi [\physrep] {10.1016/S0370-1573(02)00134-5}, \href
  {https://ui.adsabs.harvard.edu/abs/2002PhR...369..111B} {369, 111}

\bibitem[\protect\citeauthoryear{{Bolatto}, {Wolfire}  \& {Leroy}}{{Bolatto}
  et~al.}{2013}]{bolatto13}
{Bolatto} A.~D.,  {Wolfire} M.,   {Leroy} A.~K.,  2013, \mn@doi [\araa]
  {10.1146/annurev-astro-082812-140944}, \href
  {https://ui.adsabs.harvard.edu/abs/2013ARA&A..51..207B} {51, 207}

\bibitem[\protect\citeauthoryear{{Bothwell} et~al.,}{{Bothwell}
  et~al.}{2013}]{bothwell13}
{Bothwell} M.~S.,  et~al., 2013, \mn@doi [\mnras] {10.1093/mnras/sts562}, \href
  {https://ui.adsabs.harvard.edu/abs/2013MNRAS.429.3047B} {429, 3047}

\bibitem[\protect\citeauthoryear{{Bourne}, {Dunlop}, {Simpson}, {Rowland s},
  {Geach}  \& {McLeod}}{{Bourne} et~al.}{2019}]{bourne19}
{Bourne} N.,  {Dunlop} J.~S.,  {Simpson} J.~M.,  {Rowland s} K.~E.,  {Geach}
  J.~E.,   {McLeod} D.~J.,  2019, \mn@doi [\mnras] {10.1093/mnras/sty2773},
  \href {https://ui.adsabs.harvard.edu/abs/2019MNRAS.482.3135B} {482, 3135}

\bibitem[\protect\citeauthoryear{{Bower}, {Benson}, {Malbon}, {Helly}, {Frenk},
  {Baugh}, {Cole}  \& {Lacey}}{{Bower} et~al.}{2006}]{bower06}
{Bower} R.~G.,  {Benson} A.~J.,  {Malbon} R.,  {Helly} J.~C.,  {Frenk} C.~S.,
  {Baugh} C.~M.,  {Cole} S.,   {Lacey} C.~G.,  2006, \mn@doi [\mnras]
  {10.1111/j.1365-2966.2006.10519.x}, \href
  {https://ui.adsabs.harvard.edu/#abs/2006MNRAS.370..645B} {370, 645}

\bibitem[\protect\citeauthoryear{{Brogui{\`e}re}, {Blanchet}, {Chavatte},
  {Garcia}  \& {Gentaz}}{{Brogui{\`e}re} et~al.}{2020}]{broguiere20}
{Brogui{\`e}re} D.,  {Blanchet} S.,  {Chavatte} P.,  {Garcia} R.~G.,   {Gentaz}
  O.,  2020, in {Ballester} P.,  {Ibsen} J.,  {Solar} M.,   {Shortridge} K.,
  eds,  Astronomical Society of the Pacific Conference Series Vol. 522,
  Astronomical Data Analysis Software and Systems XXVII. p.~485

\bibitem[\protect\citeauthoryear{{Calistro Rivera} et~al.,}{{Calistro Rivera}
  et~al.}{2018}]{calistrorivera18}
{Calistro Rivera} G.,  et~al., 2018, \mn@doi [\apj] {10.3847/1538-4357/aacffa},
  \href {https://ui.adsabs.harvard.edu/abs/2018ApJ...863...56C} {863, 56}

\bibitem[\protect\citeauthoryear{{Carilli} \& {Walter}}{{Carilli} \&
  {Walter}}{2013}]{carilli13}
{Carilli} C.~L.,  {Walter} F.,  2013, \mn@doi [\araa]
  {10.1146/annurev-astro-082812-140953}, \href
  {https://ui.adsabs.harvard.edu/abs/2013ARA&A..51..105C} {51, 105}

\bibitem[\protect\citeauthoryear{{Carilli} et~al.,}{{Carilli}
  et~al.}{2010}]{carilli10}
{Carilli} C.~L.,  et~al., 2010, \mn@doi [\apj] {10.1088/0004-637X/714/2/1407},
  \href {https://ui.adsabs.harvard.edu/abs/2010ApJ...714.1407C} {714, 1407}

\bibitem[\protect\citeauthoryear{{Carilli}, {Hodge}, {Walter}, {Riechers},
  {Daddi}, {Dannerbauer}  \& {Morrison}}{{Carilli} et~al.}{2011}]{carilli11}
{Carilli} C.~L.,  {Hodge} J.,  {Walter} F.,  {Riechers} D.,  {Daddi} E.,
  {Dannerbauer} H.,   {Morrison} G.~E.,  2011, \mn@doi [\apjl]
  {10.1088/2041-8205/739/1/L33}, \href
  {https://ui.adsabs.harvard.edu/abs/2011ApJ...739L..33C} {739, L33}

\bibitem[\protect\citeauthoryear{{Casey} et~al.,}{{Casey}
  et~al.}{2009}]{casey09}
{Casey} C.~M.,  et~al., 2009, \mn@doi [\mnras]
  {10.1111/j.1365-2966.2009.15517.x}, \href
  {https://ui.adsabs.harvard.edu/abs/2009MNRAS.400..670C} {400, 670}

\bibitem[\protect\citeauthoryear{{Casey} et~al.,}{{Casey}
  et~al.}{2011}]{casey11}
{Casey} C.~M.,  et~al., 2011, \mn@doi [\mnras]
  {10.1111/j.1365-2966.2011.18885.x}, \href
  {https://ui.adsabs.harvard.edu/abs/2011MNRAS.415.2723C} {415, 2723}

\bibitem[\protect\citeauthoryear{{Casey}, {Narayanan}  \& {Cooray}}{{Casey}
  et~al.}{2014}]{casey14}
{Casey} C.~M.,  {Narayanan} D.,   {Cooray} A.,  2014, \mn@doi [\physrep]
  {10.1016/j.physrep.2014.02.009}, \href
  {https://ui.adsabs.harvard.edu/abs/2014PhR...541...45C} {541, 45}

\bibitem[\protect\citeauthoryear{{Cassata} et~al.,}{{Cassata}
  et~al.}{2020}]{cassata20}
{Cassata} P.,  et~al., 2020, \mn@doi [\apj] {10.3847/1538-4357/ab7452}, \href
  {https://ui.adsabs.harvard.edu/abs/2020ApJ...891...83C} {891, 83}

\bibitem[\protect\citeauthoryear{{Chapman}, {Blain}, {Smail}  \&
  {Ivison}}{{Chapman} et~al.}{2005}]{chapman05}
{Chapman} S.~C.,  {Blain} A.~W.,  {Smail} I.,   {Ivison} R.~J.,  2005, \mn@doi
  [\apj] {10.1086/428082}, \href
  {https://ui.adsabs.harvard.edu/abs/2005ApJ...622..772C} {622, 772}

\bibitem[\protect\citeauthoryear{{Chapman} et~al.,}{{Chapman}
  et~al.}{2008}]{chapman08}
{Chapman} S.~C.,  et~al., 2008, \mn@doi [\apj] {10.1086/592137}, \href
  {https://ui.adsabs.harvard.edu/abs/2008ApJ...689..889C} {689, 889}

\bibitem[\protect\citeauthoryear{{Chapman} et~al.,}{{Chapman}
  et~al.}{2015}]{chapman15}
{Chapman} S.~C.,  et~al., 2015, \mn@doi [\mnras] {10.1093/mnrasl/slv010}, \href
  {https://ui.adsabs.harvard.edu/abs/2015MNRAS.449L..68C} {449, L68}

\bibitem[\protect\citeauthoryear{{Chen} et~al.,}{{Chen} et~al.}{2017}]{chen17}
{Chen} C.-C.,  et~al., 2017, \mn@doi [\apj] {10.3847/1538-4357/aa863a}, \href
  {https://ui.adsabs.harvard.edu/abs/2017ApJ...846..108C} {846, 108}

\bibitem[\protect\citeauthoryear{{Coppin} et~al.,}{{Coppin}
  et~al.}{2008}]{coppin08}
{Coppin} K.~E.~K.,  et~al., 2008, \mn@doi [\mnras]
  {10.1111/j.1365-2966.2008.13553.x}, \href
  {https://ui.adsabs.harvard.edu/#abs/2008MNRAS.389...45C} {389, 45}

\bibitem[\protect\citeauthoryear{{Coppin} et~al.,}{{Coppin}
  et~al.}{2012}]{coppin12}
{Coppin} K.~E.~K.,  et~al., 2012, \mn@doi [\mnras]
  {10.1111/j.1365-2966.2012.21977.x}, \href
  {https://ui.adsabs.harvard.edu/abs/2012MNRAS.427..520C} {427, 520}

\bibitem[\protect\citeauthoryear{{Courteau}}{{Courteau}}{1997}]{courteau97}
{Courteau} S.,  1997, \mn@doi [\aj] {10.1086/118656}, \href
  {https://ui.adsabs.harvard.edu/abs/1997AJ....114.2402C} {114, 2402}

\bibitem[\protect\citeauthoryear{{Cowie}, {Songaila}, {Hu}  \& {Cohen}}{{Cowie}
  et~al.}{1996}]{cowie96}
{Cowie} L.~L.,  {Songaila} A.,  {Hu} E.~M.,   {Cohen} J.~G.,  1996, \mn@doi
  [\aj] {10.1086/118058}, \href
  {https://ui.adsabs.harvard.edu/\#abs/1996AJ....112..839C} {112, 839}

\bibitem[\protect\citeauthoryear{{Cowie}, {Barger}, {Hsu}, {Chen}, {Owen}  \&
  {Wang}}{{Cowie} et~al.}{2017}]{cowie17}
{Cowie} L.~L.,  {Barger} A.~J.,  {Hsu} L.~Y.,  {Chen} C.-C.,  {Owen} F.~N.,
  {Wang} W.~H.,  2017, \mn@doi [\apj] {10.3847/1538-4357/aa60bb}, \href
  {https://ui.adsabs.harvard.edu/abs/2017ApJ...837..139C} {837, 139}

\bibitem[\protect\citeauthoryear{{Daddi}, {Dannerbauer}, {Elbaz}, {Dickinson},
  {Morrison}, {Stern}  \& {Ravindranath}}{{Daddi} et~al.}{2008}]{daddi08}
{Daddi} E.,  {Dannerbauer} H.,  {Elbaz} D.,  {Dickinson} M.,  {Morrison} G.,
  {Stern} D.,   {Ravindranath} S.,  2008, \mn@doi [\apjl] {10.1086/527377},
  \href {https://ui.adsabs.harvard.edu/abs/2008ApJ...673L..21D} {673, L21}

\bibitem[\protect\citeauthoryear{{Daddi} et~al.,}{{Daddi}
  et~al.}{2009}]{daddi09}
{Daddi} E.,  et~al., 2009, \mn@doi [\apj] {10.1088/0004-637X/694/2/1517}, \href
  {https://ui.adsabs.harvard.edu/abs/2009ApJ...694.1517D} {694, 1517}

\bibitem[\protect\citeauthoryear{{Daddi} et~al.,}{{Daddi}
  et~al.}{2010}]{daddi10}
{Daddi} E.,  et~al., 2010, \mn@doi [\apj] {10.1088/0004-637X/713/1/686}, \href
  {https://ui.adsabs.harvard.edu/abs/2010ApJ...713..686D} {713, 686}

\bibitem[\protect\citeauthoryear{{Daddi} et~al.,}{{Daddi}
  et~al.}{2015}]{daddi15}
{Daddi} E.,  et~al., 2015, \mn@doi [\aap] {10.1051/0004-6361/201425043}, \href
  {https://ui.adsabs.harvard.edu/abs/2015A&A...577A..46D} {577, A46}

\bibitem[\protect\citeauthoryear{{Danielson} et~al.,}{{Danielson}
  et~al.}{2011}]{danielson11}
{Danielson} A.~L.~R.,  et~al., 2011, \mn@doi [\mnras]
  {10.1111/j.1365-2966.2010.17549.x}, \href
  {https://ui.adsabs.harvard.edu/abs/2011MNRAS.410.1687D} {410, 1687}

\bibitem[\protect\citeauthoryear{{Danielson} et~al.,}{{Danielson}
  et~al.}{2013}]{danielson13}
{Danielson} A.~L.~R.,  et~al., 2013, \mn@doi [\mnras] {10.1093/mnras/stt1775},
  \href {https://ui.adsabs.harvard.edu/abs/2013MNRAS.436.2793D} {436, 2793}

\bibitem[\protect\citeauthoryear{{Danielson} et~al.,}{{Danielson}
  et~al.}{2017}]{danielson17}
{Danielson} A.~L.~R.,  et~al., 2017, \mn@doi [\apj] {10.3847/1538-4357/aa6caf},
  \href {https://ui.adsabs.harvard.edu/abs/2017ApJ...840...78D} {840, 78}

\bibitem[\protect\citeauthoryear{{Dannerbauer}, {Daddi}, {Riechers}, {Walter},
  {Carilli}, {Dickinson}, {Elbaz}  \& {Morrison}}{{Dannerbauer}
  et~al.}{2009}]{dannerbauer09}
{Dannerbauer} H.,  {Daddi} E.,  {Riechers} D.~A.,  {Walter} F.,  {Carilli}
  C.~L.,  {Dickinson} M.,  {Elbaz} D.,   {Morrison} G.~E.,  2009, \mn@doi
  [\apjl] {10.1088/0004-637X/698/2/L178}, \href
  {https://ui.adsabs.harvard.edu/abs/2009ApJ...698L.178D} {698, L178}

\bibitem[\protect\citeauthoryear{{Dav{\'e}}, {Finlator}, {Oppenheimer},
  {Fardal}, {Katz}, {Kere{\v{s}}}  \& {Weinberg}}{{Dav{\'e}}
  et~al.}{2010}]{dave10}
{Dav{\'e}} R.,  {Finlator} K.,  {Oppenheimer} B.~D.,  {Fardal} M.,  {Katz} N.,
  {Kere{\v{s}}} D.,   {Weinberg} D.~H.,  2010, \mn@doi [\mnras]
  {10.1111/j.1365-2966.2010.16395.x}, \href
  {https://ui.adsabs.harvard.edu/abs/2010MNRAS.404.1355D} {404, 1355}

\bibitem[\protect\citeauthoryear{{Dav{\'e}}, {Finlator}  \&
  {Oppenheimer}}{{Dav{\'e}} et~al.}{2012}]{dave12}
{Dav{\'e}} R.,  {Finlator} K.,   {Oppenheimer} B.~D.,  2012, \mn@doi [\mnras]
  {10.1111/j.1365-2966.2011.20148.x}, \href
  {https://ui.adsabs.harvard.edu/abs/2012MNRAS.421...98D} {421, 98}

\bibitem[\protect\citeauthoryear{{Dessauges-Zavadsky}
  et~al.,}{{Dessauges-Zavadsky} et~al.}{2015}]{dessauges15}
{Dessauges-Zavadsky} M.,  et~al., 2015, \mn@doi [\aap]
  {10.1051/0004-6361/201424661}, \href
  {https://ui.adsabs.harvard.edu/abs/2015A&A...577A..50D} {577, A50}

\bibitem[\protect\citeauthoryear{{Dessauges-Zavadsky}
  et~al.,}{{Dessauges-Zavadsky} et~al.}{2020}]{dessauges20}
{Dessauges-Zavadsky} M.,  et~al., 2020, arXiv e-prints, \href
  {https://ui.adsabs.harvard.edu/abs/2020arXiv200410771D} {p. arXiv:2004.10771}

\bibitem[\protect\citeauthoryear{{Dole} et~al.,}{{Dole} et~al.}{2006}]{dole06}
{Dole} H.,  et~al., 2006, \mn@doi [\aap] {10.1051/0004-6361:20054446}, \href
  {https://ui.adsabs.harvard.edu/abs/2006A&A...451..417D} {451, 417}

\bibitem[\protect\citeauthoryear{{Downes} \& {Solomon}}{{Downes} \&
  {Solomon}}{1998}]{downes98}
{Downes} D.,  {Solomon} P.~M.,  1998, \mn@doi [\apj] {10.1086/306339}, \href
  {https://ui.adsabs.harvard.edu/abs/1998ApJ...507..615D} {507, 615}

\bibitem[\protect\citeauthoryear{{Dudzevi{\v{c}}i{\={u}}t{\.{e}}}
  et~al.,}{{Dudzevi{\v{c}}i{\={u}}t{\.{e}}} et~al.}{2020}]{dudzeviciute20}
{Dudzevi{\v{c}}i{\={u}}t{\.{e}}} U.,  et~al., 2020, \mn@doi [\mnras]
  {10.1093/mnras/staa769}, \href
  {https://ui.adsabs.harvard.edu/abs/2020MNRAS.494.3828D} {494, 3828}

\bibitem[\protect\citeauthoryear{{Dunne}, {Eales}, {Edmunds}, {Ivison},
  {Alexander}  \& {Clements}}{{Dunne} et~al.}{2000}]{dunne00}
{Dunne} L.,  {Eales} S.,  {Edmunds} M.,  {Ivison} R.,  {Alexander} P.,
  {Clements} D.~L.,  2000, \mn@doi [\mnras] {10.1046/j.1365-8711.2000.03386.x},
  \href {https://ui.adsabs.harvard.edu/abs/2000MNRAS.315..115D} {315, 115}

\bibitem[\protect\citeauthoryear{{Elbaz} et~al.,}{{Elbaz}
  et~al.}{2018}]{elbaz18}
{Elbaz} D.,  et~al., 2018, \mn@doi [\aap] {10.1051/0004-6361/201732370}, \href
  {https://ui.adsabs.harvard.edu/abs/2018A&A...616A.110E} {616, A110}

\bibitem[\protect\citeauthoryear{{Engel} et~al.,}{{Engel}
  et~al.}{2010}]{engel10}
{Engel} H.,  et~al., 2010, \mn@doi [\apj] {10.1088/0004-637X/724/1/233}, \href
  {https://ui.adsabs.harvard.edu/abs/2010ApJ...724..233E} {724, 233}

\bibitem[\protect\citeauthoryear{{Erb}, {Steidel}, {Shapley}, {Pettini},
  {Reddy}  \& {Adelberger}}{{Erb} et~al.}{2006}]{erb06}
{Erb} D.~K.,  {Steidel} C.~C.,  {Shapley} A.~E.,  {Pettini} M.,  {Reddy} N.~A.,
    {Adelberger} K.~L.,  2006, \mn@doi [\apj] {10.1086/504891}, \href
  {https://ui.adsabs.harvard.edu/abs/2006ApJ...646..107E} {646, 107}

\bibitem[\protect\citeauthoryear{{Fixsen}, {Bennett}  \& {Mather}}{{Fixsen}
  et~al.}{1999}]{fixsen99}
{Fixsen} D.~J.,  {Bennett} C.~L.,   {Mather} J.~C.,  1999, \mn@doi [\apj]
  {10.1086/307962}, \href
  {https://ui.adsabs.harvard.edu/abs/1999ApJ...526..207F} {526, 207}

\bibitem[\protect\citeauthoryear{{Foreman-Mackey}, {Hogg}, {Lang}  \&
  {Goodman}}{{Foreman-Mackey} et~al.}{2013}]{foreman-mackey13}
{Foreman-Mackey} D.,  {Hogg} D.~W.,  {Lang} D.,   {Goodman} J.,  2013, \mn@doi
  [\pasp] {10.1086/670067}, \href
  {https://ui.adsabs.harvard.edu/abs/2013PASP..125..306F} {125, 306}

\bibitem[\protect\citeauthoryear{{Franco} et~al.,}{{Franco}
  et~al.}{2018}]{franco18}
{Franco} M.,  et~al., 2018, \mn@doi [\aap] {10.1051/0004-6361/201832928}, \href
  {https://ui.adsabs.harvard.edu/abs/2018A&A...620A.152F} {620, A152}

\bibitem[\protect\citeauthoryear{{Frayer}, {Ivison}, {Scoville}, {Yun},
  {Evans}, {Smail}, {Blain}  \& {Kneib}}{{Frayer} et~al.}{1998}]{frayer98}
{Frayer} D.~T.,  {Ivison} R.~J.,  {Scoville} N.~Z.,  {Yun} M.,  {Evans} A.~S.,
  {Smail} I.,  {Blain} A.~W.,   {Kneib} J.~P.,  1998, \mn@doi [\apjl]
  {10.1086/311639}, \href
  {https://ui.adsabs.harvard.edu/abs/1998ApJ...506L...7F} {506, L7}

\bibitem[\protect\citeauthoryear{{Frayer} et~al.,}{{Frayer}
  et~al.}{1999}]{frayer99}
{Frayer} D.~T.,  et~al., 1999, \mn@doi [\apjl] {10.1086/311940}, \href
  {https://ui.adsabs.harvard.edu/abs/1999ApJ...514L..13F} {514, L13}

\bibitem[\protect\citeauthoryear{{Gaches}, {Offner}  \& {Bisbas}}{{Gaches}
  et~al.}{2019}]{gaches19}
{Gaches} B. A.~L.,  {Offner} S. S.~R.,   {Bisbas} T.~G.,  2019, \mn@doi [\apj]
  {10.3847/1538-4357/ab3c5c}, \href
  {https://ui.adsabs.harvard.edu/abs/2019ApJ...883..190G} {883, 190}

\bibitem[\protect\citeauthoryear{{Geach}, {Smail}, {Moran}, {MacArthur},
  {Lagos}  \& {Edge}}{{Geach} et~al.}{2011}]{geach11}
{Geach} J.~E.,  {Smail} I.,  {Moran} S.~M.,  {MacArthur} L.~A.,  {Lagos} C.
  d.~P.,   {Edge} A.~C.,  2011, \mn@doi [\apjl] {10.1088/2041-8205/730/2/L19},
  \href {https://ui.adsabs.harvard.edu/abs/2011ApJ...730L..19G} {730, L19}

\bibitem[\protect\citeauthoryear{{Genzel} et~al.,}{{Genzel}
  et~al.}{2010}]{genzel10}
{Genzel} R.,  et~al., 2010, \mn@doi [\mnras]
  {10.1111/j.1365-2966.2010.16969.x}, \href
  {https://ui.adsabs.harvard.edu/abs/2010MNRAS.407.2091G} {407, 2091}

\bibitem[\protect\citeauthoryear{{Genzel} et~al.,}{{Genzel}
  et~al.}{2015}]{genzel15}
{Genzel} R.,  et~al., 2015, \mn@doi [\apj] {10.1088/0004-637X/800/1/20}, \href
  {https://ui.adsabs.harvard.edu/abs/2015ApJ...800...20G} {800, 20}

\bibitem[\protect\citeauthoryear{{Gerin} \& {Phillips}}{{Gerin} \&
  {Phillips}}{2000}]{gerin00}
{Gerin} M.,  {Phillips} T.~G.,  2000, \mn@doi [\apj] {10.1086/309072}, \href
  {https://ui.adsabs.harvard.edu/abs/2000ApJ...537..644G} {537, 644}

\bibitem[\protect\citeauthoryear{{Greve}, {Ivison}  \& {Papadopoulos}}{{Greve}
  et~al.}{2003}]{greve03}
{Greve} T.~R.,  {Ivison} R.~J.,   {Papadopoulos} P.~P.,  2003, \mn@doi [\apj]
  {10.1086/379547}, \href
  {https://ui.adsabs.harvard.edu/abs/2003ApJ...599..839G} {599, 839}

\bibitem[\protect\citeauthoryear{{Greve} et~al.,}{{Greve}
  et~al.}{2005}]{greve05}
{Greve} T.~R.,  et~al., 2005, \mn@doi [\mnras]
  {10.1111/j.1365-2966.2005.08979.x}, \href
  {https://ui.adsabs.harvard.edu/abs/2005MNRAS.359.1165G} {359, 1165}

\bibitem[\protect\citeauthoryear{{Greve} et~al.,}{{Greve}
  et~al.}{2014}]{greve14}
{Greve} T.~R.,  et~al., 2014, \mn@doi [\apj] {10.1088/0004-637X/794/2/142},
  \href {https://ui.adsabs.harvard.edu/abs/2014ApJ...794..142G} {794, 142}

\bibitem[\protect\citeauthoryear{{Gullberg} et~al.,}{{Gullberg}
  et~al.}{2018}]{gullberg18}
{Gullberg} B.,  et~al., 2018, \mn@doi [\apj] {10.3847/1538-4357/aabe8c}, \href
  {https://ui.adsabs.harvard.edu/abs/2018ApJ...859...12G} {859, 12}

\bibitem[\protect\citeauthoryear{{Gullberg} et~al.,}{{Gullberg}
  et~al.}{2019}]{gullberg19}
{Gullberg} B.,  et~al., 2019, \mn@doi [\mnras] {10.1093/mnras/stz2835}, \href
  {https://ui.adsabs.harvard.edu/abs/2019MNRAS.490.4956G} {490, 4956}

\bibitem[\protect\citeauthoryear{{Hainline}, {Blain}, {Smail}, {Frayer},
  {Chapman}, {Ivison}  \& {Alexand er}}{{Hainline} et~al.}{2009}]{hainline09}
{Hainline} L.~J.,  {Blain} A.~W.,  {Smail} I.,  {Frayer} D.~T.,  {Chapman}
  S.~C.,  {Ivison} R.~J.,   {Alexand er} D.~M.,  2009, \mn@doi [\apj]
  {10.1088/0004-637X/699/2/1610}, \href
  {https://ui.adsabs.harvard.edu/abs/2009ApJ...699.1610H} {699, 1610}

\bibitem[\protect\citeauthoryear{{Hainline}, {Blain}, {Smail}, {Alexand er},
  {Armus}, {Chapman}  \& {Ivison}}{{Hainline} et~al.}{2011}]{hainline11}
{Hainline} L.~J.,  {Blain} A.~W.,  {Smail} I.,  {Alexand er} D.~M.,  {Armus}
  L.,  {Chapman} S.~C.,   {Ivison} R.~J.,  2011, \mn@doi [\apj]
  {10.1088/0004-637X/740/2/96}, \href
  {https://ui.adsabs.harvard.edu/abs/2011ApJ...740...96H} {740, 96}

\bibitem[\protect\citeauthoryear{{Harris} et~al.,}{{Harris}
  et~al.}{2012}]{harris12}
{Harris} A.~I.,  et~al., 2012, \mn@doi [\apj] {10.1088/0004-637X/752/2/152},
  \href {https://ui.adsabs.harvard.edu/abs/2012ApJ...752..152H} {752, 152}

\bibitem[\protect\citeauthoryear{{Hatsukade} et~al.,}{{Hatsukade}
  et~al.}{2016}]{hatsukade16}
{Hatsukade} B.,  et~al., 2016, \mn@doi [\pasj] {10.1093/pasj/psw026}, \href
  {https://ui.adsabs.harvard.edu/abs/2016PASJ...68...36H} {68, 36}

\bibitem[\protect\citeauthoryear{{Hill} et~al.,}{{Hill} et~al.}{2018}]{hill18}
{Hill} R.,  et~al., 2018, \mn@doi [\mnras] {10.1093/mnras/sty746}, \href
  {https://ui.adsabs.harvard.edu/abs/2018MNRAS.477.2042H} {477, 2042}

\bibitem[\protect\citeauthoryear{{Hodge} \& {da Cunha}}{{Hodge} \& {da
  Cunha}}{2020}]{hodge20}
{Hodge} J.~A.,  {da Cunha} E.,  2020, arXiv e-prints, \href
  {https://ui.adsabs.harvard.edu/abs/2020arXiv200400934H} {p. arXiv:2004.00934}

\bibitem[\protect\citeauthoryear{{Hodge} et~al.,}{{Hodge}
  et~al.}{2013}]{hodge13}
{Hodge} J.~A.,  et~al., 2013, \mn@doi [\apj] {10.1088/0004-637X/768/1/91},
  \href {https://ui.adsabs.harvard.edu/abs/2013ApJ...768...91H} {768, 91}

\bibitem[\protect\citeauthoryear{{Hodge} et~al.,}{{Hodge}
  et~al.}{2016}]{hodge16}
{Hodge} J.~A.,  et~al., 2016, \mn@doi [\apj] {10.3847/1538-4357/833/1/103},
  \href {https://ui.adsabs.harvard.edu/abs/2016ApJ...833..103H} {833, 103}

\bibitem[\protect\citeauthoryear{{Hodge} et~al.,}{{Hodge}
  et~al.}{2019}]{hodge19}
{Hodge} J.~A.,  et~al., 2019, \mn@doi [\apj] {10.3847/1538-4357/ab1846}, \href
  {https://ui.adsabs.harvard.edu/abs/2019ApJ...876..130H} {876, 130}

\bibitem[\protect\citeauthoryear{{Hopkins}, {Hernquist}, {Cox}  \&
  {Kere{\v{s}}}}{{Hopkins} et~al.}{2008}]{hopkins08}
{Hopkins} P.~F.,  {Hernquist} L.,  {Cox} T.~J.,   {Kere{\v{s}}} D.,  2008,
  \mn@doi [\apjs] {10.1086/524362}, \href
  {https://ui.adsabs.harvard.edu/abs/2008ApJS..175..356H} {175, 356}

\bibitem[\protect\citeauthoryear{{Hughes} et~al.,}{{Hughes}
  et~al.}{1998}]{hughes98}
{Hughes} D.~H.,  et~al., 1998, \mn@doi [\nat] {10.1038/28328}, \href
  {https://ui.adsabs.harvard.edu/\#abs/1998Natur.394..241H} {394, 241}

\bibitem[\protect\citeauthoryear{{Huynh} et~al.,}{{Huynh}
  et~al.}{2017}]{huynh17}
{Huynh} M.~T.,  et~al., 2017, \mn@doi [\mnras] {10.1093/mnras/stx156}, \href
  {https://ui.adsabs.harvard.edu/abs/2017MNRAS.467.1222H} {467, 1222}

\bibitem[\protect\citeauthoryear{{Ikarashi} et~al.,}{{Ikarashi}
  et~al.}{2015}]{ikarashi15}
{Ikarashi} S.,  et~al., 2015, \mn@doi [\apj] {10.1088/0004-637X/810/2/133},
  \href {https://ui.adsabs.harvard.edu/abs/2015ApJ...810..133I} {810, 133}

\bibitem[\protect\citeauthoryear{{Iono} et~al.,}{{Iono} et~al.}{2006}]{iono06}
{Iono} D.,  et~al., 2006, \mn@doi [\apjl] {10.1086/503290}, \href
  {https://ui.adsabs.harvard.edu/abs/2006ApJ...640L...1I} {640, L1}

\bibitem[\protect\citeauthoryear{{Ivison} et~al.,}{{Ivison}
  et~al.}{2007}]{ivison07}
{Ivison} R.~J.,  et~al., 2007, \mn@doi [\mnras]
  {10.1111/j.1365-2966.2007.12044.x}, \href
  {https://ui.adsabs.harvard.edu/abs/2007MNRAS.380..199I} {380, 199}

\bibitem[\protect\citeauthoryear{{Ivison}, {Papadopoulos}, {Smail}, {Greve},
  {Thomson}, {Xilouris}  \& {Chapman}}{{Ivison} et~al.}{2011}]{ivison11}
{Ivison} R.~J.,  {Papadopoulos} P.~P.,  {Smail} I.,  {Greve} T.~R.,  {Thomson}
  A.~P.,  {Xilouris} E.~M.,   {Chapman} S.~C.,  2011, \mn@doi [\mnras]
  {10.1111/j.1365-2966.2010.18028.x}, \href
  {https://ui.adsabs.harvard.edu/abs/2011MNRAS.412.1913I} {412, 1913}

\bibitem[\protect\citeauthoryear{{Jarugula} et~al.,}{{Jarugula}
  et~al.}{2019}]{jarugula19}
{Jarugula} S.,  et~al., 2019, \mn@doi [\apj] {10.3847/1538-4357/ab290d}, \href
  {https://ui.adsabs.harvard.edu/abs/2019ApJ...880...92J} {880, 92}

\bibitem[\protect\citeauthoryear{{Kaufman}, {Wolfire}, {Hollenbach}  \&
  {Luhman}}{{Kaufman} et~al.}{1999}]{kaufman99}
{Kaufman} M.~J.,  {Wolfire} M.~G.,  {Hollenbach} D.~J.,   {Luhman} M.~L.,
  1999, \mn@doi [\apj] {10.1086/308102}, \href
  {https://ui.adsabs.harvard.edu/abs/1999ApJ...527..795K} {527, 795}

\bibitem[\protect\citeauthoryear{{Keene}, {Blake}, {Phillips}, {Huggins}  \&
  {Beichman}}{{Keene} et~al.}{1985}]{keene85}
{Keene} J.,  {Blake} G.~A.,  {Phillips} T.~G.,  {Huggins} P.~J.,   {Beichman}
  C.~A.,  1985, \mn@doi [\apj] {10.1086/163763}, \href
  {https://ui.adsabs.harvard.edu/abs/1985ApJ...299..967K} {299, 967}

\bibitem[\protect\citeauthoryear{{Kennicutt}}{{Kennicutt}}{1998}]{kennicutt98}
{Kennicutt} Robert~C. J.,  1998, \mn@doi [\apj] {10.1086/305588}, \href
  {https://ui.adsabs.harvard.edu/abs/1998ApJ...498..541K} {498, 541}

\bibitem[\protect\citeauthoryear{{Kohandel}, {Pallottini}, {Ferrara},
  {Zanella}, {Behrens}, {Carniani}, {Gallerani}  \& {Vallini}}{{Kohandel}
  et~al.}{2019}]{kohandel19}
{Kohandel} M.,  {Pallottini} A.,  {Ferrara} A.,  {Zanella} A.,  {Behrens} C.,
  {Carniani} S.,  {Gallerani} S.,   {Vallini} L.,  2019, \mn@doi [\mnras]
  {10.1093/mnras/stz1486}, \href
  {https://ui.adsabs.harvard.edu/abs/2019MNRAS.487.3007K} {487, 3007}

\bibitem[\protect\citeauthoryear{{Koprowski} et~al.,}{{Koprowski}
  et~al.}{2016}]{koprowski16}
{Koprowski} M.~P.,  et~al., 2016, \mn@doi [\mnras] {10.1093/mnras/stw564},
  \href {https://ui.adsabs.harvard.edu/abs/2016MNRAS.458.4321K} {458, 4321}

\bibitem[\protect\citeauthoryear{{Lagos}, {Bayet}, {Baugh}, {Lacey}, {Bell},
  {Fanidakis}  \& {Geach}}{{Lagos} et~al.}{2012}]{lagos12}
{Lagos} C. d.~P.,  {Bayet} E.,  {Baugh} C.~M.,  {Lacey} C.~G.,  {Bell} T.~A.,
  {Fanidakis} N.,   {Geach} J.~E.,  2012, \mn@doi [\mnras]
  {10.1111/j.1365-2966.2012.21905.x}, \href
  {https://ui.adsabs.harvard.edu/abs/2012MNRAS.426.2142L} {426, 2142}

\bibitem[\protect\citeauthoryear{{Lagos}, {da Cunha}, {Robotham}, {Obreschkow},
  {Valentino}, {Fujimoto}, {Magdis}  \& {Tobar}}{{Lagos}
  et~al.}{2020}]{lagos20}
{Lagos} C. d.~P.,  {da Cunha} E.,  {Robotham} A. S.~G.,  {Obreschkow} D.,
  {Valentino} F.,  {Fujimoto} S.,  {Magdis} G.~E.,   {Tobar} R.,  2020, arXiv
  e-prints, \href {https://ui.adsabs.harvard.edu/abs/2020arXiv200709853L} {p.
  arXiv:2007.09853}

\bibitem[\protect\citeauthoryear{{Law}, {Steidel}, {Erb}, {Larkin}, {Pettini},
  {Shapley}  \& {Wright}}{{Law} et~al.}{2009}]{law09}
{Law} D.~R.,  {Steidel} C.~C.,  {Erb} D.~K.,  {Larkin} J.~E.,  {Pettini} M.,
  {Shapley} A.~E.,   {Wright} S.~A.,  2009, \mn@doi [\apj]
  {10.1088/0004-637X/697/2/2057}, \href
  {https://ui.adsabs.harvard.edu/abs/2009ApJ...697.2057L} {697, 2057}

\bibitem[\protect\citeauthoryear{{Le F{\`e}vre} et~al.,}{{Le F{\`e}vre}
  et~al.}{2019}]{lefevre19}
{Le F{\`e}vre} O.,  et~al., 2019, arXiv e-prints, \href
  {https://ui.adsabs.harvard.edu/abs/2019arXiv191009517L} {p. arXiv:1910.09517}

\bibitem[\protect\citeauthoryear{{Leroy} et~al.,}{{Leroy}
  et~al.}{2011}]{leroy11}
{Leroy} A.~K.,  et~al., 2011, \mn@doi [\apj] {10.1088/0004-637X/737/1/12},
  \href {https://ui.adsabs.harvard.edu/abs/2011ApJ...737...12L} {737, 12}

\bibitem[\protect\citeauthoryear{{Li}, {Narayanan}  \& {Dav{\'e}}}{{Li}
  et~al.}{2019}]{li19}
{Li} Q.,  {Narayanan} D.,   {Dav{\'e}} R.,  2019, \mn@doi [\mnras]
  {10.1093/mnras/stz2684}, \href
  {https://ui.adsabs.harvard.edu/abs/2019MNRAS.490.1425L} {490, 1425}

\bibitem[\protect\citeauthoryear{{Lilly}, {Eales}, {Gear}, {Hammer}, {Le
  F{\`e}vre}, {Crampton}, {Bond}  \& {Dunne}}{{Lilly} et~al.}{1999}]{lilly99}
{Lilly} S.~J.,  {Eales} S.~A.,  {Gear} W. K.~P.,  {Hammer} F.,  {Le F{\`e}vre}
  O.,  {Crampton} D.,  {Bond} J.~R.,   {Dunne} L.,  1999, \mn@doi [\apj]
  {10.1086/307310}, \href
  {https://ui.adsabs.harvard.edu/abs/1999ApJ...518..641L} {518, 641}

\bibitem[\protect\citeauthoryear{{Liu} et~al.,}{{Liu} et~al.}{2019}]{liu19b}
{Liu} D.,  et~al., 2019, \mn@doi [\apj] {10.3847/1538-4357/ab578d}, \href
  {https://ui.adsabs.harvard.edu/abs/2019ApJ...887..235L} {887, 235}

\bibitem[\protect\citeauthoryear{{Magdis} et~al.,}{{Magdis}
  et~al.}{2012}]{magdis12}
{Magdis} G.~E.,  et~al., 2012, \mn@doi [\apjl] {10.1088/2041-8205/758/1/L9},
  \href {https://ui.adsabs.harvard.edu/abs/2012ApJ...758L...9M} {758, L9}

\bibitem[\protect\citeauthoryear{{Magnelli} et~al.,}{{Magnelli}
  et~al.}{2012a}]{magnelli12a}
{Magnelli} B.,  et~al., 2012a, \mn@doi [\aap] {10.1051/0004-6361/201118312},
  \href {https://ui.adsabs.harvard.edu/abs/2012A&A...539A.155M} {539, A155}

\bibitem[\protect\citeauthoryear{{Magnelli} et~al.,}{{Magnelli}
  et~al.}{2012b}]{magnelli12}
{Magnelli} B.,  et~al., 2012b, \mn@doi [\aap] {10.1051/0004-6361/201118312},
  \href {https://ui.adsabs.harvard.edu/abs/2012A&A...539A.155M} {539, A155}

\bibitem[\protect\citeauthoryear{{Magnelli} et~al.,}{{Magnelli}
  et~al.}{2013}]{magnelli13}
{Magnelli} B.,  et~al., 2013, \mn@doi [\aap] {10.1051/0004-6361/201321371},
  \href {https://ui.adsabs.harvard.edu/abs/2013A&A...553A.132M} {553, A132}

\bibitem[\protect\citeauthoryear{{McAlpine} et~al.,}{{McAlpine}
  et~al.}{2019}]{mcalpine19}
{McAlpine} S.,  et~al., 2019, \mn@doi [\mnras] {10.1093/mnras/stz1692}, \href
  {https://ui.adsabs.harvard.edu/abs/2019MNRAS.488.2440M} {488, 2440}

\bibitem[\protect\citeauthoryear{{McMullin}, {Waters}, {Schiebel}, {Young}  \&
  {Golap}}{{McMullin} et~al.}{2007}]{casa}
{McMullin} J.~P.,  {Waters} B.,  {Schiebel} D.,  {Young} W.,   {Golap} K.,
  2007, in {Shaw} R.~A.,  {Hill} F.,   {Bell} D.~J.,  eds,  Astronomical
  Society of the Pacific Conference Series Vol. 376, Astronomical Data Analysis
  Software and Systems XVI. p.~127

\bibitem[\protect\citeauthoryear{{Miettinen} et~al.,}{{Miettinen}
  et~al.}{2017}]{miettinen17}
{Miettinen} O.,  et~al., 2017, \mn@doi [\aap] {10.1051/0004-6361/201730762},
  \href {https://ui.adsabs.harvard.edu/abs/2017A&A...606A..17M} {606, A17}

\bibitem[\protect\citeauthoryear{{Neri} et~al.,}{{Neri} et~al.}{2003}]{neri03}
{Neri} R.,  et~al., 2003, \mn@doi [\apjl] {10.1086/379968}, \href
  {https://ui.adsabs.harvard.edu/abs/2003ApJ...597L.113N} {597, L113}

\bibitem[\protect\citeauthoryear{{Neugebauer} et~al.,}{{Neugebauer}
  et~al.}{1984}]{neugebauer84}
{Neugebauer} G.,  et~al., 1984, \mn@doi [\apjl] {10.1086/184209}, \href
  {https://ui.adsabs.harvard.edu/abs/1984ApJ...278L...1N} {278, L1}

\bibitem[\protect\citeauthoryear{{Noeske} et~al.,}{{Noeske}
  et~al.}{2007}]{noeske07a}
{Noeske} K.~G.,  et~al., 2007, \mn@doi [\apjl] {10.1086/517926}, \href
  {https://ui.adsabs.harvard.edu/abs/2007ApJ...660L..43N} {660, L43}

\bibitem[\protect\citeauthoryear{{Omont}}{{Omont}}{2007}]{omont07}
{Omont} A.,  2007, \mn@doi [Reports on Progress in Physics]
  {10.1088/0034-4885/70/7/R03}, \href
  {https://ui.adsabs.harvard.edu/abs/2007RPPh...70.1099O} {70, 1099}

\bibitem[\protect\citeauthoryear{{Papadopoulos} \& {Greve}}{{Papadopoulos} \&
  {Greve}}{2004}]{papadopoulos04}
{Papadopoulos} P.~P.,  {Greve} T.~R.,  2004, \mn@doi [\apjl] {10.1086/426059},
  \href {https://ui.adsabs.harvard.edu/abs/2004ApJ...615L..29P} {615, L29}

\bibitem[\protect\citeauthoryear{{Papadopoulos} et~al.,}{{Papadopoulos}
  et~al.}{2014}]{papadopoulos14}
{Papadopoulos} P.~P.,  et~al., 2014, \mn@doi [\apj]
  {10.1088/0004-637X/788/2/153}, \href
  {https://ui.adsabs.harvard.edu/abs/2014ApJ...788..153P} {788, 153}

\bibitem[\protect\citeauthoryear{{Papadopoulos}, {Bisbas}  \&
  {Zhang}}{{Papadopoulos} et~al.}{2018}]{papadopoulos18}
{Papadopoulos} P.~P.,  {Bisbas} T.~G.,   {Zhang} Z.-Y.,  2018, \mn@doi [\mnras]
  {10.1093/mnras/sty1077}, \href
  {https://ui.adsabs.harvard.edu/abs/2018MNRAS.478.1716P} {478, 1716}

\bibitem[\protect\citeauthoryear{{Pope} et~al.,}{{Pope} et~al.}{2006}]{pope06}
{Pope} A.,  et~al., 2006, \mn@doi [\mnras] {10.1111/j.1365-2966.2006.10575.x},
  \href {https://ui.adsabs.harvard.edu/abs/2006MNRAS.370.1185P} {370, 1185}

\bibitem[\protect\citeauthoryear{{Puget}, {Abergel}, {Bernard}, {Boulanger},
  {Burton}, {Desert}  \& {Hartmann}}{{Puget} et~al.}{1996}]{puget96}
{Puget} J.~L.,  {Abergel} A.,  {Bernard} J.~P.,  {Boulanger} F.,  {Burton}
  W.~B.,  {Desert} F.~X.,   {Hartmann} D.,  1996, \aap, \href
  {https://ui.adsabs.harvard.edu/abs/1996A&A...308L...5P} {308, L5}

\bibitem[\protect\citeauthoryear{{Puglisi} et~al.,}{{Puglisi}
  et~al.}{2019}]{puglisi19}
{Puglisi} A.,  et~al., 2019, \mn@doi [\apjl] {10.3847/2041-8213/ab1f92}, \href
  {https://ui.adsabs.harvard.edu/abs/2019ApJ...877L..23P} {877, L23}

\bibitem[\protect\citeauthoryear{{Riechers} et~al.,}{{Riechers}
  et~al.}{2010}]{riechers10}
{Riechers} D.~A.,  et~al., 2010, \mn@doi [\apjl]
  {10.1088/2041-8205/720/2/L131}, \href
  {https://ui.adsabs.harvard.edu/abs/2010ApJ...720L.131R} {720, L131}

\bibitem[\protect\citeauthoryear{{Rosenberg} et~al.,}{{Rosenberg}
  et~al.}{2015}]{rosenberg15}
{Rosenberg} M.~J.~F.,  et~al., 2015, \mn@doi [\apj]
  {10.1088/0004-637X/801/2/72}, \href
  {https://ui.adsabs.harvard.edu/abs/2015ApJ...801...72R} {801, 72}

\bibitem[\protect\citeauthoryear{{Saintonge} et~al.,}{{Saintonge}
  et~al.}{2013}]{saintonge13}
{Saintonge} A.,  et~al., 2013, \mn@doi [\apj] {10.1088/0004-637X/778/1/2},
  \href {https://ui.adsabs.harvard.edu/abs/2013ApJ...778....2S} {778, 2}

\bibitem[\protect\citeauthoryear{{Sanders} \& {Mirabel}}{{Sanders} \&
  {Mirabel}}{1996}]{sanders96}
{Sanders} D.~B.,  {Mirabel} I.~F.,  1996, \mn@doi [\araa]
  {10.1146/annurev.astro.34.1.749}, \href
  {https://ui.adsabs.harvard.edu/abs/1996ARA&A..34..749S} {34, 749}

\bibitem[\protect\citeauthoryear{{Sanders}, {Soifer}, {Elias}, {Madore},
  {Matthews}, {Neugebauer}  \& {Scoville}}{{Sanders} et~al.}{1988}]{sanders88}
{Sanders} D.~B.,  {Soifer} B.~T.,  {Elias} J.~H.,  {Madore} B.~F.,  {Matthews}
  K.,  {Neugebauer} G.,   {Scoville} N.~Z.,  1988, \mn@doi [\apj]
  {10.1086/165983}, \href
  {https://ui.adsabs.harvard.edu/#abs/1988ApJ...325...74S} {325, 74}

\bibitem[\protect\citeauthoryear{{Sanders}, {Scoville}  \& {Soifer}}{{Sanders}
  et~al.}{1991}]{sanders91}
{Sanders} D.~B.,  {Scoville} N.~Z.,   {Soifer} B.~T.,  1991, \mn@doi [\apj]
  {10.1086/169800}, \href
  {https://ui.adsabs.harvard.edu/abs/1991ApJ...370..158S} {370, 158}

\bibitem[\protect\citeauthoryear{{Sandstrom} et~al.,}{{Sandstrom}
  et~al.}{2013}]{sandstrom13}
{Sandstrom} K.~M.,  et~al., 2013, \mn@doi [\apj] {10.1088/0004-637X/777/1/5},
  \href {https://ui.adsabs.harvard.edu/abs/2013ApJ...777....5S} {777, 5}

\bibitem[\protect\citeauthoryear{{Santini} et~al.,}{{Santini}
  et~al.}{2014}]{santini14}
{Santini} P.,  et~al., 2014, \mn@doi [\aap] {10.1051/0004-6361/201322835},
  \href {https://ui.adsabs.harvard.edu/abs/2014A&A...562A..30S} {562, A30}

\bibitem[\protect\citeauthoryear{{Schinnerer} et~al.,}{{Schinnerer}
  et~al.}{2008}]{schinnerer08}
{Schinnerer} E.,  et~al., 2008, \mn@doi [\apjl] {10.1086/595680}, \href
  {https://ui.adsabs.harvard.edu/abs/2008ApJ...689L...5S} {689, L5}

\bibitem[\protect\citeauthoryear{{Scoville}}{{Scoville}}{2013}]{scoville13}
{Scoville} N.~Z.,  2013, {Evolution of star formation and gas}.
p.~491

\bibitem[\protect\citeauthoryear{{Scoville}, {Yun}, {Windhorst}, {Keel}  \&
  {Armus}}{{Scoville} et~al.}{1997}]{scoville97}
{Scoville} N.~Z.,  {Yun} M.~S.,  {Windhorst} R.~A.,  {Keel} W.~C.,   {Armus}
  L.,  1997, \mn@doi [\apjl] {10.1086/310807}, \href
  {https://ui.adsabs.harvard.edu/abs/1997ApJ...485L..21S} {485, L21}

\bibitem[\protect\citeauthoryear{{Scoville} et~al.,}{{Scoville}
  et~al.}{2016}]{scoville16}
{Scoville} N.,  et~al., 2016, \mn@doi [\apj] {10.3847/0004-637X/820/2/83},
  \href {https://ui.adsabs.harvard.edu/abs/2016ApJ...820...83S} {820, 83}

\bibitem[\protect\citeauthoryear{{Shetty}, {Cappellari}, {McDermid},
  {Krajnovi{\'c}}, {de Zeeuw}, {Davies}  \& {Kobayashi}}{{Shetty}
  et~al.}{2020}]{shetty20}
{Shetty} S.,  {Cappellari} M.,  {McDermid} R.~M.,  {Krajnovi{\'c}} D.,  {de
  Zeeuw} P.~T.,  {Davies} R.~L.,   {Kobayashi} C.,  2020, \mn@doi [\mnras]
  {10.1093/mnras/staa1043}, \href
  {https://ui.adsabs.harvard.edu/abs/2020MNRAS.494.5619S} {494, 5619}

\bibitem[\protect\citeauthoryear{{Simpson} et~al.,}{{Simpson}
  et~al.}{2014}]{simpson14}
{Simpson} J.~M.,  et~al., 2014, \mn@doi [\apj] {10.1088/0004-637X/788/2/125},
  \href {https://ui.adsabs.harvard.edu/abs/2014ApJ...788..125S} {788, 125}

\bibitem[\protect\citeauthoryear{{Simpson} et~al.,}{{Simpson}
  et~al.}{2015}]{simpson15}
{Simpson} J.~M.,  et~al., 2015, \mn@doi [\apj] {10.1088/0004-637X/799/1/81},
  \href {https://ui.adsabs.harvard.edu/abs/2015ApJ...799...81S} {799, 81}

\bibitem[\protect\citeauthoryear{{Simpson} et~al.,}{{Simpson}
  et~al.}{2020}]{simpson20}
{Simpson} J.~M.,  et~al., 2020, arXiv e-prints, \href
  {https://ui.adsabs.harvard.edu/abs/2020arXiv200305484S} {p. arXiv:2003.05484}

\bibitem[\protect\citeauthoryear{{Smail}, {Ivison}  \& {Blain}}{{Smail}
  et~al.}{1997}]{smail97}
{Smail} I.,  {Ivison} R.~J.,   {Blain} A.~W.,  1997, \mn@doi [\apj]
  {10.1086/311017}, \href
  {https://ui.adsabs.harvard.edu/\#abs/1997ApJ...490L...5S} {490, L5}

\bibitem[\protect\citeauthoryear{{Smith}, {Collier}, {Ozaki}  \&
  {Lucey}}{{Smith} et~al.}{2019}]{smith19}
{Smith} R.~J.,  {Collier} W.~P.,  {Ozaki} S.,   {Lucey} J.~R.,  2019, arXiv
  e-prints, \href {https://ui.adsabs.harvard.edu/abs/2019arXiv191106338S} {p.
  arXiv:1911.06338}

\bibitem[\protect\citeauthoryear{{Smol{\v{c}}i{\'c}}
  et~al.,}{{Smol{\v{c}}i{\'c}} et~al.}{2012}]{smolcic12}
{Smol{\v{c}}i{\'c}} V.,  et~al., 2012, \mn@doi [\aap]
  {10.1051/0004-6361/201219368}, \href
  {https://ui.adsabs.harvard.edu/abs/2012A&A...548A...4S} {548, A4}

\bibitem[\protect\citeauthoryear{{Smol{\v{c}}i{\'c}}
  et~al.,}{{Smol{\v{c}}i{\'c}} et~al.}{2015}]{smolcic15}
{Smol{\v{c}}i{\'c}} V.,  et~al., 2015, \mn@doi [\aap]
  {10.1051/0004-6361/201424996}, \href
  {https://ui.adsabs.harvard.edu/abs/2015A&A...576A.127S} {576, A127}

\bibitem[\protect\citeauthoryear{{Solomon} \& {Vanden Bout}}{{Solomon} \&
  {Vanden Bout}}{2005}]{solomon05}
{Solomon} P.~M.,  {Vanden Bout} P.~A.,  2005, \mn@doi [\araa]
  {10.1146/annurev.astro.43.051804.102221}, \href
  {https://ui.adsabs.harvard.edu/abs/2005ARA&A..43..677S} {43, 677}

\bibitem[\protect\citeauthoryear{{Solomon}, {Rivolo}, {Barrett}  \&
  {Yahil}}{{Solomon} et~al.}{1987}]{solomon87}
{Solomon} P.~M.,  {Rivolo} A.~R.,  {Barrett} J.,   {Yahil} A.,  1987, \mn@doi
  [\apj] {10.1086/165493}, \href
  {https://ui.adsabs.harvard.edu/abs/1987ApJ...319..730S} {319, 730}

\bibitem[\protect\citeauthoryear{{Solomon}, {Downes}  \& {Radford}}{{Solomon}
  et~al.}{1992}]{solomon92}
{Solomon} P.~M.,  {Downes} D.,   {Radford} S.~J.~E.,  1992, \mn@doi [\apjl]
  {10.1086/186304}, \href
  {https://ui.adsabs.harvard.edu/abs/1992ApJ...387L..55S} {387, L55}

\bibitem[\protect\citeauthoryear{{Solomon}, {Downes}, {Radford}  \&
  {Barrett}}{{Solomon} et~al.}{1997}]{solomon97}
{Solomon} P.~M.,  {Downes} D.,  {Radford} S.~J.~E.,   {Barrett} J.~W.,  1997,
  \mn@doi [\apj] {10.1086/303765}, \href
  {https://ui.adsabs.harvard.edu/abs/1997ApJ...478..144S} {478, 144}

\bibitem[\protect\citeauthoryear{{Speagle}, {Steinhardt}, {Capak}  \&
  {Silverman}}{{Speagle} et~al.}{2014}]{speagle14}
{Speagle} J.~S.,  {Steinhardt} C.~L.,  {Capak} P.~L.,   {Silverman} J.~D.,
  2014, \mn@doi [\apjs] {10.1088/0067-0049/214/2/15}, \href
  {https://ui.adsabs.harvard.edu/abs/2014ApJS..214...15S} {214, 15}

\bibitem[\protect\citeauthoryear{{Spilker} et~al.,}{{Spilker}
  et~al.}{2014}]{spilker14}
{Spilker} J.~S.,  et~al., 2014, \mn@doi [\apj] {10.1088/0004-637X/785/2/149},
  \href {https://ui.adsabs.harvard.edu/abs/2014ApJ...785..149S} {785, 149}

\bibitem[\protect\citeauthoryear{{Stacey}, {Hailey-Dunsheath}, {Ferkinhoff},
  {Nikola}, {Parshley}, {Benford}, {Staguhn}  \& {Fiolet}}{{Stacey}
  et~al.}{2010}]{stacey10}
{Stacey} G.~J.,  {Hailey-Dunsheath} S.,  {Ferkinhoff} C.,  {Nikola} T.,
  {Parshley} S.~C.,  {Benford} D.~J.,  {Staguhn} J.~G.,   {Fiolet} N.,  2010,
  \mn@doi [\apj] {10.1088/0004-637X/724/2/957}, \href
  {https://ui.adsabs.harvard.edu/abs/2010ApJ...724..957S} {724, 957}

\bibitem[\protect\citeauthoryear{{Stach} et~al.,}{{Stach}
  et~al.}{2018}]{stach18}
{Stach} S.~M.,  et~al., 2018, \mn@doi [\apj] {10.3847/1538-4357/aac5e5}, \href
  {https://ui.adsabs.harvard.edu/#abs/2018ApJ...860..161S} {860, 161}

\bibitem[\protect\citeauthoryear{{Stach} et~al.,}{{Stach}
  et~al.}{2019}]{stach19}
{Stach} S.~M.,  et~al., 2019, \mn@doi [\mnras] {10.1093/mnras/stz1536}, \href
  {https://ui.adsabs.harvard.edu/abs/2019MNRAS.487.4648S} {487, 4648}

\bibitem[\protect\citeauthoryear{{Strandet} et~al.,}{{Strandet}
  et~al.}{2016}]{strandet16}
{Strandet} M.~L.,  et~al., 2016, \mn@doi [\apj] {10.3847/0004-637X/822/2/80},
  \href {https://ui.adsabs.harvard.edu/abs/2016ApJ...822...80S} {822, 80}

\bibitem[\protect\citeauthoryear{{Swinbank}, {Chapman}, {Smail}, {Lindner},
  {Borys}, {Blain}, {Ivison}  \& {Lewis}}{{Swinbank} et~al.}{2006}]{swinbank06}
{Swinbank} A.~M.,  {Chapman} S.~C.,  {Smail} I.,  {Lindner} C.,  {Borys} C.,
  {Blain} A.~W.,  {Ivison} R.~J.,   {Lewis} G.~F.,  2006, \mn@doi [\mnras]
  {10.1111/j.1365-2966.2006.10673.x}, \href
  {https://ui.adsabs.harvard.edu/abs/2006MNRAS.371..465S} {371, 465}

\bibitem[\protect\citeauthoryear{{Swinbank} et~al.,}{{Swinbank}
  et~al.}{2010}]{swinbank10}
{Swinbank} A.~M.,  et~al., 2010, \mn@doi [\nat] {10.1038/nature08880}, \href
  {https://ui.adsabs.harvard.edu/abs/2010Natur.464..733S} {464, 733}

\bibitem[\protect\citeauthoryear{{Swinbank} et~al.,}{{Swinbank}
  et~al.}{2012}]{swinbank12}
{Swinbank} A.~M.,  et~al., 2012, \mn@doi [\mnras]
  {10.1111/j.1365-2966.2012.22048.x}, \href
  {https://ui.adsabs.harvard.edu/abs/2012MNRAS.427.1066S} {427, 1066}

\bibitem[\protect\citeauthoryear{{Swinbank} et~al.,}{{Swinbank}
  et~al.}{2014}]{swinbank14}
{Swinbank} A.~M.,  et~al., 2014, \mn@doi [\mnras] {10.1093/mnras/stt2273},
  \href {https://ui.adsabs.harvard.edu/abs/2014MNRAS.438.1267S} {438, 1267}

\bibitem[\protect\citeauthoryear{{Tacconi} et~al.,}{{Tacconi}
  et~al.}{2006}]{tacconi06}
{Tacconi} L.~J.,  et~al., 2006, \mn@doi [\apj] {10.1086/499933}, \href
  {https://ui.adsabs.harvard.edu/abs/2006ApJ...640..228T} {640, 228}

\bibitem[\protect\citeauthoryear{{Tacconi} et~al.,}{{Tacconi}
  et~al.}{2008}]{tacconi08}
{Tacconi} L.~J.,  et~al., 2008, \mn@doi [\apj] {10.1086/587168}, \href
  {https://ui.adsabs.harvard.edu/abs/2008ApJ...680..246T} {680, 246}

\bibitem[\protect\citeauthoryear{{Tacconi} et~al.,}{{Tacconi}
  et~al.}{2010}]{tacconi10}
{Tacconi} L.~J.,  et~al., 2010, \mn@doi [\nat] {10.1038/nature08773}, \href
  {https://ui.adsabs.harvard.edu/abs/2010Natur.463..781T} {463, 781}

\bibitem[\protect\citeauthoryear{{Tacconi} et~al.,}{{Tacconi}
  et~al.}{2018}]{tacconi18}
{Tacconi} L.~J.,  et~al., 2018, \mn@doi [\apj] {10.3847/1538-4357/aaa4b4},
  \href {https://ui.adsabs.harvard.edu/abs/2018ApJ...853..179T} {853, 179}

\bibitem[\protect\citeauthoryear{{Toft} et~al.,}{{Toft} et~al.}{2014}]{toft14}
{Toft} S.,  et~al., 2014, \mn@doi [\apj] {10.1088/0004-637X/782/2/68}, \href
  {https://ui.adsabs.harvard.edu/abs/2014ApJ...782...68T} {782, 68}

\bibitem[\protect\citeauthoryear{{Valentino} et~al.,}{{Valentino}
  et~al.}{2020}]{valentino20}
{Valentino} F.,  et~al., 2020, arXiv e-prints, \href
  {https://ui.adsabs.harvard.edu/abs/2020arXiv200612521V} {p. arXiv:2006.12521}

\bibitem[\protect\citeauthoryear{{Vieira} et~al.,}{{Vieira}
  et~al.}{2013}]{vieria13}
{Vieira} J.~D.,  et~al., 2013, \mn@doi [\nat] {10.1038/nature12001}, \href
  {https://ui.adsabs.harvard.edu/abs/2013Natur.495..344V} {495, 344}

\bibitem[\protect\citeauthoryear{{Walter} et~al.,}{{Walter}
  et~al.}{2012}]{walter12}
{Walter} F.,  et~al., 2012, \mn@doi [\nat] {10.1038/nature11073}, \href
  {https://ui.adsabs.harvard.edu/abs/2012Natur.486..233W} {486, 233}

\bibitem[\protect\citeauthoryear{{Walter} et~al.,}{{Walter}
  et~al.}{2016}]{walter16}
{Walter} F.,  et~al., 2016, \mn@doi [\apj] {10.3847/1538-4357/833/1/67}, \href
  {https://ui.adsabs.harvard.edu/abs/2016ApJ...833...67W} {833, 67}

\bibitem[\protect\citeauthoryear{{Wardlow} et~al.,}{{Wardlow}
  et~al.}{2011}]{wardlow11}
{Wardlow} J.~L.,  et~al., 2011, \mn@doi [\mnras]
  {10.1111/j.1365-2966.2011.18795.x}, \href
  {https://ui.adsabs.harvard.edu/#abs/2011MNRAS.415.1479W} {415, 1479}

\bibitem[\protect\citeauthoryear{{Wardlow} et~al.,}{{Wardlow}
  et~al.}{2018}]{wardlow18}
{Wardlow} J.~L.,  et~al., 2018, \mn@doi [\mnras] {10.1093/mnras/sty1526}, \href
  {https://ui.adsabs.harvard.edu/#abs/2018MNRAS.479.3879W} {479, 3879}

\bibitem[\protect\citeauthoryear{{Wei{\ss}}, {Henkel}, {Downes}  \&
  {Walter}}{{Wei{\ss}} et~al.}{2003}]{weiss03}
{Wei{\ss}} A.,  {Henkel} C.,  {Downes} D.,   {Walter} F.,  2003, \mn@doi [\aap]
  {10.1051/0004-6361:20031337}, \href
  {https://ui.adsabs.harvard.edu/abs/2003A&A...409L..41W} {409, L41}

\bibitem[\protect\citeauthoryear{{Wei{\ss}}, {Ivison}, {Downes}, {Walter},
  {Cirasuolo}  \& {Menten}}{{Wei{\ss}} et~al.}{2009}]{weiss09}
{Wei{\ss}} A.,  {Ivison} R.~J.,  {Downes} D.,  {Walter} F.,  {Cirasuolo} M.,
  {Menten} K.~M.,  2009, \mn@doi [\apjl] {10.1088/0004-637X/705/1/L45}, \href
  {https://ui.adsabs.harvard.edu/abs/2009ApJ...705L..45W} {705, L45}

\bibitem[\protect\citeauthoryear{{Wei{\ss}} et~al.,}{{Wei{\ss}}
  et~al.}{2013}]{weiss13}
{Wei{\ss}} A.,  et~al., 2013, \mn@doi [\apj] {10.1088/0004-637X/767/1/88},
  \href {https://ui.adsabs.harvard.edu/abs/2013ApJ...767...88W} {767, 88}

\bibitem[\protect\citeauthoryear{{Whitaker}, {van Dokkum}, {Brammer}  \&
  {Franx}}{{Whitaker} et~al.}{2012}]{whitaker12}
{Whitaker} K.~E.,  {van Dokkum} P.~G.,  {Brammer} G.,   {Franx} M.,  2012,
  \mn@doi [\apjl] {10.1088/2041-8205/754/2/L29}, \href
  {https://ui.adsabs.harvard.edu/abs/2012ApJ...754L..29W} {754, L29}

\bibitem[\protect\citeauthoryear{{Yan} et~al.,}{{Yan} et~al.}{2010}]{yan10}
{Yan} L.,  et~al., 2010, \mn@doi [\apj] {10.1088/0004-637X/714/1/100}, \href
  {https://ui.adsabs.harvard.edu/abs/2010ApJ...714..100Y} {714, 100}

\bibitem[\protect\citeauthoryear{{da Cunha}, {Charlot}  \& {Elbaz}}{{da Cunha}
  et~al.}{2008}]{dacunha08}
{da Cunha} E.,  {Charlot} S.,   {Elbaz} D.,  2008, \mn@doi [\mnras]
  {10.1111/j.1365-2966.2008.13535.x}, \href
  {https://ui.adsabs.harvard.edu/abs/2008MNRAS.388.1595D} {388, 1595}

\bibitem[\protect\citeauthoryear{{da Cunha} et~al.,}{{da Cunha}
  et~al.}{2015}]{dacunha15}
{da Cunha} E.,  et~al., 2015, \mn@doi [\apj] {10.1088/0004-637X/806/1/110},
  \href {https://ui.adsabs.harvard.edu/abs/2015ApJ...806..110D} {806, 110}

\bibitem[\protect\citeauthoryear{{van der Werf} et~al.,}{{van der Werf}
  et~al.}{2010}]{vanderwerf10}
{van der Werf} P.~P.,  et~al., 2010, \mn@doi [\aap]
  {10.1051/0004-6361/201014682}, \href
  {https://ui.adsabs.harvard.edu/abs/2010A&A...518L..42V} {518, L42}

\makeatother
\end{thebibliography}





\bsp	
\label{lastpage}

\end{document}